\documentclass[format=acmsmall, review=false, screen=true]{acmart}

\usepackage[ruled,norelsize]{algorithm2e} 
\renewcommand{\algorithmcfname}{ALGORITHM}
\SetAlFnt{\small}
\SetAlCapFnt{\small}
\SetAlCapNameFnt{\small}
\SetAlCapHSkip{0pt}
\IncMargin{-\parindent}
\SetKwProg{Fn}{Function}{}{}

\usepackage{graphicx}
\usepackage{subfig}
\usepackage{tikz}
\usetikzlibrary{shapes.misc, positioning}
\usetikzlibrary{fit}
\usepackage{amsmath}
\usepackage{multirow}
\usepackage{booktabs}
\usepackage[retainorgcmds]{IEEEtrantools}
\usepackage{ifthen}

\usepackage{enumitem}
\usepackage{rotating}
\usepackage{array}
\usepackage[T1]{fontenc}

\newcommand{\defn}[1]{\ifmmode\text{\emph{\textbf{#1}}}\else\emph{\textbf{#1}}\fi}
\ifdefined\textln\relax\else\newcommand{\textln}[1]{#1}\fi
\ifdefined\shortcite\relax\else\newcommand{\shortcite}[1]{\cite{#1}}\fi

\newcolumntype{x}[1]{>{\centering\arraybackslash\hspace{0pt}}p{#1}}

\acmJournal{TOPC}
\acmVolume{9}
\acmNumber{4}
\acmArticle{39}
\acmYear{2017}
\acmMonth{9}
\copyrightyear{2016}

\acmDOI{0000001.0000001}

\received{January 2017}
\received[revised]{May 2017}
\received[accepted]{May 2017}

\begin{document}

\title{GPU Multisplit: an extended study of a parallel algorithm}
\author{Saman Ashkiani}
\affiliation{%
  \institution{University of California, Davis}
  \department{Electrical and Computer Engineering}
  \streetaddress{}
  \city{}
  \state{}
  \postcode{}
  \country{}
}
\author{Andrew Davidson}
\affiliation{%
  \institution{University of California, Davis}
  \department{Electrical and Computer Engineering}
  \streetaddress{}
  \city{}
  \state{}
  \postcode{}
  \country{}
}
\author{Ulrich Meyer}
\affiliation{%
  \institution{Goethe-Universit\"{a}t Frankfurt am Main}
  \department{Institute for Computer Science}
  \streetaddress{}
  \city{}
  \state{}
  \postcode{}
  \country{}
}
\author{John D. Owens}
\affiliation{%
  \institution{University of California, Davis}
  \department{Electrical and Computer Engineering}
  \streetaddress{}
  \city{}
  \state{}
  \postcode{}
  \country{}
}

\begin{abstract}
Multisplit is a broadly useful parallel primitive that permutes its input data into contiguous \emph{buckets} or \emph{bins}, where the function that categorizes an element into a bucket is provided by the programmer.
Due to the lack of an efficient multisplit on GPUs, programmers often choose to implement multisplit with a sort.
One way is to first generate an auxiliary array of bucket IDs and then sort input data based on it.
In case smaller indexed buckets possess smaller valued keys, another way for multisplit is to directly sort input data.
Both methods are inefficient and require more work than necessary: the former requires more expensive data movements while the latter spends unnecessary effort in sorting elements within each bucket.
In this work, we provide a parallel model and multiple implementations for the multisplit problem. Our principal focus is multisplit for a small (up to 256) number of buckets.
We use warp-synchronous programming models and emphasize warp-wide communications to avoid branch divergence and reduce memory usage.
We also hierarchically reorder input elements to achieve better coalescing of global memory accesses.
On a GeForce GTX 1080 GPU, we can reach a peak throughput of 18.93 Gkeys/s (or 11.68 Gpairs/s) for a  key-only (or key-value) multisplit.
Finally, we demonstrate how multisplit can be used as a building block for radix sort. In our multisplit-based sort implementation, we achieve comparable performance to the fastest GPU sort routines, sorting 32-bit keys (and key-value pairs) with a throughput of 3.0 G~keys/s (and 2.1 Gpair/s).

\end{abstract}

%
%






\maketitle
\graphicspath{{Figure/}}
\section{Introduction}\label{sec:intro}

This paper studies the multisplit primitive for GPUs.
\footnote{This paper is an extended version of initial results published at PPoPP 2016~\cite{Ashkiani:2016:GM}. The source code is available at \url{https://github.com/owensgroup/GpuMultisplit}.} 
Multisplit divides a set of items (keys or key-value pairs) into contiguous buckets, where each bucket contains items whose keys satisfy a programmer-specified criterion (such as falling into a particular range). Multisplit is broadly useful in a wide range of applications, some of which we will cite later in this introduction. But we begin our story by focusing on one particular example, the delta-stepping formulation of single-source shortest path (SSSP)\@.

The traditional (and work-efficient) serial approach to SSSP is Dijkstra's algorithm~\shortcite{Dijkstra:1959:ANO}, which considers one vertex per iteration---the vertex with the lowest weight. The traditional parallel approach (Bellman-Ford-Moore~\cite{Bang-Jensen:2009:DTA}) considers all vertices on each iteration, but as a result incurs more work than the serial approach. On the GPU, the recent SSSP work of Davidson et al.~\shortcite{Davidson:2014:WPG:nourl} instead built upon the delta-stepping work of Meyer and Sanders~\shortcite{Meyer:2003:DAP}, which on each iteration classifies candidate vertices into \emph{buckets} or \emph{bins} by their weights and then processes the bucket that contains the vertices with the lowest weights. Items within a bucket are unordered and can be processed in any order.

Delta-stepping is a good fit for GPUs. It avoids the inherent serialization of Dijkstra's approach and the extra work of the fully parallel Bellman-Ford-Moore approach. At a high level, delta-stepping divides up a large amount of work into multiple buckets and then processes all items within one bucket in parallel at the same time. How many buckets? Meyer and Sanders describe how to choose a bucket size that is ``large enough to allow for sufficient parallelism and small enough to keep the algorithm work-efficient''~\shortcite{Meyer:2003:DAP}. Davidson et al.\ found that 10 buckets was an appropriate bucket count across their range of datasets. More broadly, for modern parallel architectures, this design pattern is a powerful one: expose just enough parallelism to fill the machine with work, then choose the most efficient algorithm to process that work. (For instance, Hou et al.\ use this strategy in efficient GPU-based tree traversal~\shortcite{Hou:2011:MGS}.)

Once we've decided the bucket count, how do we efficiently classify vertices into buckets? Davidson et al.\ called the necessary primitive \emph{multisplit}. Beyond SSSP, multisplit has significant utility across a range of GPU applications.
Bucketing is a key primitive in one implementation of radix sort on GPUs~\cite{Merrill:2010:RSF}, where elements are reordered iteratively based on a group of their bits in their binary representation%
; as the first step in building a GPU hash table~\cite{Alcantara:2009:RPH:nourl}; in hash-join for relational databases to group low-bit keys~\cite{Diamos:2012:ERA}; in string sort for singleton compaction and elimination~\cite{Deshpande:2013:CGS}; in suffix array construction to organize the lexicographical rank of characters~\cite{Deo:2013:PSA}; in a graphics voxelization pipeline for splitting tiles based on their descriptor (dominant axis)~\cite{Pantaleoni:2011:VAP}; in the shallow stages of $k$-d tree construction~\cite{Wu:2011:SKC}; in Ashari et al.'s sparse-matrix dense-vector multiplication work, which bins rows by length~\cite{Ashari:2014:FSM}; and in probabilistic top-$k$ selection, whose core multisplit operation is three bins around two pivots~\cite{Monroe:2011:RSO}. And while multisplit is a crucial part of each of these and many other GPU applications, it has received little attention to date in the literature. The work we present here addresses this topic with a comprehensive look at efficiently implementing multisplit as a general-purpose parallel primitive.

The approach of Davidson et al.\ to implementing multisplit reveals the need for this focus. If the number of buckets is 2, then a scan-based ``split'' primitive~\cite{Harris:2007:PPS:nourl} is highly efficient on GPUs. Davidson et al.\ built both a 2-bucket (``Near-Far'') and 10-bucket implementation. Because they lacked an efficient multisplit, they were forced to recommend their theoretically-less-efficient 2-bucket implementation:

\begin{quote}\vspace{-2pt}
  The missing primitive on GPUs is a high-performance \emph{multisplit} that separates primitives based on key value (bucket id); in our implementation, we instead use a sort; in the absence of a more efficient multisplit, we recommend utilizing our Near-Far work-saving strategy for most graphs.~\cite[Section~7]{Davidson:2014:WPG:nourl}\vspace{-2pt}
\end{quote}

Like Davidson et al., we could implement multisplit on GPUs with a sort. Recent GPU sorting implementations~\cite{Merrill:2010:RSF} deliver high throughput, but are overkill for the multisplit problem: unlike sort, multisplit has no need to order items within a bucket. In short, sort does more work than necessary. For Davidson et al., reorganizing items into buckets after each iteration with a sort is too expensive: ``the overhead of this reorganization is significant: on average, with our bucketing implementation, the reorganizational overhead takes 82\% of the runtime.''~~\cite[Section~7]{Davidson:2014:WPG:nourl}

In this paper we design, implement, and analyze numerous approaches to multisplit, and make the following contributions:
\begin{itemize}
\item On modern GPUs, ``global'' operations (that require global communication across the whole GPU) are more expensive than ``local'' operations that can exploit faster, local GPU communication mechanisms. Straightforward implementations of multisplit primarily use global operations. Instead, we propose a parallel model under which the multisplit problem can be factored into a sequence of local, global, and local operations better suited for the GPU's memory and computational hierarchies.
\item We show that reducing the cost of global operations, even by significantly increasing the cost of local operations, is critical for achieving the best performance.
  We base our model on a hierarchical divide and conquer, where at the highest level each subproblem is small enough to be easily solved locally in parallel, and at the lowest level we have only a small number of operations to be performed globally.
\item We locally reorder input elements before global operations, trading more work (the reordering) for better memory performance (greater coalescing) for an overall improvement in performance.
\item We promote the warp-level privatization of local resources as opposed to the more traditional thread-level privatization. This decision can contribute to an efficient implementation of our local computations by using warp-synchronous schemes to avoid branch divergence, reduce shared memory usage, leverage warp-wide instructions, and minimize intra-warp communication.
\item We design a novel voting scheme using only binary ballots. We use this scheme to efficiently implement our warp-wide local computations (e.g., histogram computations).
\item We use these contributions to implement a high-performance multisplit targeted to modern GPUs. We then use our multisplit as an effective building block to achieve the following:
  \begin{itemize}
  \item We build an alternate radix sort competitive with CUB (the current fastest GPU sort library). Our implementation is particularly effective with key-value sorts (Section~\ref{subsec:multisplit_sort}).
  \item We demonstrate a significant performance improvement in the delta-stepping formulation of the SSSP algorithm (Section~\ref{sec:app_sssp}).
  \item We build an alternate device-wide histogram procedure competitive with CUB\@. Our implementation is particularly suitable for a small number of bins (Section~\ref{subsec:multisplit_histogram}).
  \end{itemize}
\end{itemize}

\section{Related Work and Background}\label{sec:related}
\subsection{The Graphics Processing Unit (GPU)}
The GPU of today is a highly parallel, throughput-focused programmable processor. GPU programs (``kernels'') launch over a \emph{grid} of numerous \emph{blocks}; the GPU hardware maps blocks to available parallel cores. Each block typically consists of dozens to thousands of individual \emph{threads}, which are arranged into 32-wide \emph{warps}. Warps run under SIMD control on the GPU hardware. While blocks cannot directly communicate with each other within a kernel, threads within a block can, via a user-programmable 48~kB \emph{shared-memory}, and threads within a warp additionally have access to numerous warp-wide instructions. The GPU's global memory (DRAM), accessible to all blocks during a computation, achieves its maximum bandwidth only when neighboring threads access neighboring locations in the memory; such accesses are termed \emph{coalesced}.
In this work, when we use the term ``\emph{global}'', we mean an operation of device-wide scope. Our term ``\emph{local}'' refers to an operation limited to smaller scope (e.g., within a thread, a warp, a block, etc.), which we will specify accordingly. The major difference between the two is the cost of communication: global operations must communicate through global DRAM, whereas local operations can communicate through lower-latency, higher-bandwidth mechanisms like shared memory or warp-wide intrinsics.
Lindholm et al.~\shortcite{Lindholm:2008:NTA} and Nickolls et al.~\shortcite{Nickolls:2008:SPP} provide more details on GPU hardware and the GPU programming model, respectively.

We use NVIDIA's CUDA as our programming language in this work~\cite{NVIDIA:2016:CUDA}. CUDA provides several warp-wide voting and shuffling instructions for intra-warp communication of threads. All threads within a warp can see the result of a user-specified predicate in a bitmap variable returned by \texttt{\_\_ballot(predicate)}~\cite[Ch.~B13]{NVIDIA:2016:CUDA}. Any set bit in this bitmap denotes the predicate  being non-zero for the corresponding thread. Each thread can also access registers from other threads in the same warp with \texttt{\_\_shfl(register\_name, source\_thread)}~\cite[Ch.~B14]{NVIDIA:2016:CUDA}. Other shuffling functions such as \texttt{\_\_shfl\_up()} or \texttt{\_\_shfl\_xor()} use relative addresses to specify the source thread.
In CUDA, threads also have access to some efficient integer intrinsics, e.g., \texttt{\texttt{\_\_popc()}} for counting the number of set bits in a register.

\subsection{Parallel primitive background}
In this paper we leverage numerous standard parallel primitives, which we briefly describe here. A \emph{reduction} inputs a vector of elements and applies a binary associative operator (such as addition) to reduce them to a single element; for instance, sum-reduction simply adds up its input vector.
The \emph{scan} operator takes a vector of input elements and an associative binary operator, and returns an output vector of the same size as the input vector.
In exclusive (resp., inclusive) scan, output location $i$ contains the reduction of input elements 0 to $i-1$ (resp., 0 to $i$).
Scan operations with binary addition as their operator are also known as \emph{prefix-sum}~\cite{Harris:2007:PPS:nourl}.
Any reference to a multi- operator (multi-reduction, multi-scan) refers to running multiple instances of that operator in parallel on separate inputs. \emph{Compaction} is an operation that filters a subset of its input elements into a smaller output array while preserving the order.

\subsection{Multisplit and Histograms}
Many multisplit implementations, including ours, depend heavily on knowledge of the total number of elements within each bucket (bin), i.e., histogram computation.
Previous competitive GPU histogram implementations share a common philosophy: divide the problem into several smaller sized subproblems and assign each subproblem to a thread, where each thread sequentially processes its subproblem and keeps track of its own \emph{privatized} local histogram.
Later, the local histograms are aggregated to produce a globally correct histogram.
There are two common approaches to this aggregation: 1) using atomic operations to correctly add bin counts together (e.g., Shams and Kennedy~\shortcite{Shams:2007:EHA}), 2)~storing per-thread sequential histogram computations and combining them via a global reduction (e.g., Nugteren et al.~\shortcite{Nugteren:2011:HPP}).
The former is suitable when the number of buckets is large; otherwise atomic contention is the bottleneck.
The latter avoids such conflicts by using more memory (assigning exclusive memory units per-bucket and per-thread), then performing device-wide reductions to compute the global histogram.

The hierarchical memory structure of NVIDIA GPUs, as well as NVIDIA's more recent addition of faster but local shared memory atomics (among all threads within a thread block), provides more design options to the programmer.
With these features, the aggregation stage could be performed in multiple rounds from thread-level to block-level and then to device-level (global) results.
Brown et al.~\shortcite{Brown:2012:MFH} implemented both Shams's and Nugteren's 
aforementioned methods, as well as a variation of their own, focusing only on 8-bit data, considering careful optimizations that make the best use of the GPU, including loop unrolling, thread coarsening, and subword parallelism, as well as others.
Recently, NVIDIA's CUDA Unbound (CUB)~\cite{Merrill:2015:CUB} library has included an efficient and consistent histogram implementation that carefully uses a minimum number of shared-memory atomics to combine per-thread privatized histograms per thread-block, followed by aggregation via global atomics. CUB's histogram supports any data type (including multi-channel 8-bit inputs) with any number of bins.

Only a handful of papers have explored multisplit as a standalone primitive. He et al.~\cite{He:2008:RJG} implemented multisplit by reading multiple elements with each thread, sequentially computing their histogram and local offsets (their order among all elements within the same bucket and processed by the same thread), then storing all results (histograms and local offsets) into memory. Next, they performed a device-wide scan operation over these histogram results and scattered each item into its final position. Their main bottlenecks were the limited size of shared memory, an expensive global scan operation, and random non-coalesced memory accesses.%
\footnote{On an NVIDIA 8800 GTX GPU, for 64 buckets, He et al.\ reported 134~Mkeys/sec. As a very rough comparison, our GeForce GTX 1080 GPU has 3.7x the memory bandwidth, and our best 64-bucket implementation runs 126 times faster.}

Patidar~\cite{Patidar:2009:SPD} proposed two methods with a particular focus on a large number of buckets (more than 4k): one based on heavy usage of shared-memory atomic operations (to compute block level histogram and intra-bucket orders), and the other by iterative usage of basic binary split for each bucket (or groups of buckets). Patidar used a combination of these methods in a hierarchical way to get his best results.%
\footnote{On an NVIDIA GTX280 GPU, for 32 buckets, Patidar reported 762~Mkeys/sec. As a very rough comparison, our GeForce GTX 1080 GPU has 2.25x the memory bandwidth, and our best 32-bucket implementation runs 23.5 times faster.}
Both of these multisplit papers focus only on key-only scenarios, while data movements and privatization of local memory become more challenging with key-value pairs.

\section{Multsiplit and Common Approaches}\label{sec:init_approaches}
In this section, we first formally define the multisplit as a primitive algorithm.
Next, we describe some common approaches for performing the multisplit algorithm, which form a baseline for the comparison to our own methods, which we then describe in Section~\ref{sec:algorithm}.

\subsection{The multisplit primitive}\label{sec:multisplit}

We informally characterize multisplit as follows:

\begin{itemize}
\item Input: An unordered set of keys or key-value pairs. ``Values'' that are larger than the size of a pointer use a pointer to the value in place of the actual value.
\item Input: A function, specified by the programmer, that inputs a key and outputs the bucket corresponding to that key (\emph{bucket identifier}).
For example, this function might classify a key into a particular numerical range, or divide keys into prime or composite buckets.
\item Output: Keys or key-value pairs separated into $m$ buckets. Items within each output bucket must be contiguous but are otherwise unordered. Some applications may prefer output order within a bucket that preserves input order; we call these multisplit implementations ``stable''.
\end{itemize}

\noindent
More formally, let $\mathbf{u}$ and $\mathbf{v}$ be vectors of $n$ \emph{key} and \emph{value} elements, respectively.
Altogether $m$ buckets $B_0, B_1, \dots, B_{m-1}$  partition the entire key domain such that each key element uniquely belongs to one and only one bucket.
Let $\defn{f}(\cdot)$ be an arbitrary bucket identifier that assigns a bucket ID to each input key (e.g., $\defn{f}(u_i) = j$ if and only if $u_i \in B_j$).
Throughout this paper, $m$ always refers to the total number of buckets.
For any input key vector, we define \emph{multisplit} as a permutation of that input vector into an output vector. The output vector is densely packed and has two properties: (1) All output elements within the same bucket are stored contiguously in the output vector, and (2) All output elements are stored contiguously in a vector in ascending order by their bucket IDs\@. Optionally, the beginning index of each bucket in the output vector can also be stored in an array of size $m$.
Our main focus in this paper is on 32-bit keys and values (of any data type).

\begin{figure}
  \centering
  \input{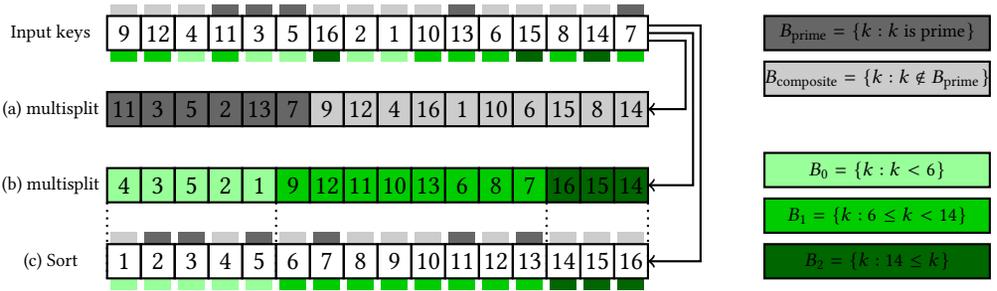}
  \caption{Multisplit examples. (a) Stable multisplit over two buckets ($B_\text{prime}$ and $B_\text{composite}$). (b) Stable multisplit over three range-based buckets ($B_0, B_1, B_2$). (c) Sort can implement multisplit over ordered buckets (e.g., for $B_0, B_1, B_3$), but not for any general buckets (e.g., $B_\text{prime}$ and $B_\text{composite}$); note that this multisplit implementation is not stable (initial intra bucket orders are not preserved).\label{fig:multisplit_example}}
\end{figure}

This multisplit definition allows for a variety of implementations. It places no restrictions on the order of elements within each bucket before and after the multisplit (intra-bucket orders); buckets with larger indices do not necessarily have larger elements. In fact, key elements may not even be comparable entities, e.g., keys can be strings of names with buckets assigned to male names, female names, etc. We do require that buckets are assigned to consecutive IDs and will produce buckets ordered in this way.
Figure~\ref{fig:multisplit_example} illustrates some multisplit examples.
Next, we consider some common approaches for dealing with non-trivial multisplit problems.

\subsection{Iterative and Recursive scan-based splits}\label{sec:scan_split}
The first approach is based on binary split. Suppose we have two buckets. We identify buckets in a binary flag vector, and then compact keys (or key-value pairs) based on the flags. We also compact the complemented binary flags from right to left, and store the results. Compaction can be efficiently implemented by a scan operation, and in practice we can concurrently do both left-to-right and right-to-left compaction with a single scan operation.

With more buckets, we can take two approaches.
One is to iteratively perform binary splits and reduce our buckets one by one. For example, we can first split based on $B_0$ and all remaining buckets ($\cup_{j=1}^{m-1} B_j$). Then we can split the remaining elements based on $B_1$ and $\cup_{j=2}^{m-1} B_j$. After $m$ rounds the result will be equivalent to a multisplit operation.
Another approach is that we can recursively perform binary splits; on each round we split key elements into two groups of buckets.
We continue this process for at most $\lceil \log m\rceil$ rounds and in each round we perform twice number of multisplits and in the end we will have a stable multisplit.
Both of these scan-based splits require multiple global operations (e.g., scan) over all elements, and may also have load-balancing issues if the distribution of keys is non-uniform.
As we will later see in Section~\ref{subsec:perf_references}, on modern GPUs and with just two buckets this approach is not efficient enough.

\subsection{Radix sort}\label{sec:radix}
It should be clear by now that sorting is not a general solution to a multisplit problem.
However, it is possible to achieve a non-stable multisplit by directly sorting our input elements under the following condition: if buckets with larger IDs have larger elements (e.g., all elements in $B_0$ are less than all elements in $B_1$, and so on).
Even in this case, this is not a work-efficient solution as it unnecessarily sorts all elements within each bucket as well.
On average, as the number of buckets ($m$) increases, this performance gap should decrease because there are fewer elements within each bucket and hence less extra effort to sort them. As a result, at some point we expect the multisplit problem to converge to a regular sort problem, when there are large enough number of buckets.

Among all sorting algorithms, there is a special connection between radix sort and multisplit.
Radix sort iteratively sorts key elements based on selected groups of bits in keys. The process either starts from the least significant bits (``LSB sort'') or from the most significant bits (``MSB sort'').
In general MSB sort is more common because, compared to LSB sort, it requires less intermediate data movement when keys vary significantly in length (this is more of an issue for string sorting).
MSB sort ensures data movements become increasingly localized for later  iterations, because keys will not move between buckets (``bucket'' here refers to the group of keys with the same set of considered bits from previous iterations).
However, for equal width key types (such as 32-bit variables, which are our focus in this paper) and with a uniform distribution of keys in the key domain (i.e., an equivalently uniform distribution of bits across keys), there will be less difference between the two methods.

\subsection{Reduced-bit sort}\label{sec:reduced_bit}
Because sorting is an efficient primitive on GPUs, we modify it to be specific to multisplit: here we introduce our \emph{reduced-bit sort} method (RB-sort), which is based on sorting bucket IDs and permuting the original key-value pairs afterward. For multisplit, this method is superior to a full radix sort because we expect the number of significant bits across all bucket IDs is less than the number of significant bits across all keys.
Current efficient GPU radix sorts (such as CUB) provide an option of sorting only a subset of bits in keys. This results in a significant performance improvement for RB-sort, because we only sort bucket IDs (with $\log m$ bits instead of 32-bit keys as in a full radix sort).

\paragraph{Key-only} In this scenario, we first make a \emph{label} vector containing each key's bucket ID\@. Then we sort (label, key) pairs based on label values. Since labels are all less than $m$, we can limit the number of bits in the radix sort to be $\lceil \log m \rceil$.
\paragraph{Key-value} In this scenario, we similarly make a label vector from key elements. Next, we would like to permute (key, value) pairs by sorting labels. One approach is to sort (label, (key, value)) pairs all together, based on label. To do so, we first pack our original key-value pairs into a single 64-bit variable and then do the sort.\footnote{For data types that are larger than 32 bits, we need further modifications for the RB-sort method to work, because it may no longer be possible to pack each key-value pair into a single 64-bit variable and use the current already-efficient 64-bit GPU sorts for it. For such cases, we first sort the array of indexes, then manually permute the arbitrary sized key-value pairs.}
In the end we unpack these elements to form the final results. Another way is to sort (label, index) pairs and then manually permute key-value pairs based on the permuted indices. We tried both approaches and the former seems to be more efficient. The latter requires non-coalesced global memory accesses and gets worse as $m$ increases, while the former reorders for better coalescing internally and scales better with $m$.

The main problem with the reduced-bit sort method is its extra overhead (generating labels, packing original key-value pairs, unpacking the results), which makes the whole process less efficient.
Another inefficiency with the reduced-bit sort method is that it requires more expensive data movements than an ideal solution.
For example, to multisplit on keys only, RB-sort performs a radix sort on (label, key) pairs.

Today's fastest sort primitives do not currently provide APIs for user-specified computations (e.g., bucket identifications) to be integrated as functors directly into sort's kernels; while this is an intriguing area of future work for the designers of sort primitives, we believe that our reduced-bit sort appears to be the best solution today for multisplit using current sort primitives.

\section{Algorithm Overview}\label{sec:algorithm}

In analyzing the performance of methods from the previous section, we make two observations:

\begin{enumerate}
\item Global computations (such as a global scan) are expensive, and approaches to multisplit that require many rounds, each with a global computation, are likely to be uncompetitive. Any reduction in the cost of global computation is desirable.
\item After we derive the permutation, the cost of permuting the elements with a global scatter (consecutive input elements going into arbitrarily distant final destinations) is also expensive.
This is primarily because of the non-coalesced memory accesses associated with the scatter. Any increase in memory locality associated with the scatter is also desirable.
\end{enumerate}

The key design insight in this paper is that we can reduce the cost of both global computation and global scatter at the cost of doing more local work, and that doing so is beneficial for overall performance. We begin by describing and analyzing a framework for the different approaches we study in this paper, then discuss the generic structure common to all our implementations.

\subsection{Our parallel model}\label{subsec:parallel_model}
Multisplit cannot be solved by using only local operations; i.e., we cannot divide a multisplit problem into two independent subparts and solve each part locally without any communication between the two parts. We thus assume any viable implementation must include at least a single global operation to gather necessary global information from all elements (or group of elements).
We generalize the approaches we study in this paper into a series of $N$ rounds, where each round has 3 stages: a set of local operations (which run in parallel on independent subparts of the global problem); a global operation (across all subparts); and another set of local operations. In short: \textbraceleft local, global, local\textbraceright, repeated $N$ times; in this paper we refer to these three stages as \textbraceleft prescan, scan, postscan\textbraceright.

The approaches from Section~\ref{sec:init_approaches} all fit this model. Scan-based split starts by making a flag vector (where the local level is per-thread), performing a global scan operation on all flags, and then ordering the results into their final positions (thread-level local). The iterative (or recursive) scan-based split with $m$ buckets repeats the above approach for $m$ (or $\lceil \log m \rceil$) rounds.
Radix sort also requires several rounds. Each round starts by identifying a bit (or a group of bits) from its keys (local), running a global scan operation, and then locally moving data such that all keys are now sorted based on the selected bit (or group of bits).
In radix sort literature, these stages are mostly known as up-sweep, scan and down-sweep.
Reduced-bit sort is derived from radix sort; the main differences are that in the first round, the label vector and the new packed values are generated locally (thread-level), and in the final round, the packed key-value pairs are locally unpacked (thread-level) to form the final results.

\subsection{Multisplit requires a global computation}
Let's explore the global and local components of stable multisplit, which together compute a unique permutation of key-value pairs into their final positions.
Suppose we have $m$ buckets $B_0, B_1, \dots, B_{m-1}$, each with $h_0, h_1, \dots, h_{m-1}$ elements respectively ($\sum_i{h_i} = n$, where $n$ is the total number of elements).
If $u_i \in B_j$ is the $i$th element in key vector $\mathbf{u}$, then its final permuted position $p(i)$ should be (from $u_i$'s perspective):
\begin{equation}\label{eq:permutation}
        p(i) = \underbrace{\sum_{k = 0}^{j-1}h_k}_{\text{global offset}} + \underbrace{\left| \{u_r \in B_j: r < i\}\right|}_{\text{local offset ($u_i$'s bucket)}},
\end{equation}
where $|\cdot|$ is the cardinality operator that denotes the number of elements within its set argument. The left term is the total number of key elements that belong to the preceding buckets, and the right term is the total number of preceding elements (with respect to $u_i$) in $u_i$'s bucket, $B_j$.
Computing both of these terms in this form and for all elements (for all $i$) requires global operations (e.g., computing a histogram of buckets).

\subsection{Dividing multisplit into subproblems}\label{subsec:2levels}
Equation~\eqref{eq:permutation} clearly shows what we need in order to compute each permutation (i.e., final destinations for a stable multisplit solution): a histogram of buckets among all elements ($h_k$) as well as local offsets for all elements within the same bucket (the second term). However, it lacks intuition about how we should compute each term.
Both terms in equation~\eqref{eq:permutation}, at their core, answer the following question: to which bucket does each key belong? If we answer this question for every key and for all buckets (hypothetically, for each bucket we store a binary bitmap variable of length $n$ to show all elements that belong to that bucket), then each term can be computed intuitively as follows: 1)~histograms are equal to counting all elements in each bucket (reduction of a specific bitmap); 2)~local offsets are equivalent to counting all elements from the beginning to that specific index and within the same bucket (scan operation on a specific bitmap).
This intuition is closely related to our definition of the scan-based split method in Section~\ref{sec:scan_split}. However, it is practically not competitive because it requires storing huge bitmaps (total of $mn$ binary variables) and then performing global operations on them.

Although the above solution seems impractical for a large number of keys, it seems more favorable for input problems that are small enough.
As an extreme example, suppose we wish to perform multisplit on a single key. Each bitmap variable becomes just a single binary bit. Performing reduction and scan operations become as trivial as whether a single bit is set or not.
Thus, a divide-and-conquer approach seems like an appealing solution to solve equation~\eqref{eq:permutation}: we would like to divide our main problem into small enough subproblems such that solving each subproblem is ``easy'' for us. By an easy computation we mean that it is either small enough so that we can afford to process it sequentially, or that instead we can use an efficient parallel hardware alternative (such as the GPU's ballot instruction). When we solve a problem directly in this way, we call it a \emph{direct solve}.
Next, we formalize our divide-and-conquer formulation.

Let us divide our input key vector $\mathbf{u}$ into $L$ contiguous subproblems: $\mathbf{u} = [\mathbf{u}_{0}, \mathbf{u}_{1}, \dots, \mathbf{u}_{L-1}]$. Suppose each subvector $\mathbf{u}_\ell$ has $h_{0,\ell}, h_{1,\ell}, \dots, h_{m-1,\ell}$ elements in buckets $B_0, B_1, \dots B_{m-1}$ respectively.
For example, for arbitrary values of $i$, $s$, and $j$ such that key item $u_i \in \mathbf{u}_s$ and $u_i$ is in bucket $B_j$, equation~\eqref{eq:permutation} can be rewritten as (from $u_i$'s perspective):
\begin{equation}\label{eq:permutation2}
p(i) = \underbrace{\overbrace{\sum_{k=0}^{j-1}\left(\sum_{\ell = 0}^{L-1}h_{k,\ell}\right)}^{\text{previous buckets}}+\overbrace{\sum_{\ell=0}^{s-1}h_{j,\ell}}^\text{$u_i$'s bucket}}_\text{global offset}
 + \underbrace{\left| \{u_r \in \mathbf{u}_s: (u_r \in B_j) \land (r < i)\}\right|}_{\text{local offset within $u_i$'s subproblem}}.
 \end{equation}
This formulation has two separate parts. The first and second terms require global computation (first: the element count of all preceding buckets across all subproblems, and second: the element count of the same bucket in all preceding subproblems). The third term can be computed locally within each subproblem. Note that equation~\eqref{eq:permutation} and~\eqref{eq:permutation2}'s first terms are equivalent (total number of previous buckets), but the second term in~\eqref{eq:permutation} is broken into the second and third terms in~\eqref{eq:permutation2}.

The first and second terms can both be computed with a global histogram computed over $L$ local histograms. A global histogram is generally implemented with global scan operations (here, exclusive prefix-sum). We can characterize this histogram as a scan over a 2-dimensional matrix  $\mathbf{H} = [h_{i,\ell}]_{m\times L}$, where the ``height'' of the matrix is the bucket count $m$ and the ``width'' of the matrix is the number of subproblems $L$\@.
The second term can be computed by a scan operation of size $L$ on each row (total of $m$ scans for all buckets). The first term will be a single scan operation of size $m$ over the reduction of all rows (first reduce each row horizontally to compute global histograms and then scan the results vertically). Equivalently, both terms can be computed by a single scan operation of size $mL$ over a row-vectorized $\mathbf{H}$.
Either way, the cost of our global operation is roughly proportional to both $m$ and $L$\@. We see no realistic way to reduce $m$. Thus we concentrate on reducing $L$\@.

\subsection{Hierarchical approach toward multisplit localization}\label{subsec:localization}
We prefer to have small enough subproblems ($\bar{n}$) so that our local computations are ``easy'' for a direct solve. For any given subproblem size, we will have $L = n/\bar{n}$ subproblems to be processed globally as described before.
On the other hand, we want to minimize our global computations as well, because they require synchronization among all subproblems and involve (expensive) global memory accesses.
So, with a fixed input size and a fixed number of buckets ($n$ and $m$), we would like to both decrease our subproblem size and number of subproblems, which is indeed paradoxical.

Our solution is a hierarchical approach. We do an arbitrary number of levels of divide-and-conquer, until at the last level, subproblems are small enough to be solved easily and directly (our preferred $\bar{n}$).
These results are then appropriately combined together to eventually reach the first level of the hierarchy, where now we have a reasonable number of subproblems to be combined together using global computations (our preferred $L$).

Another advantage of such an approach is that, in case our hardware provides a memory hierarchy with smaller but faster local memory storage (as GPUs do with register level and shared memory level hierarchies, as opposed to the global memory), we can potentially perform all computations related to all levels except the first one in our local memory hierarchies without any global memory interaction.
Ideally, we would want to use all our available register and shared memory with our subproblems to solve them locally, and then combine the results using global operations.
In practice, however, since our local memory storage options are very limited, such solution may still lead to a large number of subproblems to be combined with global operations (large $L$).
As a result, by adding more levels of hierarchy (than the available memory hierarchies in our device) we can systematically organize the way we fill our local memories, process them locally, store intermediate results, and then proceed to the next batch, which overall reduces our global operations.
 Next, we will theoretically consider such a hierarchical approach (\emph{multi-level localization}) and explore the changes to equation~\eqref{eq:permutation2}.

\begin{figure}
  \centering
  \begin{tikzpicture}[every node/.style={thick,rectangle,inner sep=0pt}]
\def \dx {0.1}
\def \dy {0.2}
\def \recx {0.15}
\def \recX {5*\recx + 6*\dx}
\def \recXX {3 * \recX + 26 * \dx}
\def \recy {0.75}
\def \recY {\recy + 3 * \dy}
\def \recYY {\recY + 3 * \dy}

\draw [black] (0 + 0, 0 + 0) rectangle (0 + \recx, 0 + \recy);
\draw [black] (0 + \recx + \dx, 0) rectangle (0 + 2*\recx + \dx, 0 + \recy);
\draw [black] (0 + 2*\recx + 2*\dx, 0 + 0) rectangle (0 + 3*\recx + 2*\dx, 0 + \recy);
\draw [black] (0 + 3*\recx + 3*\dx, 0 + 0) rectangle (0 + 4*\recx + 3*\dx, 0 + \recy);
\draw [black] (0 + 4*\recx + 4*\dx, 0 + 0) rectangle (0 + 5*\recx + 4*\dx, 0 + \recy);

\draw[thick, black] (0-2*\dx, 0-\dy) rectangle (0-2*\dx + \recX + 2 * \dx, 0-\dy + \recY);
\node () at (0 + 2 * \recx + 3 * \dx, \recy + \dy) {$L_2$};

\draw [black] (\recX + 4*\dx + 0, 0 + 0) rectangle (\recX + 4*\dx + \recx, 0 + \recy);
\draw [black] (\recX + 4*\dx + \recx + \dx, 0) rectangle (\recX + 4*\dx + 2*\recx + \dx, 0 + \recy);
\draw [black] (\recX + 4*\dx + 2*\recx + 2*\dx, 0 + 0) rectangle (\recX + 4*\dx + 3*\recx + 2*\dx, 0 + \recy);
\draw [black] (\recX + 4*\dx + 3*\recx + 3*\dx, 0 + 0) rectangle (\recX + 4*\dx + 4*\recx + 3*\dx, 0 + \recy);
\draw [black] (\recX + 4*\dx + 4*\recx + 4*\dx, 0 + 0) rectangle (\recX + 4*\dx + 5*\recx + 4*\dx, 0 + \recy);

\draw[thick, black] (\recX + 4*\dx-2*\dx, 0-\dy) rectangle (\recX + 4*\dx-2*\dx + \recX + 2 * \dx, 0-\dy + \recY);
\node () at (\recX + 4*\dx + 2 * \recx + 3 * \dx, \recy + \dy) {$L_2$};

\draw [black] (2*\recX + 14*\dx + 0, 0 + 0) rectangle (2*\recX + 14*\dx + \recx, 0 + \recy);
\draw [black] (2*\recX + 14*\dx + \recx + \dx, 0) rectangle (2*\recX + 14*\dx + 2*\recx + \dx, 0 + \recy);
\draw [black] (2*\recX + 14*\dx + 2*\recx + 2*\dx, 0 + 0) rectangle (2*\recX + 14*\dx + 3*\recx + 2*\dx, 0 + \recy);
\draw [black] (2*\recX + 14*\dx + 3*\recx + 3*\dx, 0 + 0) rectangle (2*\recX + 14*\dx + 4*\recx + 3*\dx, 0 + \recy);
\draw [black] (2*\recX + 14*\dx + 4*\recx + 4*\dx, 0 + 0) rectangle (2*\recX + 14*\dx + 5*\recx + 4*\dx, 0 + \recy);

\draw[thick, black] (2*\recX + 14*\dx-2*\dx, 0-\dy) rectangle (2*\recX + 14*\dx-2*\dx + \recX + 2 * \dx, 0-\dy + \recY);
\node () at (2*\recX + 14*\dx + 2 * \recx + 3 * \dx, \recy + \dy) {$L_2$};

\draw [thick, black] (0-4*\dx, 0-2*\dy) rectangle (0-4*\dx + \recXX, 0-2*\dy + \recYY);
\node () at (\recX + 4*\dx + 2 * \recx + 3 * \dx, \recy + 3* \dy) {$L_1$};

\draw [black] (\recXX + 4*\dx + 0, 0 + 0) rectangle (\recXX + 4 * \dx + \recx, 0 + \recy);
\draw [black] (\recXX + 4*\dx + \recx + \dx, 0) rectangle (\recXX + 4 * \dx + 2*\recx + \dx, 0 + \recy);
\draw [black] (\recXX + 4*\dx + 2*\recx + 2*\dx, 0 + 0) rectangle (\recXX + 4 * \dx + 3*\recx + 2*\dx, 0 + \recy);
\draw [black] (\recXX + 4*\dx + 3*\recx + 3*\dx, 0 + 0) rectangle (\recXX + 4 * \dx + 4*\recx + 3*\dx, 0 + \recy);
\draw [black] (\recXX + 4*\dx + 4*\recx + 4*\dx, 0 + 0) rectangle (\recXX + 4 * \dx + 5*\recx + 4*\dx, 0 + \recy);

\draw[thick, black] (\recXX + 4*\dx-2*\dx, 0-\dy) rectangle (\recXX + 4 * \dx-2*\dx + \recX + 2 * \dx, 0-\dy + \recY);
\node () at (\recXX + 4*\dx + 3*\dx + 2 * \recx, \recy + \dy) {$L_2$};

\draw [black] (\recXX + 4*\dx + \recX + 4*\dx + 0, 0 + 0) rectangle (\recXX + 4*\dx + \recX + 4*\dx + \recx, 0 + \recy);
\draw [black] (\recXX + 4*\dx + \recX + 4*\dx + \recx + \dx, 0) rectangle (\recXX + 4*\dx + \recX + 4*\dx + 2*\recx + \dx, 0 + \recy);
\draw [black] (\recXX + 4*\dx + \recX + 4*\dx + 2*\recx + 2*\dx, 0 + 0) rectangle (\recXX + 4*\dx + \recX + 4*\dx + 3*\recx + 2*\dx, 0 + \recy);
\fill [green] (\recXX + 4*\dx + \recX + 4*\dx + 3*\recx + 3*\dx, 0 + 0) rectangle (\recXX + 4*\dx + \recX + 4*\dx + 4*\recx + 3*\dx, 0 + \recy);
\draw [black] (\recXX + 4*\dx + \recX + 4*\dx + 3*\recx + 3*\dx, 0 + 0) rectangle (\recXX + 4*\dx + \recX + 4*\dx + 4*\recx + 3*\dx, 0 + \recy);
\draw [black] (\recXX + 4*\dx + \recX + 4*\dx + 4*\recx + 4*\dx, 0 + 0) rectangle (\recXX + 4*\dx + \recX + 4*\dx + 5*\recx + 4*\dx, 0 + \recy);

\draw[thick, black] (\recXX + 4*\dx + \recX + 4*\dx-2*\dx, 0-\dy) rectangle (\recXX + 4*\dx + \recX + 4*\dx-2*\dx + \recX + 2 * \dx, 0-\dy + \recY);
\node () at (\recXX + 4*\dx + \recX + 4*\dx + 2 * \recx + 3 * \dx, \recy + \dy) {$L_2$};

\draw [black] (\recXX + 4*\dx + 2*\recX + 14*\dx + 0, 0 + 0) rectangle (\recXX + 4*\dx + 2*\recX + 14*\dx + \recx, 0 + \recy);
\draw [black] (\recXX + 4*\dx + 2*\recX + 14*\dx + \recx + \dx, 0) rectangle (\recXX + 4*\dx + 2*\recX + 14*\dx + 2*\recx + \dx, 0 + \recy);
\draw [black] (\recXX + 4*\dx + 2*\recX + 14*\dx + 2*\recx + 2*\dx, 0 + 0) rectangle (\recXX + 4*\dx + 2*\recX + 14*\dx + 3*\recx + 2*\dx, 0 + \recy);
\draw [black] (\recXX + 4*\dx + 2*\recX + 14*\dx + 3*\recx + 3*\dx, 0 + 0) rectangle (\recXX + 4*\dx + 2*\recX + 14*\dx + 4*\recx + 3*\dx, 0 + \recy);
\draw [black] (\recXX + 4*\dx + 2*\recX + 14*\dx + 4*\recx + 4*\dx, 0 + 0) rectangle (\recXX + 4*\dx + 2*\recX + 14*\dx + 5*\recx + 4*\dx, 0 + \recy);

\draw[thick, black] (\recXX + 4*\dx + 2*\recX + 14*\dx-2*\dx, 0-\dy) rectangle (\recXX + 4*\dx + 2*\recX + 14*\dx-2*\dx + \recX + 2 * \dx, 0-\dy + \recY);
\node () at (\recXX + 4*\dx + 2*\recX + 14*\dx + 2 * \recx + 3 * \dx, \recy + \dy) {$L_2$};

\draw [thick, black] (\recXX + 4*\dx + 0-4*\dx, 0-2*\dy) rectangle (\recXX + 4*\dx + 0-4*\dx + \recXX, 0-2*\dy + \recYY);
\node () at (\recXX + 4*\dx + \recX + 4*\dx + 2 * \recx + 3 * \dx, \recy + 3* \dy) {$L_1$};

\draw [thick, black] (0-6*\dx, 0 - 3 * \dy) rectangle (0-6*\dx + 3*\recXX + 12*\dx + 6*\dx, 0-3*\dy + \recYY + 3*\dy);
\node () at (\recXX -2* \dx, \recYY -\dy) {$L_0$};

\end{tikzpicture}
  \caption{An example for our localization terminology. Here we have 3 levels of localization with $L_0 = 2$, $L_1 = 3$ and $L_2 = 5$. For example, the marked subproblem (in green) can be addressed by a tuple $(1,1,3)$ where each index respectively denotes its position in the hierarchical structure.\label{fig:localization}}
\end{figure}
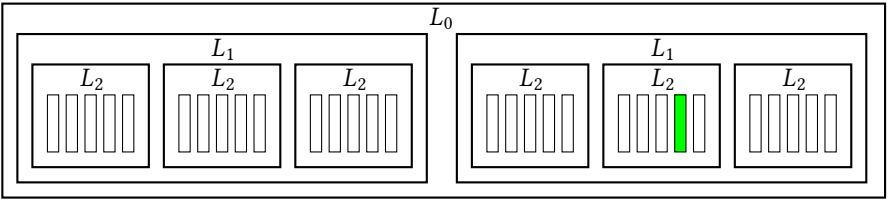

\paragraph{$\lambda$-level localization}
For any given set of arbitrary non-zero integers $\{L_0, L_1, \dots, L_{\lambda - 1}\}$, we can perform $\lambda$ levels of localizations as follows: Suppose we initially divide our problem into $L_0$ smaller parts. These divisions form our first level (i.e., the \emph{global level}).
Next, each subproblem at the first level is divided into $L_1$ smaller subproblems to form the second level. We continue this process until the $\lambda$th level (with $L_{\lambda-1}$ subproblems each). Figure~\ref{fig:localization} shows an example of our hierarchical division of the multisplit problem.
There are total of $L_\text{total} = L_0 \times L_1 \times \dots L_{\lambda - 1}$ smaller problems and their results should be hierarchically added together to compute the final permutation $p(i)$.
Let $(\ell_0,\ell_1,\dots,\ell_{\lambda-1})$ denote a subproblem's position in our hierarchical tree: $\ell_0$th branch from the first level, $\ell_1$th branch from the second level, and so forth until the last level.
Among all elements within this subproblem, we count those that belong to bucket $B_i$ as $h_{i,(\ell_0,\dots,\ell_{\lambda-1})}$.
Similar to our previous permutation computation, for an arbitrary $i$, $j$, and $(s_0,\dots,s_{\lambda-1})$, suppose $u_i \in \mathbf{u}_{(s_0, \dots, s_{\lambda-1})}$ and $u_i \in B_j$.
We can write $p(i)$ as follows (from $u_i$'s perspective):

\begin{IEEEeqnarray}{rCl}\label{eq:permutation3}
p(i) & = & \sum_{k = 0}^{j-1}\left(\sum_{\ell_0 = 0}^{L_0 - 1} \sum_{\ell_1 = 0}^{L_1 - 1} \dots \sum_{\ell_{\lambda - 1} = 0}^{L_{\lambda - 1} - 1}h_{k, (\ell_0, \ell_1, \dots, \ell_{\lambda-1})}\right)
\hfill \rightarrow \text{\scriptsize previous buckets in the whole problem} \nonumber \\
&& +\: \sum_{\ell_0 = 0}^{s_0 - 1}\left(\sum_{\ell_1 = 0}^{L_1-1}\dots\sum_{\ell_{\lambda-1}=0}^{L_{\lambda-1} - 1}{h_{j,(\ell_0,\ell_1,\dots,\ell_{\lambda-1})}}\right)
\hfill\rightarrow \text{\scriptsize $u_i$'s bucket, first level, previous subproblems} \nonumber \\
&& +\: \sum_{\ell_1 = 0}^{s_1 - 1}\left(\sum_{\ell_2 = 0}^{L_2-1}\dots\sum_{\ell_{\lambda-1}=0}^{L_{\lambda-1} - 1}{h_{j,(s_0,\ell_1,\ell_2,\dots,\ell_{\lambda-1})}}\right)
\hfill\rightarrow \text{\scriptsize $u_i$'s bucket, second level, previous subproblems} \nonumber \\
&& \vdots \hfill\rightarrow \text{\scriptsize $u_i$'s bucket, previous levels, previous subproblems} \nonumber \\
&& +\: \sum_{\ell_{\lambda-1} = 0}^{s_{\lambda-1} - 1}h_{j,(s_0,\dots,s_{\lambda-2},\ell_{\lambda-1})}
\hfill\rightarrow \text{\scriptsize $u_i$'s bucket, last level, previous subproblems} \nonumber \\
&& +\: \left| \left\{u_r \in \mathbf{u}_{(s_0, \dots, s_{\lambda-1})}: (u_r \in B_j) \land (r < i)\right\}\right|.
\hfill\rightarrow \text{\scriptsize $u_i$'s bucket, $u_i$'s subproblem} \nonumber \\
\end{IEEEeqnarray}

There is an important resemblance between this equation and equation~\eqref{eq:permutation2}. The first and second terms (from top to bottom) are similar, with the only difference that each $h_{k,\ell}$ is now further broken into $L_1\times \dots \times L_{\lambda-1}$ subproblems (previously it was just $L = L_0$ subproblems). The other terms of \eqref{eq:permutation3} can be seen as a hierarchical disintegration of the local offset in \eqref{eq:permutation2}.

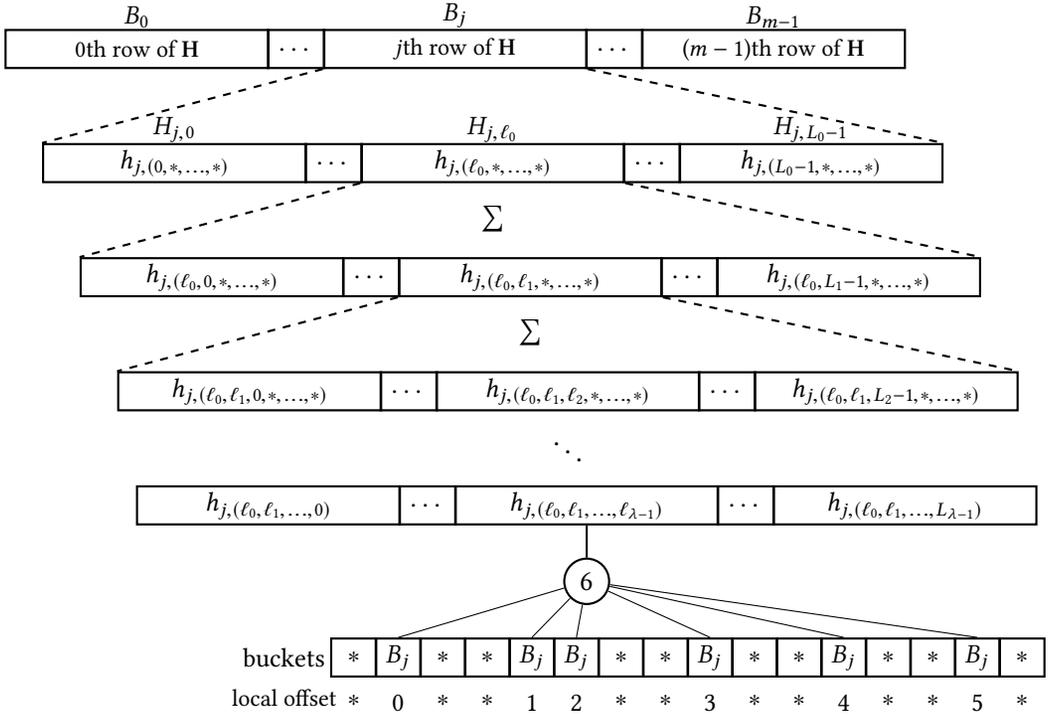
\begin{figure}
  \centering
  \begin{tikzpicture}[every node/.style={thick,rectangle,inner sep=0pt}]
\def\recB{3.5}
\def\recX{3.5}
\def\recXX{3.5}
\def\recXXX{3.5}
\def\aXX{0.5}
\def\recY{0.5}
\def\midX{0.75}
\def \a {0.6}
\node (d1) at (0 * \recB, 1.5) [draw=black,rectangle,minimum width = \recB cm, minimum height = \recY cm] {\small$0$th row of $\mathbf{H}$};
\node (md1) [right=-1pt of d1, draw=black,rectangle,minimum width = \midX cm, minimum height = \recY cm] {$\dots$};
\node (d2) [right=-1pt of md1,draw=black,rectangle,minimum width = \recB cm, minimum height = \recY cm] {\small$j$th row of $\mathbf{H}$};
\node (md2) [right=-1pt of d2, draw=black,rectangle,minimum width = \midX cm, minimum height = \recY cm] {$\dots$};
\node (d3) [right=-1pt of md2,draw=black,rectangle,minimum width = \recB cm, minimum height = \recY cm] {\small($m-1$)th row of $\mathbf{H}$};
\node at (d1.north) [anchor=south] {$B_0$};
\node at (d2.north) [anchor=south] {$B_j$};
\node at (d3.north) [anchor=south] {$B_{m-1}$};

\node (a1) at (0 * \recX + \aXX, 0) [draw=black,rectangle,minimum width = \recX cm, minimum height = \recY cm] {$h_{j, (0, \ast, \dots, \ast)}$};
\node (m1) [right=-1pt of a1, draw=black,rectangle,minimum width = \midX cm, minimum height = \recY cm] {$\dots$};
\node (a2) [right=-1pt of m1,draw=black,rectangle,minimum width = \recX cm, minimum height = \recY cm] {$h_{j, (\ell_0, \ast, \dots, \ast)}$};
\node (m2) [right=-1pt of a2, draw=black,rectangle,minimum width = \midX cm, minimum height = \recY cm] {$\dots$};
\node (a3) [right=-1pt of m2,draw=black,rectangle,minimum width = \recX cm, minimum height = \recY cm] {$h_{j, (L_{0} - 1, \ast, \dots, \ast)}$};

\node (sum1) [below=8pt of a2] {\large$\sum$};
\node at (a1.north) [anchor=south] {$H_{j,0}$};
\node at (a2.north) [anchor=south] {$H_{j,\ell_0}$};
\node at (a3.north) [anchor=south] {$H_{j, L_0-1}$};

\node (b1) at (0 * \recXX + 2*\aXX, -1.5) [draw=black,rectangle,minimum width = \recXX cm, minimum height = \recY cm] {$h_{j, (\ell_0, 0, \ast, \dots, \ast)}$};
\node (mb1) [right=-1pt of b1, draw=black,rectangle,minimum width = \midX cm, minimum height = \recY cm] {$\dots$};
\node (b2) [right=-1pt of mb1,draw=black,rectangle,minimum width = \recXX cm, minimum height = \recY cm] {$h_{j, (\ell_0, \ell_1, \ast, \dots, \ast)}$};
\node (mb2) [right=-1pt of b2, draw=black,rectangle,minimum width = \midX cm, minimum height = \recY cm] {$\dots$};
\node (b3) [right=-1pt of mb2,draw=black,rectangle,minimum width = \recXX cm, minimum height = \recY cm] {$h_{j, (\ell_0, L_{1} - 1, \ast, \dots, \ast)}$};

\node (sum2) [below=8pt of b2] {\large$\sum$};

\draw [thick, black, dashed] (a2.south west) -- (b1.north west);
\draw [thick, black, dashed] (a2.south east) -- (b3.north east);

\node (c1) at (0 * \recXXX + 3*\aXX, -3) [draw=black,rectangle,minimum width = \recXXX cm, minimum height = \recY cm] {$h_{j, (\ell_0, \ell_1, 0,  \ast, \dots, \ast)}$};
\node (mc1) [right=-1pt of c1, draw=black,rectangle,minimum width = \midX cm, minimum height = \recY cm] {$\dots$};
\node (c2) [right=-1pt of mc1,draw=black,rectangle,minimum width = \recXXX cm, minimum height = \recY cm] {$h_{j, (\ell_0, \ell_1, \ell_2,\ast,  \dots, \ast)}$};
\node (mc2) [right=-1pt of c2, draw=black,rectangle,minimum width = \midX cm, minimum height = \recY cm] {$\dots$};
\node (c3) [right=-1pt of mc2,draw=black,rectangle,minimum width = \recXXX cm, minimum height = \recY cm] {$h_{j, (\ell_0, \ell_1, L_{2} - 1, \ast, \dots, \ast)}$};

\draw [thick, black, dashed] (b2.south west) -- (c1.north west);
\draw [thick, black, dashed] (b2.south east) -- (c3.north east);

\draw [thick, black, dashed] (d2.south west) -- (a1.north west);
\draw [thick, black, dashed] (d2.south east) -- (a3.north east);

\node (dots1) [below=4pt of c2] {\large$\ddots$};

\node (e1) at (0 * \recXXX + 3.5*\aXX, -4.5) [draw=black,rectangle,minimum width = \recXXX cm, minimum height = \recY cm] {$h_{j, (\ell_0, \ell_1, \dots, 0)}$};
\node (me1) [right=-1pt of e1, draw=black,rectangle,minimum width = \midX cm, minimum height = \recY cm] {$\dots$};
\node (e2) [right=-1pt of me1,draw=black,rectangle,minimum width = \recXXX cm, minimum height = \recY cm] {$h_{j, (\ell_0, \ell_1, \dots, \ell_{\lambda-1})}$};
\node (me2) [right=-1pt of e2, draw=black,rectangle,minimum width = \midX cm, minimum height = \recY cm] {$\dots$};
\node (e3) [right=-1pt of me2,draw=black,rectangle,minimum width = \recXXX cm, minimum height = \recY cm] {$h_{j, (\ell_0, \ell_1, \dots, L_{\lambda-1})}$};

\node (sum3) [below=12pt of e2,draw,circle, inner sep=3pt]{\large $6$};

\def\inputKeys{{"*","B_j","*","*","B_j","B_j","*","*","B_j","*","*","B_j","*","*","B_j","*"}}

\def\localOffset{{"*","0","*","*","1","2","*","*","3","*","*","4","*","*","5","*"}}

\node (A) at (0 * \recXXX + 4*\aXX, -6.5) [minimum height = \recY cm] {\text{buckets    }};
\node (l) [below=0pt of A, minimum height = \recY cm] {\small\text{local offset  }};
\foreach \x in {0,...,15}{
	\node (A) [right=-1pt of A, draw=black,rectangle,minimum width = \a cm, minimum height = \recY cm] {$\pgfmathparse{\inputKeys[\x]}\pgfmathresult$};
	\node (l) [below=6pt of A] {$\pgfmathparse{\localOffset[\x]}\pgfmathresult$};
	\ifthenelse{\x=1 \OR \x=4 \OR \x=5 \OR \x=8 \OR \x=11 \OR \x=14}{\draw(A.north)--(sum3);}{}	
}

\draw [thick,black] (sum3.north) -- (e2.south);
\end{tikzpicture}
  \caption{Each row of $\mathbf{H}$ belongs to a different bucket. Results from different subproblems in different levels are added together to form a lower level bucket count. This process is continued until reaching the first level where $\mathbf{H}$ is completed.\label{fig:localization_matrix}}
\end{figure}

Similar to Section~\ref{subsec:2levels}, we can form a matrix $\mathbf{H} = [h_{j,\ell_0}]_{m\times L_0}$ for global computations where
\begin{equation}
h_{j,\ell_0} = \sum_{\ell_1 = 0}^{L_1-1}\dots\sum_{\ell_{\lambda-1} = 0}^{L_{\lambda-1}-1} h_{j, (\ell_0, \ell_1,\dots,\ell_{\lambda-1})}.
\end{equation}
Figure~\ref{fig:localization_matrix} depicts an schematic example of multiple levels of localization.
At the highest level, for any arbitrary subproblem $(\ell_0,\ell_1\dots,\ell_{\lambda-1})$, local offsets per key are computed as well as all bucket counts. Bucket counts are then summed and sent to a lower level to form the bucket count for more subproblems ($L_{\lambda-1}$ consecutive subproblems). This process is continued until reaching the first level (the global level) where we have bucket counts for each $L_1\times\dots\times L_{\lambda-1}$ consecutive subproblems. This is where $\mathbf{H}$ is completed, and we can proceed with our global computation. Next, we discuss the way we compute local offsets.

\subsection{Direct solve: Local offset computation}
At the very last level of our localization,
each element must compute its own local offset, which represents the number of elements in its subproblem (with our preferred size $\bar{n}$) that both precede it and share its bucket.
To compute local offsets of a subproblem of size $\bar{n}$, we  make a new binary matrix $\bar{\mathbf{H}}_{m\times \bar{n}}$, where each row represents a bucket and each column represents a key element.
Each entry of this new matrix is one if the corresponding key element belongs to that bucket, and zero otherwise.
Then by performing an exclusive scan on each row, we can compute local offsets for all elements belonging to that row (bucket). So each subproblem requires the following computations:
\begin{enumerate}
        \item Mark all elements in each bucket (making local $\bar{\mathbf{H}}$)
        \item $m$ local reductions over the rows of $\bar{\mathbf{H}}$ to compute local histograms (a column in $\mathbf{H}$)
        \item $m$ local exclusive scans on rows of $\bar{\mathbf{H}}$ (local offsets)
\end{enumerate}
For clarity, we separate steps 2 and 3 above, but we can achieve both with a single local scan operation.
Step 2 provides all histogram results that we need in equations~\eqref{eq:permutation2} or \eqref{eq:permutation3} (i.e., all $h_{k, (\ell_0, \dots, \ell_{\lambda-1})}$ values) and step 3 provides the last term in either equation (Fig.~\ref{fig:localization_matrix}).

It is interesting to note that as an extreme case of localization, we can have $L = n$ subproblems, where we divide our problem so much that in the end each subproblem is a single element. In such case, $\bar{\mathbf{H}}$ is itself a binary value. Thus, step 2's result is either 0 or 1. The local offset (step 3) for such a singleton matrix is always a zero (there is no other element within that subproblem).

\subsection{Our multisplit algorithm}

Now that we've outlined the different computations required for the multisplit, we can present a high-level view of the algorithmic skeleton we use in this paper. We require three steps:

\begin{enumerate}
\item \emph{Local}. For each subproblem at the highest level of our localization, for each bucket, count the number of items in the subproblem that fall into that bucket (direct solve for bucket counts).
Results are then combined hierarchically and locally (based on equation~\eqref{eq:permutation3}) until we have bucket counts per subproblem for the first level (global level).
\item \emph{Global}. Scan the bucket counts for each bucket across all subproblems in the first level (global level), then scan the bucket totals. Each subproblem now knows both a)~for each bucket, the total count for all its previous buckets across the whole input vector (term 1 in equation~\eqref{eq:permutation3}) and b)~for each bucket, the total count from the previous subproblems  (term 2 in equation~\eqref{eq:permutation3}).
\item \emph{Local}. For each subproblem at the highest level of our localization, for each item, recompute bucket counts and compute the local offset for that item's bucket (direct solve).
Local results for each level are then appropriately combined together with the global results from the previous levels (based on equation~\eqref{eq:permutation3}) to compute final destinations.
We can now write each item in parallel into its location in the output vector.
\end{enumerate}

\subsection{Reordering elements for better locality}\label{subsec:alg_reordering}
After computing equation~\eqref{eq:permutation3} for each key element, we can move key-value pairs to their final positions in global memory. However, in general, any two consecutive key elements in the original input do not belong to the same bucket, and thus their final destination might be far away from each other (i.e., a global scatter).
Thus, when we write them back to memory, our memory writes are poorly coalesced, and our achieved memory bandwidth during this global scatter is similarly poor. This results in a huge performance bottleneck. How can we increase our coalescing and thus the memory bandwidth of our final global scatter?
Our solution is to \emph{reorder} our elements within a subproblem at the lowest level (or any other higher level) before they are scattered back to memory.
Within a subproblem, we attempt to place elements from the same bucket next to each other, while still preserving order within a bucket (and thus the stable property of our multisplit implementation). We do this reordering at the same time we compute local offsets in equation~\eqref{eq:permutation2}. How do we group elements from the same bucket together? A local multisplit within the subproblem!

We have already computed histogram and local offsets for each element in each subproblem. We only need to perform another local exclusive scan on local histogram results to compute new positions for each element in its subproblem (computing equation~\eqref{eq:permutation} for each subproblem).
We emphasize that performing this additional stable multisplit on each subproblem does \emph{not} change its histogram and local offsets, and hence does not affect any of our computations described previously from a global perspective; the final multisplit result is identical. But, it has a significant positive impact on the locality of our final data writes to global memory.

It is theoretically better for us to perform reordering in our largest subproblems (first level) so that there are potentially more candidate elements that might have consecutive/nearby final destinations.
However, in practice, we may prefer to reorder elements in higher levels, not because they provide better locality but for purely practical limitations (such as limited available local memory to contain all elements within that subproblem).



\section{Implementation Details}\label{sec:impl_details}
So far we have discussed our high level ideas for implementing an efficient multisplit algorithm for GPUs. In this section we thoroughly describe our design choices and implementation details.
We first discuss existing memory and computational hierarchies in GPUs, and conventional localization options available on such devices.
Then we discuss traditional design choices for similar problems such as multisplit, histogram, and radix sort.
We follow this by our own design choices and how they differ from previous work.
Finally, we propose three implementation variations of multisplit, each with its own localization method and computational details.

\subsection{GPU memory and computational hierarchies}\label{subsec:gpu_hierarchy}
As briefly discussed in Section~\ref{sec:related}, GPUs offer three main memory storage options: 1)~registers dedicated to each thread, 2)~shared memory dedicated to all threads within a thread-block, 3)~global memory accessible by all threads in the device.\footnote{There are other types of memory units in GPUs as well, such as local, constant, and texture memory. However, these are in general special-purpose memories and hence we have not targeted them in our design.}
From a computational point of view there are two main computational units: 1)~threads have direct access to arithmetic units and perform register-based computations, 2)~all threads within a warp can perform a limited but useful set of hardware based warp-wide intrinsics (e.g., ballots, shuffles, etc.). Although the latter is not physically a new computational unit, its inter-register communication among threads opens up new computational capabilities (such as parallel voting).

Based on memory and computational hierarchies discussed above, there are four primary ways of solving a problem on the GPU: 1)~thread-level, 2)~warp-level, 3)~block-level, and 4)~device-level (global).
Traditionally, most efficient GPU programs for multisplit, histogram and radix sort~\cite{He:2008:RJG, Bell:2011:TAP, Merrill:2015:CUB} start from  thread-level computations, where each thread processes a group of input elements and performs local computations (e.g., local histograms).
These thread-level results are then usually combined to form a block-level solution, usually to benefit from the block's shared memory.
Finally, block-level results are combined together to form a global solution.
If implemented efficiently, these methods are capable of achieving high-quality performance from available GPU resources (e.g., CUB's high efficiency in histogram and radix sort).

In contrast, we advocate another way of solving these problems, based on a warp granularity. We start from a warp-level solution and then proceed up the hierarchy to form a device-wide (global) solution (we may bypass the block-level solution as well).
Consequently, we target two major implementation alternatives to solve our multisplit problem: 1)~warp-level~$\rightarrow$ device-level, 2)~warp-level~$\rightarrow$ block-level~$\rightarrow$ device-level
(in Section~\ref{subsec:ms_privatization}, we discuss the costs and benefits of our approaches compared to a thread-level approach).
Another algorithmic option that we outlined in Section~\ref{subsec:alg_reordering} was to reorder elements to get better (coalesced) memory accesses.
As a result of combining these two sets of alternatives, there will be four possible variations that we can explore.
However, if we neglect reordering, our block-level solution will be identical to our warp-level solution, which leaves us with three main final options that all start with warp-level subproblem solutions and end up with a device-level global solution: 1)~no reordering, 2)~with reordering and bypassing a block-level solution, 3)~with reordering and including a block-level solution.
Next, we describe these three implementations and show how they fit into the multi-level localization model we described in Section~\ref{subsec:localization}.

\subsection{Our proposed multisplit algorithms}
So far we have seen that we can reduce the size and cost of our global operation (size of $\mathbf{H}$) by doing more local work (based on our multi-level localization and hierarchical approach). This is a complex tradeoff, since we prefer a small number of subproblems in our first level (global operations), as well as small enough subproblem sizes in our last levels so that they can easily be solved within a warp.
What remains is to choose the number and size of our localization levels and where to perform reordering. All these design choices should be made based on a set of complicated factors such as available shared memory and registers, achieved occupancy, required computational load, etc.

In this section we describe three novel and efficient multisplit implementations that explore different points in the design space, using the terminology that we introduced in Section~\ref{subsec:localization} and Section~\ref{subsec:gpu_hierarchy}.

\begin{description}
\item[Direct Multisplit] Rather than split the problem into subproblems across threads, as in traditional approaches~\cite{He:2008:RJG}, Direct Multisplit (DMS) splits the problem into subproblems across warps (warp-level approach), leveraging efficient warp-wide intrinsics to perform the local computation.
\item[Warp-level Multisplit] Warp-level Multisplit (WMS) also uses a warp-level approach, but additionally reorders elements within each subproblem for better locality.
\item[Block-level Multisplit] Block-level Multisplit (BMS) modifies WMS to process larger-sized subproblems with a block-level approach that includes reordering, offering a further reduction in the cost of the global step at the cost of considerably more complex local computations.
\end{description}

We now discuss the most interesting aspects of our implementations of these three approaches, separately describing how we divide the problem into smaller pieces (our localization strategies), compute histograms and local offsets for larger subproblem sizes, and reorder final results before writing them to global memory to increase coalescing.
\subsection{Localization and structure of our multisplit}\label{subsec:ms_localization}
In Section~\ref{sec:algorithm} we described our parallel model in solving the multisplit problem.
Theoretically, we would like to both minimize our global computations as well as maximize our hardware utilization.
However, in practice designing an efficient GPU algorithm is more complicated.
There are various factors that need to be considered, and sometimes even be smartly sacrificed in order to satisfy a more important goal: efficiency of the whole algorithm.
For example, we may decide to recompute the same value multiple times in different kernel launches, just so that we do not need to store them in global memory for further reuse.

In our previous work~\cite{Ashkiani:2016:GM}, we implemented our multisplit algorithms with a straightforward localization strategy: Direct and Warp-level Multisplit divided problems into warp-sized subproblems (two levels of localization), and Block-level Multisplit used block-sized subproblems (three levels of localization) to extract more locality by performing more complicated computations needed for reordering.
In order to have better utilization of available resources, we assigned multiple similar tasks to each launched warp/block (so each warp/block processed multiple independent subproblems).
Though this approach was effective, we still faced relatively expensive global computations, and did not extract enough locality from our expensive reordering step.
Both of these issues could be remedied by using larger subproblems within the same localization hierarchy. However, larger subproblems require more complicated computations and put more pressure on the limited available GPU resources (registers, shared memory, memory bandwidth, etc.). Instead, we redesigned our implementations to increase the number of levels of localization. This lets us have larger subproblems, while systematically coordinating our computational units (warps/blocks) and available resources to achieve better results.

\paragraph{Direct Multisplit} Our DMS implementation has three levels of localizations: 
Each warp is assigned to a chunk of consecutive elements (first level).
This chunk is then divided into a set of consecutive windows of warp-width ($N_\text{thread}=32$) size (second level).
For each window, we multisplit without any reordering (third level).
\paragraph{Warp-level Multisplit} WMS is similar to DMS, but it also performs reordering to get better locality. In order to get better resource utilization, we add another level of localization compared to DMS (total of four). Each warp performs reordering over only a number of processed windows (a \emph{tile}), and then continues to process the next tile. In general, each warp is in charge of a chunk of consecutive elements (first level). Each chunk is divided into several consecutive tiles (second level). Each tile is processed by a single warp and reordered by dividing it into several consecutive windows (third level). Each window is then directly processed by warp-wide methods (fourth level).
The reason that we add another level of localization for each tile is simply because we do not have sufficient shared memory per warp to store the entire subproblem.

\paragraph{Block-level Multisplit} BMS has five levels of localization. Each thread-block is in charge of a chunk of consecutive elements (first level). Each chunk is divided into a consecutive number of tiles (second level). Each tile is processed by all warps within a block (third level) and reordering happens in this level. Each warp processes multiple consecutive windows of input data within the tile (fourth level). In the end each window is processed directly by using warp-wide methods (fifth level). Here, for a similar reason as in WMS (limited shared memory), we added another level of localization per tile.

\begin{figure}
  \centering

  \begin{tabular}{ccc}
  \subfloat[DMS localization]{
          \begin{tikzpicture}[every node/.style={thick,rectangle,inner sep=0pt}]

\def \dx {0.05}
\def \dy {0.2}
\def \recx {0.15}
\def \recX {9*\recx + 10*\dx}
\def \recy {0.75}
\def \recY {\recy + 3 * \dy}

\draw [black] (0 + 0, 0 + 0) rectangle (0 + \recx, 0 + \recy);
\draw [black] (0 + \recx + \dx, 0) rectangle (0 + 2*\recx + \dx, 0 + \recy);
\draw [black] (0 + 2*\recx + 2*\dx, 0 + 0) rectangle (0 + 3*\recx + 2*\dx, 0 + \recy);
\draw [black] (0 + 3*\recx + 3*\dx, 0 + 0) rectangle (4*\recx + 3*\dx, 0 + \recy);
\draw [black] (0 + 4*\recx + 4*\dx, 0 + 0) rectangle (5*\recx + 4*\dx, 0 + \recy);
\draw [black] (0 + 5*\recx + 5*\dx, 0 + 0) rectangle (6*\recx + 5*\dx, 0 + \recy);
\draw [black] (0 + 6*\recx + 6*\dx, 0 + 0) rectangle (7*\recx + 6*\dx, 0 + \recy);
\draw [black] (0 + 7*\recx + 7*\dx, 0 + 0) rectangle (8*\recx + 7*\dx, 0 + \recy);
\draw [black] (0 + 8*\recx + 8*\dx, 0 + 0) rectangle (9*\recx + 8*\dx, 0 + \recy);

\draw[thick, black] (0-13*\dx, 0-\dy) rectangle (0-13*\dx + \recX + 23 * \dx, 0-\dy + \recY);
\node () at (0 + 5 * \recx + 2 * \dx, \recy + \dy) {\small Subproblem per warp};

\end{tikzpicture}
  } &
  \subfloat[WMS localization]{
                \begin{tikzpicture}[every node/.style={thick,rectangle,inner sep=0pt}]

\def \dx {0.05}
\def \dy {0.2}
\def \recx {0.15}
\def \recX {3*\recx + 4*\dx}
\def \recXX {3 * \recX + 22 * \dx}
\def \recy {0.75}
\def \recY {\recy + 3 * \dy}
\def \recYY {\recY + 3 * \dy}

\draw [black] (0 + 0, 0 + 0) rectangle (0 + \recx, 0 + \recy);
\draw [black] (0 + \recx + \dx, 0) rectangle (0 + 2*\recx + \dx, 0 + \recy);
\draw [black] (0 + 2*\recx + 2*\dx, 0 + 0) rectangle (0 + 3*\recx + 2*\dx, 0 + \recy);

\draw[thick, black] (0-2*\dx, 0-\dy) rectangle (0-2*\dx + \recX + 2 * \dx, 0-\dy + \recY);
\node () at (0 + 1 * \recx + 2 * \dx, \recy + \dy) {\tiny tile 0};

\draw [black] (\recX + 4*\dx + 0, 0 + 0) rectangle (\recX + 4*\dx + \recx, 0 + \recy);
\draw [black] (\recX + 4*\dx + \recx + \dx, 0) rectangle (\recX + 4*\dx + 2*\recx + \dx, 0 + \recy);
\draw [black] (\recX + 4*\dx + 2*\recx + 2*\dx, 0 + 0) rectangle (\recX + 4*\dx + 3*\recx + 2*\dx, 0 + \recy);

\draw[thick, black] (\recX + 4*\dx-2*\dx, 0-\dy) rectangle (\recX + 4*\dx-2*\dx + \recX + 2 * \dx, 0-\dy + \recY);
\node () at (\recX + 4*\dx + 1 * \recx + 2 * \dx, \recy + \dy) {\tiny tile 1};

\draw [black] (2*\recX + 12*\dx + 0, 0 + 0) rectangle (2*\recX + 12*\dx + \recx, 0 + \recy);
\draw [black] (2*\recX + 12*\dx + \recx + \dx, 0) rectangle (2*\recX + 12*\dx + 2*\recx + \dx, 0 + \recy);
\draw [black] (2*\recX + 12*\dx + 2*\recx + 2*\dx, 0 + 0) rectangle (2*\recX + 12*\dx + 3*\recx + 2*\dx, 0 + \recy);

\draw[thick, black] (2*\recX + 12*\dx-2*\dx, 0-\dy) rectangle (2*\recX + 12*\dx-2*\dx + \recX + 2 * \dx, 0-\dy + \recY);
\node () at (2*\recX + 12*\dx + 1 * \recx + 2 * \dx, \recy + \dy) {\tiny tile 2};

\draw [thick, black] (0-8*\dx, 0-2*\dy) rectangle (0 + \recXX, 0-2*\dy + \recYY);
\node () at (\recX + 4*\dx + 1 * \recx + 2 * \dx, \recy + 3* \dy) {\small Subproblem per warp};

\end{tikzpicture}
  } &
  \subfloat[BMS localization]{
                \begin{tikzpicture}[every node/.style={thick,rectangle,inner sep=0pt}]
\def \dx {0.05}
\def \dy {0.2}
\def \recx {0.15}
\def \recX {3*\recx + 4*\dx}
\def \recXX {3 * \recX + 22 * \dx}
\def \recy {0.75}
\def \recY {\recy + 3 * \dy}
\def \recYY {\recY + 3 * \dy}

\draw [black] (0 + 0, 0 + 0) rectangle (0 + \recx, 0 + \recy);
\draw [black] (0 + \recx + \dx, 0) rectangle (0 + 2*\recx + \dx, 0 + \recy);
\draw [black] (0 + 2*\recx + 2*\dx, 0 + 0) rectangle (0 + 3*\recx + 2*\dx, 0 + \recy);

\draw[thick, black] (0-2*\dx, 0-\dy) rectangle (0-2*\dx + \recX + 2 * \dx, 0-\dy + \recY);
\node () at (0 + 1 * \recx + 2 * \dx, \recy + \dy) {\tiny warp 0};

\draw [black] (\recX + 4*\dx + 0, 0 + 0) rectangle (\recX + 4*\dx + \recx, 0 + \recy);
\draw [black] (\recX + 4*\dx + \recx + \dx, 0) rectangle (\recX + 4*\dx + 2*\recx + \dx, 0 + \recy);
\draw [black] (\recX + 4*\dx + 2*\recx + 2*\dx, 0 + 0) rectangle (\recX + 4*\dx + 3*\recx + 2*\dx, 0 + \recy);

\draw[thick, black] (\recX + 4*\dx-2*\dx, 0-\dy) rectangle (\recX + 4*\dx-2*\dx + \recX + 2 * \dx, 0-\dy + \recY);
\node () at (\recX + 4*\dx + 1 * \recx + 2 * \dx, \recy + \dy) {\tiny warp 1};

\draw [black] (2*\recX + 12*\dx + 0, 0 + 0) rectangle (2*\recX + 12*\dx + \recx, 0 + \recy);
\draw [black] (2*\recX + 12*\dx + \recx + \dx, 0) rectangle (2*\recX + 12*\dx + 2*\recx + \dx, 0 + \recy);
\draw [black] (2*\recX + 12*\dx + 2*\recx + 2*\dx, 0 + 0) rectangle (2*\recX + 12*\dx + 3*\recx + 2*\dx, 0 + \recy);

\draw[thick, black] (2*\recX + 12*\dx-2*\dx, 0-\dy) rectangle (2*\recX + 12*\dx-2*\dx + \recX + 2 * \dx, 0-\dy + \recY);
\node () at (2*\recX + 12*\dx + 1 * \recx + 2 * \dx, \recy + \dy) {\tiny warp 2};

\draw [thick, black] (0-4*\dx, 0-2*\dy) rectangle (0-4*\dx + \recXX, 0-2*\dy + \recYY);
\node () at (\recX + 4*\dx + 1 * \recx + 2 * \dx, \recy + 3* \dy) {\small tile 0};

\draw [black] (\recXX + 2*\dx + 0, 0 + 0) rectangle (\recXX +2*\dx+ \recx, 0 + \recy);
\draw [black] (\recXX + 2*\dx + \recx + \dx, 0) rectangle (\recXX +2*\dx+ 2*\recx + \dx, 0 + \recy);
\draw [black] (\recXX + 2*\dx + 2*\recx + 2*\dx, 0 + 0) rectangle (\recXX +2*\dx+ 3*\recx + 2*\dx, 0 + \recy);

\draw[thick, black] (\recXX, 0-\dy) rectangle (\recXX + \recX + 2 * \dx, 0-\dy + \recY);
\node () at (\recXX + 2*\dx +  1 * \recx + 2 * \dx, \recy + \dy) {\tiny warp 0};

\draw [black] (\recXX + 2*\dx+\recX + 4*\dx + 0, 0 + 0) rectangle (\recXX + 2*\dx+\recX + 4*\dx + \recx, 0 + \recy);
\draw [black] (\recXX + 2*\dx+\recX + 4*\dx + \recx + \dx, 0) rectangle (\recXX + 2*\dx+\recX + 4*\dx + 2*\recx + \dx, 0 + \recy);
\draw [black] (\recXX + 2*\dx+\recX + 4*\dx + 2*\recx + 2*\dx, 0 + 0) rectangle (\recXX + 2*\dx+\recX + 4*\dx + 3*\recx + 2*\dx, 0 + \recy);

\draw[thick, black] (\recXX + 2*\dx+\recX + 4*\dx-2*\dx, 0-\dy) rectangle (\recXX + 2*\dx+\recX + 4*\dx-2*\dx + \recX + 2 * \dx, 0-\dy + \recY);
\node () at (\recXX + 2*\dx+\recX + 4*\dx + 1 * \recx + 2 * \dx, \recy + \dy) {\tiny warp 1};

\draw [black] (\recXX + 2*\dx + 2*\recX + 12*\dx + 0, 0 + 0) rectangle (\recXX + 2*\dx + 2*\recX + 12*\dx + \recx, 0 + \recy);
\draw [black] (\recXX + 2*\dx + 2*\recX + 12*\dx + \recx + \dx, 0) rectangle (\recXX + 2*\dx + 2*\recX + 12*\dx + 2*\recx + \dx, 0 + \recy);
\draw [black] (\recXX + 2*\dx + 2*\recX + 12*\dx + 2*\recx + 2*\dx, 0 + 0) rectangle (\recXX + 2*\dx + 2*\recX + 12*\dx + 3*\recx + 2*\dx, 0 + \recy);

\draw[thick, black] (\recXX + 2*\dx + 2*\recX + 12*\dx-2*\dx, 0-\dy) rectangle (\recXX + 2*\dx + 2*\recX + 12*\dx-2*\dx + \recX + 2 * \dx, 0-\dy + \recY);
\node () at (\recXX + 2*\dx + 2*\recX + 12*\dx + 1 * \recx + 2 * \dx, \recy + \dy) {\tiny warp 2};

\draw [thick, black] (\recXX-2*\dx, 0-2*\dy) rectangle (\recXX-2*\dx + \recXX, 0-2*\dy + \recYY);
\node () at (\recXX  + 2*\dx + \recX + 4*\dx + 1 * \recx + 2 * \dx, \recy + 3* \dy) {\small tile 1};

\draw [thick, black] (0-6*\dx, 0 - 3 * \dy) rectangle (0-6*\dx + 3*\recXX + 5*\dx, 0-3*\dy + \recYY + 3*\dy);
\node () at (\recXX -2* \dx, \recYY -\dy) {\small Subproblem per block};
\end{tikzpicture}
  }
  \end{tabular}
  \caption{Different localizations for DMS, WMS and BMS are shown schematically. Assigned indices are just for illustration. Each small rectangle denotes a window of 32 consecutive elements. Reordering takes place per tile, but global offsets are computed per subproblem. \label{fig:multisplit_levels}}
\end{figure}
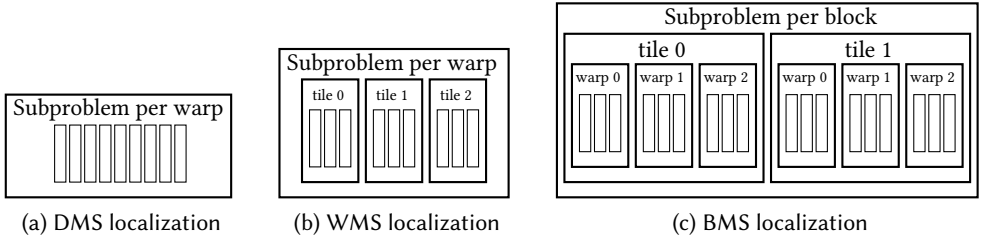

Figure~\ref{fig:multisplit_levels} shows a schematic example of our three methods next to each other. Note that we have flexibility to tune subproblem sizes in each implementation by changing the sizing parameters in Table~\ref{table:subproblems}.

\begin{table}
\centering
\small
\begin{tabular}{ll}
\toprule
Algorithm & subproblem size ($\bar{n}$) \\
\midrule
DMS     & $N_\text{(window/warp)}N_\text{thread}$ \\
WMS     & $N_\text{(tile/warp)}N_\text{(window/tile)}N_\text{thread}$ \\
BMS     & $N_\text{(tile/block)}N_\text{(warp/tile)}N_\text{(window/warp)}N_\text{thread}$ \\
\bottomrule
\end{tabular}
\caption{Size of subproblems for each multisplit algorithm. Total size of our global computations will then be the size of $\mathbf{H}$ equal to $mn/\bar{n}$.}\label{table:subproblems}
\end{table}

Next we briefly highlight the general structure of our computations (based on our model in Section~\ref{subsec:parallel_model}). We use DMS to illustrate:

\paragraph{Pre-scan (local)} Each warp reads a window of key elements (of size $N_\text{thread} = 32$), generates a local matrix $\bar{\mathbf{H}}$, and computes its histogram (reducing each row). Next, histogram results are stored locally in registers. Then, each warp continues to the next window and repeats the process, adding histogram results to the results from previous windows.
In the end, each warp has computed a single column of $\mathbf{H}$ and stores its results into global memory.
\paragraph{Scan (global)} We perform an exclusive scan operation over the row-vectorized $\mathbf{H}$ and store the result back into global memory (e.g., matrix $\mathbf{G} = [g_{i,\ell}]_{m\times L_0}$).
\paragraph{Post-scan (local)} Each warp reads a window of key-value pairs, generates its local matrix $\bar{\mathbf{H}}$ again,\footnote{Note that we compute $\bar{\mathbf{H}}$ a second time rather than storing and reloading the results from the computation in the first step. This is deliberate. We find that the recomputation is cheaper than the cost of global store and load.} and computes local offsets (with a local exclusive scan on each row).
Similar to the pre-scan stage, we store histogram results into registers.
We then compute final positions by using the global base addresses from $\mathbf{G}$, warp-wide base addresses (warp-wide histogram results up until that window), and local offsets. Then we write key-value pairs directly to their storage locations in the output vector. For example, if key $u \in B_i$ is read by warp $\ell$ and its local offset is equal to $k$, its final position will be $g_{i,\ell} + h_i + k$, where $h_i$ is the warp-wide histogram result up until that window (referring to equation~\eqref{eq:permutation3} is helpful to visualize how we compute final addresses with a multi-level localization).

Algorithm~\ref{alg:dms} shows a simplified pseudo-code of the DMS method (with less than 32 buckets). Here, we can identify each key's bucket by using a \texttt{bucket\_identifier()} function. We compute warp histogram and local offsets with \texttt{warp\_histogram()} and \texttt{warp\_offsets()} procedures, which we describe in detail later in this section (Alg.~\ref{alg:warp_histogram}~and~\ref{alg:warp_offsets}).

\begin{algorithm}
\KwIn {\text{key\_in[]}, \text{value\_in[]}, \text{bucket\_identifier()}: keys, values and a bucket identifier function.}
\KwOut {\text{key\_out[]}, \text{value\_out[]}: keys and values after multisplit.}
// {key\_in[], value\_in[], key\_out[], value\_out[], H, and G are all stored in global memory.}
// {L: number of subproblems}

// {\bf ====== Pre-scan stage:}

\For{each warp \textnormal{i=0:L-1} {\bf parallel device}}{
        histo[0:m-1] = 0\;
        \For{each window \textnormal{j=0:N\_window-1}}{
    \text{bucket\_id[0:31] = \text{bucket\_identifier(key\_in[32*i*N\_window + 32*j + (0:31)])\;}}

    \text{histo[0:m-1]} += \text{warp\_histogram(bucket\_id[0:31])\;}
  }
  \text{H[0:m-1][i] = histo[0:m-1]\;}
}

// {\bf ====== Scan stage:}

\text{H\_row = [H[0][0:L-1],H[1][0:L-1], \ldots, H[m-1][0:L-1]]\;}

\text{G\_row = exclusive\_scan(H\_row)\;}

// \text{[G[0][0:L-1],G[1][0:L-1], \ldots, G[m-1][0:L-1]] = G\_row\;}

\For{\textnormal{i = 0:m-1} and \textnormal{j = 0:L-1}}{
  G[i][j] = \text{G\_row[i $\ast$ m + j]\;}
}

// {\bf ====== Post-scan stage:}

\For {each warp \textnormal{i=0:L-1} {\bf parallel device}}{
        histo[0:m-1] = 0\;
        \For{each window \textnormal{j=0:N\_window-1}}{
                read\_key = key\_in[32*i*N\_window + 32*j + (0:31)]\;
                read\_value = value\_in[32*i*N\_window + 32*j + (0:31)]\;
          \text{bucket\_id[0:31] = \text{bucket\_identifier(read\_key)\;}}

          \text{offsets[0:31]} = \text{warp\_offsets(bucket\_id[0:31])\;}

    \For{each thread \textnormal{k=0:31} {\bf parallel warp}}{
        \text{final\_position[k] = G[bucket\_id[k]][i] + histo[bucket\_id[k]] + offsets[k]\;}

        \text{key\_out[final\_position[k]] = read\_key\;}

        \text{value\_out[final\_position[k]] = read\_value\;}
    }

    \text{histo[0:m-1]} += \text{warp\_histogram(bucket\_id[0:31])\;} // \emph{updating histograms}    
  }
}
\caption{The Direct Multisplit (DMS) algorithm}\label{alg:dms}
\end{algorithm}

\subsection{Ballot-based voting}\label{subsec:ballot}
In this section, we momentarily change our direction into exploring a theoretical problem about voting.
We then use this concept to design and implement our warp-wide histograms (Section~\ref{subsec:histogram}).
We have previously emphasized our design decision of a warp-wide granularity. This decision is enabled by the efficient warp-wide intrinsics of NVIDIA GPUs.
In particular, NVIDIA GPUs support a warp-wide intrinsic \texttt{\_\_ballot(predicate)}, which performs binary voting across all threads in a warp. More specifically, each thread evaluates a local Boolean predicate, and depending on the result of that predicate (true or false), it toggles a specific bit corresponding to its position in the warp (i.e., lane ID from 0 to 31).
With a 32-element warp, this ballot fits in a 32-bit register, so that the $i$th thread in a warp toggles the $i$th bit.
After the ballot is computed, every participant thread can access the ballot result (as a bitmap) and see the voting result from all other threads in the same warp.

Now, the question we want to answer within our multisplit implementation is a generalization to the voting problem: Suppose there are $m$ arbitrary agents (indexed from $0$ to $n-1$), each evaluating a personalized non-binary predicate (a vote for a candidate) that can be any value from $0$ to $m-1$. 
How is it possible to perform the voting so that any agent can know all the results (who voted for whom)?

A naive way to solve this problem is to perform $m$ separate binary votes. For each round $0 \leq i < m$, we just ask if anyone wants to vote for $i$. Each vote has a binary result (either voting for $i$ or not). After $m$ votes, any agent can look at the $m\times n$ ballots and know which agent voted for which of the $m$ candidates. 

We note that each agent's vote ($0 \leq v_i < n$) can be represented by $\log m$ binary digits. So a more efficient way to solve this problem requires just $\log m$ binary ballots per agent. Instead of directly asking for a vote per candidate ($m$ votes/bitmaps), we can ask for consecutive bits of each agent's vote (a bit at a time) and store them as a bitmap (for a total of $\log m$ bitmaps, each bitmap of size $n$).
For the $j$th bitmap ($0 \leq j < \lceil\log m\rceil$), every $i$th bit is one if only the $i$th agent have voted to a candidate whose $j$th bit in its binary representation was also one (e.g., the 0th bitmap includes all agents who voted for an odd-numbered candidate).  

As a result, these $\lceil \log m \rceil$ bitmaps together contain all information to reconstruct every agent's vote. All we need to do is to imagine each bitmap as a row of a $m\times n$ matrix.
Each column represent the binary representation of the vote of that specific agent. 
Next, we use this scheme to perform some of our warp-wide computations, only using NVIDIA GPU's binary ballots.
\subsection{Computing Histograms and Local Offsets}\label{subsec:histogram}
The previous subsections described why and how we create a hierarchy of localizations. Now we turn to the problem of computing a direct solve of histograms and local offsets on a warp-sized (DMS or WMS) or a block-sized (BMS) problem. In our implementation, we leverage the balloting primitives we described in Section~\ref{subsec:ballot}.
We assume throughout this section that the number of buckets does not exceed the warp width ($m \leq N_\text{thread}$). Later we extend our discussion to any number of buckets in Section~\ref{subsec:more_buckets}.

\subsubsection{Warp-level Histogram}\label{subsubsec:warp_histogram}
Previously, we described our histogram computations in each subproblem as forming a binary matrix $\bar{\mathbf{H}}$ and doing certain computations on each row (reduction and scan).
Instead of explicitly forming the binary matrix $\bar{\mathbf{H}}$, each thread generates its own version of the rows of this matrix and stores it in its local registers as a binary bitmap. Then per-row reduction is equivalent to a population count operation (\texttt{\_\_popc}), and exclusive scan equates to first masking corresponding bits and then reducing the result. We now describe both in more detail.

To compute warp-level histograms, we assign each bucket (each row of $\bar{\mathbf{H}}$) to a thread. That thread is responsible for counting the elements of the warp (of the current read window of input keys) that fall into that bucket.
We described in Section~\ref{subsec:ballot} how each agent (here, each thread) can know about all the votes (here, the bucket to which each key belongs).
Since we assigned each thread to count the results of a particular bucket (here, the same as its lane ID ranging from 0 to 31), each thread must only count the number of other threads that voted for a bucket equal to its lane ID (the bucket to which it is assigned).
For cases where there are more buckets than the warp width, we assign $\lceil m/32 \rceil$ buckets to each thread.
First, we focus on $m\leq 32$; Algorithm~\ref{alg:warp_histogram} shows the detailed code.

\begin{algorithm}
\DontPrintSemicolon
\Fn{warp\_histogram(bucket\_id[0:31])}{
\KwIn {\text{bucket\_id[0:31]} // \emph{a warp-wide array of bucket IDs}}
\KwOut {\text{histo[0:m-1]} // \emph{number of elements within each $m$ buckets}}
        \For{each thread \textnormal{i = 0:31} {\bf parallel warp}}{
            \text{histo\_bmp[i] = 0xFFFFFFFF\;}
            \For{\textnormal{(int k = 0; k < ceil(log2(m)); k++)}}{
              \text{temp\_buffer = \_\_ballot((bucket\_id[i] >> k) \& 0x01);}

              \lIf{\textnormal{((i >> k) \& 0x01})}{
                      \text{histo\_bmp[i] \&= temp\_buffer;}
              }\lElse{
                \text{histo\_bmp[i] \&= $\sim$ temp\_buffer;}
              }
                        }
            \text{histo[i] = \_\_popc(histo\_bmp[i]);} // \emph{counting number of set bits}
  }
  \Return{\textnormal{histo[0:m-1];}}
}
\caption{Warp-level histogram computation}\label{alg:warp_histogram}
\end{algorithm}

Each thread $i$ is in charge of the bucket with an index equal to its lane ID (0--31). Thread $i$ reads a key, computes that key's bucket ID (0--31), and initializes a warp-sized bitvector (32 bits) to all ones. This bitvector  corresponds to threads (keys) in the warp that might have a bucket ID equal to this thread's assigned bucket. Then each thread broadcasts the least significant bit (LSB) of its observed bucket ID, using the warp-wide ballot instruction. Thread $i$ then zeroes out the bits in its local bitmap that correspond to threads that are broadcasting a LSB that is incompatible with $i$'s assigned bucket. This process continues with all other bits of the observed bucket IDs (for $m$ buckets, that's $\log m$ rounds). When all rounds are complete, each thread has a bitmap that indicates which threads in the warp have a bucket ID corresponding to its assigned bucket. The histogram result is then a reduction over these set bits, which is computed with a single population count (\texttt{\_\_popc}) instruction.

For $m > 32$, there will still be $\lceil \log(m) \rceil$ rounds of ballots. However, each thread will need $\lceil m/32 \rceil$ 32-bit registers to keep binary bitvectors (for multiple \texttt{histo\_bmp} registers per thread). Each of those registers is dedicated to the same lane IDs within that warp, but one for buckets 0--31, the second for buckets 32--63, and so forth.

\subsubsection{Warp-level Local Offsets}\label{subsubsec:warp_offset}
Local offset computations follow a similar structure to histograms (Algorithm~\ref{alg:warp_offsets}). In local offset computations, however, each thread is only interested in keeping track of ballot results that match its item's \emph{observed} bucket ID, rather than the bucket ID to which it has been assigned. Thus we compute a bitmap that corresponds to threads whose items share our same bucket, mask away all threads with higher lane IDs, and use  the population count instruction to compute the local offset.

For $m > 32$, we can use the same algorithm (Algorithm~\ref{alg:warp_offsets}).
The reason for this, and the main difference with our histogram computation, is that regardless of the number of buckets, there are always a fixed number of threads within a warp. 
So, we still need to perform $\lceil \log m\rceil$ ballots but our computations are the same as before, to look for those that share the same bucket ID as ours (only 32 other potential threads). 
This is unlike our histogram computations (Algorithm~\ref{alg:warp_histogram}), that we needed to keep track of $\lceil m/32 \rceil$ different buckets because there were more buckets than existing threads within a warp.

Histogram and local offset computations are shown in two separate procedures (Alg.~\ref{alg:warp_histogram}~and~\ref{alg:warp_offsets}), but since they share many common operations they can be merged into a single procedure if necessary (by sharing the same ballot results). For example, in our implementations, we only require histogram computation in the pre-scan stage, but we need both a histogram and local offsets in the post-scan stage.

\begin{algorithm}
\DontPrintSemicolon
\Fn{warp\_offset(bucket\_id[0:31])}{
\KwIn {\text{bucket\_id[0:31]} // \emph{a warp-wide array of bucket IDs}}
\KwOut {\text{local\_offset[0:31]} // \emph{for each element, number of preceding elements within the same bucket}}
        \For{each thread \textnormal{i = 0:31} {\bf parallel warp}}{
            \text{offset\_bmp[i] = 0xFFFFFFFF\;}
            \For{\textnormal{(int k = 0; k < ceil(log2(m)); k++)}}{
              \textnormal{temp\_buffer = \_\_ballot((bucket\_id[i] >> k) \& 0x01);}

              \lIf{\textnormal{((i >> k) \& 0x01})}{
                      \text{offset\_bmp[i] \&= temp\_buffer;}
              }\lElse{
                \textnormal{offset\_bmp[i] \&= $\sim$ temp\_buffer;}
              }
                        }
            \text{local\_offset[i] = \_\_popc(offset\_bmp[i]\&(0xFFFFFFFF>>(31-i))) - 1;} // \emph{counting number of preceding set bits}
  }
  \Return{\textnormal{local\_offset[0:31];}}
}
\caption{Warp-level local offset computation}\label{alg:warp_offsets}
\end{algorithm}

\subsubsection{Block-level Histogram}\label{subsubsec:block_histogram}
For our Block-level MS, we perform the identical computation as Direct MS and Warp-level MS, but over a block rather than a warp. If we chose explicit local computations (described in Section~\ref{sec:algorithm}) to compute histograms and local offsets, the binary matrix $\bar{\mathbf{H}}$ would be large, and we would have to reduce it over rows (for histograms) and scan it over rows (for local offsets). Because of this complexity, and because our warp-level histogram computation is quite efficient, we pursue the second option: the hierarchical approach.
Each warp reads consecutive windows of keys and computes and aggregates their histograms.
Results are then stored into consecutive parts of shared memory such that all results for the first bucket ($B_0$) are next to each other: results from the first warp followed by results from the second warp, up to the last warp. Then, the same happens for the second bucket ($B_1$) and so forth until the last bucket $B_{m-1}$.
Now our required computation is a segmented (per-bucket) reduction over $m$ segments of $N_\text{warp}$ histogram results.

In our implementation, we choose to always use a power-of-two number of warps per block. As a result, after writing these intermediate warp-wide histograms into shared memory and syncing all threads, we can make each warp read a consecutive segment of 32 elements that will include $32/N_\text{warp}$ segments (buckets). Then, segmented reduction can be performed with exactly $\log N_\text{warp}$ rounds of \texttt{\_\_shfl\_xor()}.\footnote{This is a simple modification of a warp-wide reduction example given in the CUDA Programming guide~\cite[Chapter~B14]{NVIDIA:2016:CUDA}.} 
This method is used in our BMS's pre-scan stage, where each segment's result (which is now a tile's histogram) is written into global memory (a column of $\mathbf{H}$). The process is then continued for the next tile in the same thread-block.

\subsubsection{Block-level Local Offsets}\label{subsubsec:block_offset}
Our block-wide local offset computation is similar to our warp-level local offsets with some additional offset computations.
First, for each warp within a thread-block, we use both  Alg.~\ref{alg:warp_histogram} and Alg.~\ref{alg:warp_offsets} to compute histogram and local offsets per window of input keys. By using the same principle showed in our hierarchical computations in equation~\ref{eq:permutation3}, we use these intermediate results to compute block-wide local offsets (the tile's local offset in BMS). To compute the local offset of each element within its own bucket, we require three components:

\begin{enumerate}
        \item local offsets for each window (same tile, same bucket, same window)
        \item total number of elements processed by the same warp, same bucket, but from previous windows
        \item total number of elements in the same tile, same bucket, but from previous warps (each with multiple windows)
\end{enumerate}

To illustrate, suppose, $h_{j,(x,y)}$ shows the histogram results of bucket $B_j$ for warp $x$ and window $y$. Then for a key $u_i$ which is read by warp $X$ in its window $Y$, we have:
\begin{IEEEeqnarray}{rCl}\label{eq:local_offset}
\text{local offset}(u_i) & = & \left| \left\{u_r \in (\text{warp } X, \text{window } Y): (u_r \in B_j) \land (r < i)\right\}\right| \nonumber \\
& + & \sum_{y=0}^{Y-1}h_{j,(X,y)} + \sum_{x=0}^{X-1}\sum_{y=0}^{N_\text{window}-1}h_{j,(x,y)}.
\end{IEEEeqnarray}
The first term (item) is exactly the outcome of Alg.~\ref{alg:warp_offsets}, computed per window of input keys.
By having all histogram results per window, each warp can locally compute the second term. However, each thread only has access to the counts corresponding to its in-charge bucket (equal to its lane ID). By using a single shuffle instruction, each thread can ask for this value from a thread that has the result already computed (within the same warp).
For computing the third term, we replicate what we did in Section~\ref{subsubsec:block_histogram}. Histogram results (aggregated results for all windows per warp) are stored into consecutive segments of shared memory (total of $32/N_\text{warp}$ segments).
Then, we perform a segmented scan by using $\log (N_\text{warp})$ rounds of \texttt{\_\_shfl\_up()} (instead of \texttt{\_\_shfl\_xor}).
All results are then stored back into shared memory, so that all threads can access appropriate values that they need (to compute equation~\ref{eq:local_offset}) based on their warp ID\@.
As a result, we have computed the local offset of each key within each thread-block.
\subsection{Reordering for better locality}\label{subsec:reordering}
\begin{figure}
\centering
\subfloat[Key distribution with 2 buckets]
{
        \includegraphics[width = 0.5\linewidth]{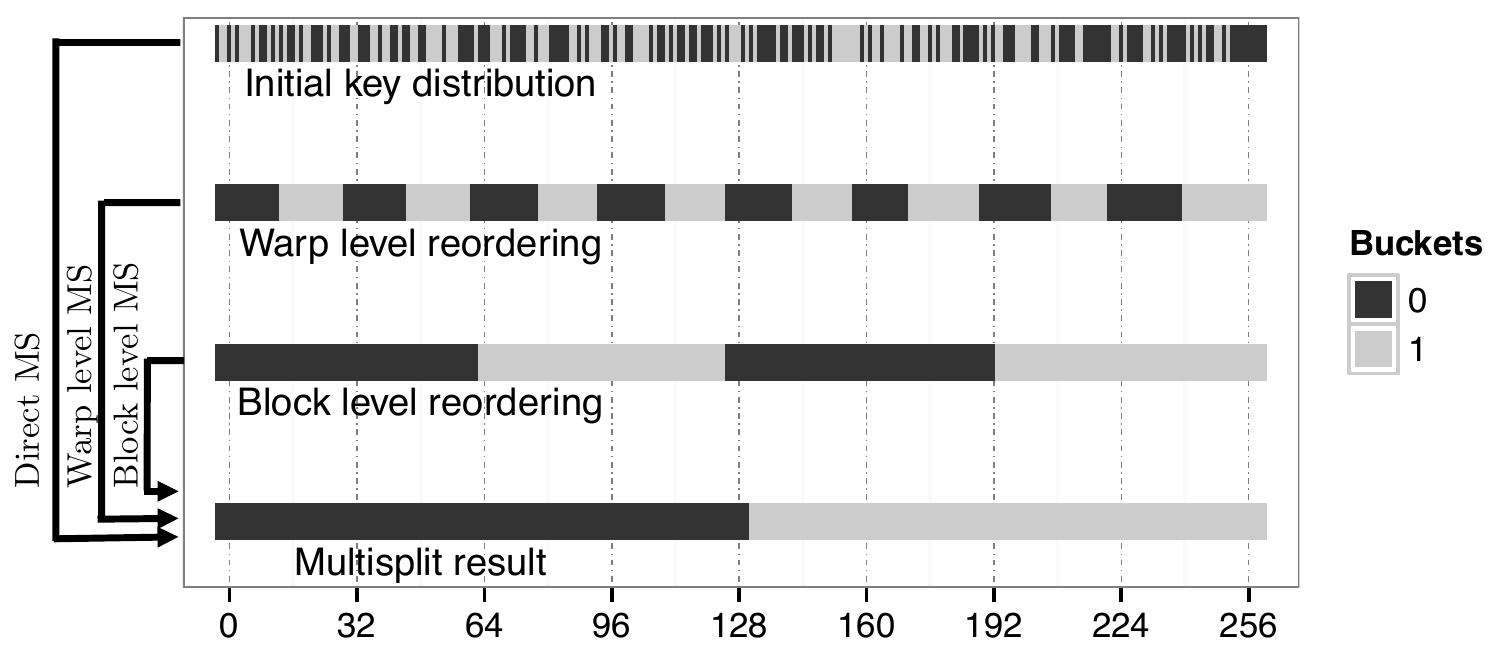}
        \label{fig:dist_2}
}
\subfloat[Key distribution with 8 buckets]
{
        \includegraphics[width = 0.5\linewidth]{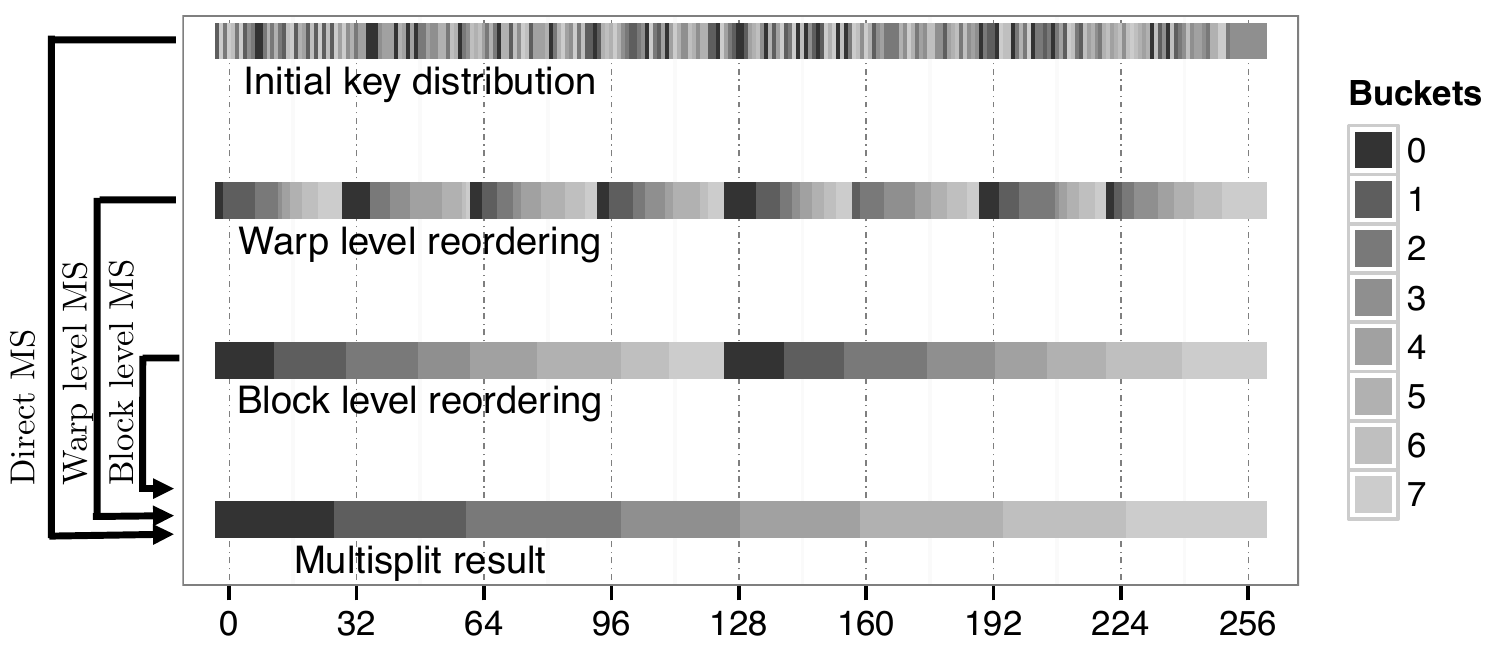}
        \label{fig:dist_8}
}
\caption{Key distributions for different multisplit methods and different number of buckets. Key elements are initially uniformly distributed among different buckets. This window shows an input key vector of length 256; each warp is 32 threads wide and each block has 128 threads. }\label{fig:dist}
\end{figure}

As described in Section~\ref{sec:algorithm}, one of the main bottlenecks in a permutation like multisplit is the random scatter in its final data movement. Figure~\ref{fig:dist} shows an example of such a case. As we suggested previously, we can improve scatter performance by reordering elements locally in each subproblem such that in the final scatter, we get better coalescing behavior (i.e., consecutive elements are written to consecutive locations in global memory).

However, while a higher achieved write memory bandwidth will improve our runtime, it comes at the cost of more local work to reorder elements. Warp-level reordering requires the fewest extra computations, but it may not be able to give us enough locality as the number of buckets increases (Fig.~\ref{fig:dist}). We can achieve better locality, again at the cost of more computation, by reordering across warps within a block.

\subsubsection{Warp-level Reordering}\label{subsubsec:warp_reordering}
WMS extends DMS by reordering each tile (a group of consecutive windows in WMS) before the final write for better memory coalescing behavior.
Our first question was whether we prefer to perform the reordering in our pre-scan stage or our post-scan stage. We know that in order to compute the new index for each element in a tile, we need to know about its histogram and we need to perform a local (warp-level) exclusive scan over the results. We have already computed the warp level histogram in the pre-scan stage, but we do not have it in the post-scan stage and thus would either have to reload it or recompute it.

However, if we reorder key-value pairs in the pre-scan stage, we must perform two coalesced global reads (reading key-value pairs) and two coalesced global writes (storing the reordered key-value pairs before our global operation) per thread and for each key-value pair. Recall that in DMS, we only required one global read (just the key) per thread and per key in its pre-scan stage.

In the end, the potential cost of the additional global reads was significantly more expensive than the much smaller cost of recomputing our efficient warp-level histograms. As a result, we reorder in the post-scan stage and require fewer global memory accesses overall.

The main difference between DMS and WMS is in post-scan, where we compute both warp-level histogram and local offsets (Algorithm~\ref{alg:warp_histogram}~and~\ref{alg:warp_offsets}).
As we described before, in WMS each subproblem is divided into several tiles. Each tile is assigned to a warp and is processed in a consecutive number of windows of length 32 each.
Each warp, however, only performs reordering for all elements within a tile (not among all tiles).
This decision is mostly because of limited available shared memory per block.

Algorithms~\ref{alg:warp_histogram}~and~\ref{alg:warp_offsets} provide histogram and local offsets per window of read data. So, in order to perform warp-level reordering (performing a local multisplit within a tile), we need the following items to be computed for each key-value pair to be able to compute their new positions in shared memory:
\begin{enumerate}
        \item local offsets for each window (same tile, same bucket, same window)
        \item total number of elements processed by the same warp (same tile), from the same bucket, but from previous windows
        \item total number of elements in the same tile, but from previous buckets
\end{enumerate}
The first item is exactly the warp-level local offset computed in Section~\ref{subsubsec:warp_offset}.
The second item can be computed as follows: each thread within a warp directly computes all histogram results for all its windows. So, it can locally compute an exclusive scan of these results. However, each thread only has access to the counts per window of the  bucket for which it is in charge (i.e., the bucket equal to its lane ID). As a result, it is enough for each thread to ask for the appropriate value of~(2) by asking the thread who owns it (via a single shuffle instruction).

For the third item, we do the following: each thread within a warp has already computed the total number of elements per window (for specific buckets). So, it can easily add them together to form the histogram for all the windows (tile histogram).
All that remains is to perform another exclusive scan among these values such that each thread has the total number of items from previous buckets within the same tile.
This result too can be provided to threads by using a single shuffle instruction.
We can now perform reordering for key-value pairs into shared memory.

After performing the warp-level reordering, we need to write back results into global memory and into their final positions. To do so, all threads within a warp once again read key-value pairs from shared memory (it will be in $N_\text{window}$ coalesced reads).
Since items are reordered, previous threads are not necessarily reading the same items as before and hence they should identify their buckets once again.
Since elements are already reordered (consecutive elements belong to the same bucket), the new local offset among all other keys within the same bucket and same tile can be recomputed in a much easier way: each item's index (within shared memory) minus the offset computed in step (3)~of reordering (which can be accessed by a single shuffle instruction).
Finally we can add the computed local tile-offsets to the global offset offsets computed by the scan-stage, and perform final data movements in a (more likely) coalesced way.

\subsubsection{Block-level Reordering}\label{subsubsec:block_reordering}
The benefit from warp-level reordering is rather modest, particularly as the number of buckets grows, because we only see a small number of elements per warp (a WMS's tile) that belong to the same bucket. For potentially larger gains in coalescing, our BMS reorders entire blocks (larger tiles by a factor of $N_\text{warp}$).
That being said, an important advantage of our WMS is that almost everything can be computed within a warp, and since warps perform in lockstep, there will not be any need for further synchronizations among different warps.
In contrast, any coordination among warps within a thread-block requires proper synchronization.

As we mentioned before, each subproblem in BMS is assigned to a thread-block and is divided into several tiles. Each tile is assigned to multiple warps within a block. Each warp divides its share into multiple consecutive windows of 32 elements each.
Final positions can be computed in a hierarchical approach with 5 levels of localization, as described in equation~\ref{eq:permutation3}.

In BMS, although each block can process multiple consecutive tiles, we only perform reordering per tile (mostly because of limited available shared memory per block).
Reordering is equivalent to performing local multisplit over each tile. We can summarize the required computations for this reordering with the following two items:
\begin{enumerate}
        \item block-level local offsets
        \item total number of keys in the same tile, from previous buckets
\end{enumerate}
The first item can be computed as described in Section~\ref{subsubsec:block_offset}.
During the above computation, at some point we stored results from a segmented exclusive scan into shared memory.
Since our shuffle-based segmented scan is initially an inclusive scan, each thread has to subtract its initial bucket count from it to make it an exclusive scan.  
So, during that computation, with minimal extra effort, we could also store results for the sum of all elements within each segment (i.e., the total number of elements within each bucket in that tile) into another location in shared memory as well (for a total of $m$ elements). We refer to this result as our tile histogram.
Now, here each warp can reload the tile histogram from shared memory and perform a warp-wide exclusive scan operation on it.
In order to avoid extra synchronizations, every warp performs this step independently (as opposed to the option that a single warp computes it and puts it into shared memory, thus making it available for all other warps).
Thus, each thread can easily ask the required value for item~(2) by using a single shuffle instruction to fetch the value from the thread that owns that bucket's result.

After all threads within a block finish storing reordered key-value pairs into shared memory (for a single tile), we perform the final data movements.
Threads read the reordered tile (different key-value pairs than before), identify their buckets, and compute the final positions based on the following items:
\begin{enumerate}[label=(\roman*)]
        \item The new block-level local offset
        \item total number of elements from previous subproblems (the whole device) and from previous buckets
\end{enumerate}
Since elements are now already reordered, consecutive elements belong to the same bucket. As a result, the first item is equivalent to the index of that element (within shared memory) minus the starting index for that specific bucket (which is exactly item (2)~in reordering).
The second item is also already computed and available from our scan stage.
So, threads can proceed to perform the final data movement.
\subsection{More buckets than the warp width}\label{subsec:more_buckets}
\paragraph{Warp-level histogram and local offset computation}
Throughout the previous discussion, our primary way of computing a histogram in each warp (which processes a group of windows one by one) is to make each thread responsible to count the total number of elements with the bucket ID equal to its lane ID\@.
Since the current generation of GPUs has $N_\text{thread}=32$ threads per warp, if the number of buckets is larger than the warp width, we must put each thread in charge of multiple buckets (each thread is in charge of $\lceil m/32\rceil$ buckets as described in Section~\ref{subsubsec:warp_histogram}).
The total number of ballots required for our warp-level histogram procedure scales logarithmically ($\lceil\log m\rceil$).
Local offset computations are also as before (with $\lceil \log m\rceil$ ballots).

\paragraph{Block-level histogram and local offsets computations}
If $m>32$, besides the changes described above, to perform our segmented scan and reductions (Section~\ref{subsubsec:block_histogram}), 
each thread should participate in the computations required by $\lceil m/32\rceil$ segments.
Previously, for $m > 32$ buckets we used CUB's block-wide scan operation~\cite{Ashkiani:2016:GM}.
However, although CUB is a very efficient and high-performance algorithm in performing scan, it uses a lot of registers to achieve its goal. As a result, we prefer a more register-friendly approach to these block-level operations, and hence implemented our own segmented scan and reductions by simply iterating over multiple segments, processing each as described in Section~\ref{subsubsec:block_histogram} and Section~\ref{subsubsec:block_offset}.
\subsection{Privatization}\label{subsec:ms_privatization}
Traditionally, multisplit~\cite{He:2008:RJG} (as well as histogram and radix sort~\cite{Bell:2011:TAP,Merrill:2015:CUB} with some similarities) were implemented with a thread-level approach with thread-level memory \emph{privatization}: each thread was responsible for a (possibly contiguous) portion of input data.
Intermediate results (such as local histograms, local offsets, etc.) were computed and stored in parts of a memory (either in register or shared memory) exclusively dedicated to that thread.
Privatization eliminates contention among parallel threads at the expense of register/shared memory usage (valuable resources).
Memory accesses are also more complicated since the end goal is to have a sequence of memory units per thread (commonly known as \emph{vectorized} access), as opposed to natural coalesced accesses where consecutive memory units are accessed by consecutive threads. That being said, this situation can be remedied by careful pipelining and usage of shared memory (initially coalescing reads from global memory, writing the results into shared memory, followed by vectorized reads from shared memory).

In contrast, in this work, we advocate a \emph{warp-level} strategy that assigns different warps to consecutive segments of the input elements and stores intermediate results in portions of memory (distributed across all registers of a warp, or in shared memory) that are exclusively assigned to that particular warp~\cite{Ashkiani:2016:GM}.
An immediate advantage of a warp-level strategy is a reduction in shared memory requirements per thread-block (by a factor of warp-width).
Another advantage is that there is no more need for vectorized memory accesses, relieving further pressure on shared memory usage.
The chief disadvantage is the need for warp-wide intrinsics for our computations. These intrinsics may be less capable or deliver less performance. On recent NVIDIA GPUs, in general, using warp-wide intrinsics are faster than regular shared memory accesses but slower than register-level accesses.
A more detailed comparison is not possible since it would heavily depend on the specific task at hand.
However, efficient warp-wide intrinsics open up new possibilities for warp-wide computations as fundamental building blocks, allowing algorithm designers to consider using both thread and warp granularities when constructing and tuning their algorithms.

\section{Performance Evaluation}\label{sec:perf_eval}
In this section we evaluate our multisplit methods and analyze their performance. First, we discuss a few characteristics in our simulations:

\paragraph{Simulation Framework}
All experiments are run on a NVIDIA K40c with the Kepler architecture, and a NVIDIA GeForce GTX 1080 with the Pascal architecture (Table~\ref{table:gpus}).
All programs are compiled with NVIDIA's nvcc compiler (version~8.0.44).
The authors have implemented all codes except for device-wide scan operations and radix sort, which are from CUB (version~1.6.4).
All experiments are run over 50 independent trials.
Since the main focus of this paper is on multisplit as a GPU primitive within the context of a larger GPU application, we assume that all required data is already in the GPU's memory and hence no transfer time is included.

Some server NVIDIA GPUs (such as Tesla K40c) provide an error correcting code (ECC) feature to decrease occurrence of unpredictable memory errors (mostly due to physical noise perturbations within the device in long-running applications).
ECCs are by default enabled in these devices, which means that hardware dedicates a portion of its bandwidth to extra parity bits to make sure all memory transfers are handled correctly (with more probability).
Some developers prefer to disable ECC to get more bandwidth from these devices.
In this work, in order to provide a more general discussion, we opt to consider three main hardware choices: 1)~Tesla K40c with ECC enabled (default), 2)~Tesla K40c with ECC disabled, and 3)~GeForce GTX 1080 (no ECC option).

\begin{table}
\centering
\scriptsize
\begin{tabular}{lcc}
\toprule
NVIDIA GPU & Tesla K40c & GeForce GTX 1080 \\
\midrule
Architecture    & Kepler & Pascal \\
Compute capability & 3.5 & 6.1 \\
Number of SMs & 15 & 20 \\
Global Memory size & 12~GB & 8~GB \\
Global memory bandwidth & 288~GB/s & 320~GB/s \\
Shared memory per SM & 48~KB & 96~KB \\
\bottomrule
\end{tabular}
\caption{Hardware characteristics of the NVIDIA GPUs that we used in this paper.}\label{table:gpus}
\end{table}
\paragraph{Bucket identification}
The choice of bucket identification directly impacts performance results of any multisplit method, including ours.
We support user-defined bucket identifiers.
These can be as simple as unary functions, or complicated functors with arbitrary local arguments. For example, one could utilize a functor which determines whether a key is prime or not.
Our implementation is simple enough to let users easily change the bucket identifiers as they please.

In this section, we assume a simple user-defined bucket identifier as follows: buckets are assumed to be of equal width $\Delta$ and to partition the whole key domain (\emph{delta-buckets}). For example, for an arbitrary key $u$, bucket IDs can be computed by a single integer division (i.e., $\defn{f}(u) = \lfloor u/ \Delta\rfloor$).
Later, in Section~\ref{subsec:multisplit_sort} we will consider a simpler bucket identifier (\emph{identity buckets}): where keys are equal to their bucket IDs (i.e., $\defn{f}(u) = u$).
This is particularly useful when we want to use our multisplit algorithm to implement a radix sort.
In Section~\ref{subsec:multisplit_histogram} we use more complicated identifiers as follows: given a set of arbitrary splitters $s_0< s_1 < \dots < s_{m-1}$, for each key $s_0 < u < s_{m-1}$, finding those splitters (i.e., bucket $B_j$) such that $s_j \leq u < s_{j+1}$. This type of identification requires performing a binary search over all splitters per input key.

\paragraph{Key distribution}
Throughout this paper we assume uniform distribution of keys among buckets (except in Section~\ref{subsec:perf_distribution} where we consider other distributions), meaning that keys are randomly generated such that there are, on average, equal number of elements within each bucket.
For delta-buckets and identity buckets (or any other linear bucket identifier), this criteria results in uniform distribution of keys in the key domain as well.
For more complicated nonlinear bucket identifiers this does not generally hold true.

\paragraph{Parameters} In all our methods and for every GPU architecture we have used either: 1)~four warps per block (128 threads per block), where each warp processes 7 consecutive windows, or 2)~eight warps per block where each warp processes 4 consecutive windows.
Our key-only BMS for up to $m \leq 32$ uses the former, while every other case uses the latter (including WMS and BMS for both key-only and key-value pairs).
These options were chosen because they gave us the best performance experimentally.

This is a trade off between easier inter-warp computations (fewer warps) versus easier intra-warp communications (fewer windows).
By having fewer warps per block, all our inter-warp computations in BMS (segmented scans and reductions in Section~\ref{subsubsec:block_histogram} and \ref{subsubsec:block_offset}) are directly improved, because each segment will be smaller-sized and hence fewer rounds of shuffles are required ($\log{N_\text{warp}}$ rounds).
On the other hand, we use subword optimizations to pack the intermediate results of 4 processed windows into a single 32-bit integer (a byte per bucket per window). This lets us communicate among threads within a warp by just using a single shuffle per 4 windows.
Thus, by having fewer warps per block, if we want to load enough input keys to properly hide memory access latency, we would need more than 4 windows to be read by each warp (here we used 7), which doubles the total number of shuffles that we use.

It is a common practice for GPU libraries, such as in CUB's radix sort, to choose their internal parameters at runtime based on the GPU's compute capability and architecture.
These parameters may include the number of threads per block and the number of consecutive elements to be processed by a single thread.
The optimal parameters may substantially differ on one architecture compared to the other.
In our final API, we hid these internal parameters from the user; however, our experiments on the two GPUs we used (Tesla K40c with \texttt{sm\_35} and GeForce GTX 1080 with \texttt{sm\_61}) exhibited little difference between the optimal set of parameters for the best performance on each architecture.

In our algorithms, we always use as many threads per warp as allowed on NVIDIA  hardware ($N_\text{thread} = 32$).
Based on our reliance on warp-wide ballots and shuffles to perform our local computations (as discussed in Section~\ref{subsec:histogram}), using smaller-sized logical warps would mean having smaller sized subproblems (reducing potential local work and increasing global computations), which is unfavorable.
On the other hand, providing a larger-sized warp in future hardware with efficient ballots and shuffles (e.g., performing ballot over 64 threads and storing results as a 64-bit bitvector) would directly improve all our algorithms.
\subsection{Common approaches and performance references}\label{subsec:perf_references}
\paragraph{Radix sort} As we described in Section~\ref{sec:init_approaches}, not every multisplit problem can be solved by directly sorting input keys (or key-values). However, in certain scenarios where keys that belong to a lower indexed bucket are themselves smaller than keys belonging to larger indexed buckets (e.g., in delta-buckets), direct sorting results in a non-stable multisplit solution.
In this work, as a point of reference, we compare our performance to a full sort (over 32-bit keys or key-values).
Currently, the fastest GPU sort is provided by CUB's radix sort (Table~\ref{table:reference}).
With a uniform distribution of keys, radix sort's performance is independent of the number of buckets; instead, it only depends on the number of significant bits.

\paragraph{Reduced-bit sort} Reduced-bit sort (\emph{RB-sort}) was introduced in Section~\ref{sec:init_approaches} as the most competitive conventional-GPU-sort-based multisplit method.
In this section, we will compare all our methods against RB-sort. We have implemented our own kernels to perform labeling (generating an auxiliary array of bucket IDs) and possible packing/unpacking (for key-value multisplit). For its sorting stage, we have used CUB's radix sort.

\paragraph{Scan-based splits}
Iterative scan-based split can be used on any number of buckets. For this method, we ideally have a completely balanced distribution of keys, which means in each round we run twice the number of splits as the previous round over half-sized subproblems.
So, we can assume that in the best-case scenario, recursive (or iterative) scan-based split's average running time is lower-bounded by $\log(m)$ (or $m$) times the runtime of a single scan-based split method. This ideal lower bound is not competitive for any of our scenarios, and thus we have not implemented this method for more than two buckets.

\begin{table}
  \centering
  \small
  \resizebox{\columnwidth}{!}{
  \begin{tabular}{lcc|cc|cc}
    \toprule
    & \multicolumn{2}{c}{Tesla K40c (ECC on)} & \multicolumn{2}{c}{Tesla K40c (ECC off)} & \multicolumn{2}{c}{GeForce GTX 1080} \\
    \cmidrule(r){2-3}\cmidrule{4-5} \cmidrule(l){6-7}
    Method & time & rate & time & rate & time & rate\\
    \midrule
    Radix sort (key-only) & 25.99~ms & 1.29~Gkeys/s
    & 19.41~ms & 1.73~Gkeys/s & 9.84~ms & 3.40~Gkeys/s \\
    Radix sort (key-value) & 43.70~ms & 0.77~Gpairs/s
    & 28.60~ms & 1.17~Gpairs/s & 17.59~ms & 1.90~Gpairs/s\\
    \midrule
    \midrule
    {\scriptsize Scan-based split (key-only)} & 5.55~ms & 6.05~Gkeys/s
    & 4.91~ms & 6.84~Gkeys/s  & 3.98~ms & 8.44~Gkeys/s \\
    {\scriptsize Scan-based split (key-value)} & 6.96~ms & 4.82~Gpairs/s
    & 5.97~ms & 5.62~Gpairs/s  & 5.13~ms & 6.55~Gpairs/s \\
    \bottomrule
  \end{tabular}
  }
  \caption{On the top: CUB's radix sort. Average running time (ms) and processing rate (billion elements per second), over $2^{25}$ randomly generated 32-bit inputs (keys or key-value pairs). On the bottom: our scan-based split. Average running time (ms) and processing rate (billion elements per second), over $2^{25}$ randomly generated 32-bit inputs uniformly distributed into two buckets.\label{table:reference}}
\end{table}
\subsection{Performance versus number of buckets: \texorpdfstring{$m\leq 256$}{m less than or equal to 256}}\label{sec:perf_avg_time}
In this section we analyze our performance as a function of the number of buckets ($m \leq 256$).
Our methods differ in three principal ways: 1)~how expensive are our local computations, 2)~how expensive are our memory accesses, and 3)~how much locality can be extracted by reordering.

In general, our WMS method is faster for a small number of buckets and BMS is faster for a large number of buckets. Both are generally faster than RB-sort.
There is a crossover between WMS and BMS (a number of buckets such that BMS becomes superior for all larger numbers of buckets) that may differ based on 1) whether multisplit is key-only or key-value, and 2) the GPU architecture and its available hardware resources.
Key-value scenarios require more expensive data movements and hence benefit more from reordering (for better coalesced accesses).
That being said, BMS requires more computational effort for its reordering (because of multiple synchronizations for communications among warps), but it is more effective after reordering (because it reorders larger sized subproblems compared to WMS).
As a result, on each device, we expect to see this crossover with a smaller number of buckets for key-value multisplit vs.\ key-only.

\subsubsection{Average running time}
Table~\ref{table:timing} shows the average running time of different stages in each of our three approaches, and the reduced bit sort (RB-sort) method.
All of our proposed methods have the same basic computational core, warp-wide local histogram, and local offset computations. Our methods differ in performance as the number of buckets increases for three major reasons (Table~\ref{table:timing}):
\begin{description}
        \item [Reordering process] Reordering keys (key-values) requires extra computation and shared memory accesses. Reordering is always more expensive for BMS as it also requires inter-warp communications.
        These negative costs mostly depend on the number of buckets $m$, the number of warps per block $N_\text{warp}$, and the number of threads per warp $N_\text{thread}$.
        \item [Increased locality from reordering] Since block level subproblems have more elements than warp level subproblems, BMS is always superior to WMS in terms of locality.
        On average and for both methods, our achieved gain from locality decreases by $\frac{1}{m}$ as $m$ increases.
        \item [Global operations] As described before, by increasing $m$, the height of the matrix $\mathbf{H}$ increases. However, since BMS's subproblem sizes are relatively larger (by a factor of $N_\text{warp}$), BMS requires fewer global operations compared to DMS and WMS (because the smaller width of its $\mathbf{H}$).
        As a result, scan operations for both the DMS and WMS get significantly more expensive, compared to other stages, as $m$ increases (as $m$ doubles, the cost of scan for all methods also doubles).
\end{description}
\newlength{\runtimefirstcol}
\setlength{\runtimefirstcol}{0.4in}
\begin{table}
  \centering
  \scriptsize
  \begin{tabular}{p{\runtimefirstcol}lccc|ccc || ccc|ccc}
  	& & \multicolumn{6}{c}{Tesla K40c (ECC on)} & \multicolumn{6}{c}{GeForce GTX 1080} \\ 
  	\cmidrule(r){3-8}\cmidrule(l){9-14}
    & & \multicolumn{3}{c}{Key-only} & \multicolumn{3}{c}{Key-value} & \multicolumn{3}{c}{Key-only} & \multicolumn{3}{c}{Key-value}\\
    \cmidrule(r){3-5} \cmidrule(r){6-8} \cmidrule(l){9-11} \cmidrule(r){12-14}
    & & \multicolumn{6}{c}{Number of buckets (m)} & \multicolumn{6}{c}{Number of buckets (m)} \\
    \cmidrule(r){3-8} \cmidrule(l){9-14}
    Algorithm & Stage 
    & 2 & 8 & 32 & 2 & 8 & 32 
    & 2 & 8 & 32 & 2 & 8 & 32 \\
    \toprule
    \multirow{4}{*}{\parbox{\runtimefirstcol}{DMS}} 
    & Pre-scan 
    	& 1.40 & 1.53 & 3.98 & 1.40 & 1.53 & 3.98 
    	& 0.61 & 0.72 & 1.80 & 0.61 & 0.72 & 1.80 \\ 
    & Scan 
    	& 0.13 & 0.39 & 1.47 & 0.13 & 0.39 & 1.47
	    & 0.10 & 0.31 & 1.16 & 0.09  & 0.31  & 1.16 \\  
    & Post-scan 
    	& 2.29  & 2.94 & 4.85 & 3.34  & 4.05  & 11.84 
    	& 1.19  & 2.02  & 3.10 & 2.29  & 3.71  & 6.60 \\  
    & Total 
    	& 3.82 & 4.86 & 10.29 & 4.87  & 5.97  & 17.28 
    	& 1.90 & 3.05  & 6.06 & 3.00  & 4.74  & 9.56 \\ 
    \midrule    
    \multirow{4}{*}{\parbox{\runtimefirstcol}{WMS}} 
    & Pre-scan 
        & 0.79 & 0.93 & 1.38 & 0.89 & 0.97 & 1.39 
        & 0.58 & 0.60 & 0.93 & 0.59 & 0.62 & 0.93 \\ 
    & Scan 
        & 0.05 & 0.08 & 0.40 & 0.06 & 0.13 & 0.39
        & 0.04 & 0.06 & 0.31 & 0.04 & 0.10 & 0.31 \\  
    & Post-scan 
        & 1.85 & 2.38 & 2.66 & 3.09 & 4.06 & 5.53 
        & 1.15 & 1.20 & 1.51 & 2.32 & 2.38 & 2.94 \\  
    & Total 
        & \textbf{2.69} & \textbf{3.39} & \textbf{4.43} & \textbf{4.04} & \textbf{5.16} & 7.31
        & \textbf{1.77} & \textbf{1.87} & 2.75 & 2.95 & 3.11 & 4.17 \\ 
	\midrule    
    \multirow{4}{*}{\parbox{\runtimefirstcol}{BMS}} 
    & Pre-scan 
        & 0.88 & 0.84 & 1.11 & 0.83 & 0.93 & 1.35  
        & 0.57 & 0.58 & 0.62 & 0.57 & 0.58 & 0.62 \\ 
    & Scan 
        & 0.04 & 0.05 & 0.08 & 0.04 & 0.05 & 0.08  
        & 0.03 & 0.04 & 0.06 & 0.03 & 0.04 & 0.06 \\  
    & Post-scan 
        & 3.04 & 3.28 & 3.97 & 3.78 & 4.37 & 5.08  
        & 1.22 & 1.27 & 1.33 & 2.27 & 2.29 & 2.36 \\  
    & Total 
        & 3.96 & 4.17 & 5.15 & 4.65 & 5.35 & \textbf{6.52}  
        & 1.82 & 1.89 & \textbf{2.02} & \textbf{2.88} & \textbf{2.90} & \textbf{3.04} \\ 
	\midrule    
    \multirow{4}{*}{\parbox{\runtimefirstcol}{RB-sort}} 
    & Labeling 
        & 1.69 & 1.67 & 1.67 & 1.69 & 1.67 & 1.67 
        & 1.16 & 1.15 & 1.14 & 1.13 & 1.15 & 1.13 \\ 
    & Sorting 
        & 4.39 & 4.87 & 6.98 & 5.81 & 7.17 & 10.58 
        & 2.97 & 3.00 & 3.11 & 4.11 & 4.16 & 4.38 \\  
    & (un)Packing 
        & -- & -- & -- & 5.66 & 5.67 & 5.67 
        & -- & -- & -- & 4.51 & 4.50 & 4.52 \\  
    & Total 
        & 6.08 & 6.53 & 8.65 & 13.13 & 14.51 & 17.92
        & 4.13 & 4.15 & 4.24 & 9.75 & 9.81 & 10.04 \\ 
    \bottomrule
  \end{tabular}
  \caption{Average running time (ms) for different stages of our multisplit approaches and reduced-bit sort, with $n=2^{25}$ and a varying number of buckets.}\label{table:timing}
\end{table}

\begin{figure}
\centering
\subfloat[Key-only: K40c (ECC on)]{
        \includegraphics[width=0.32\linewidth]{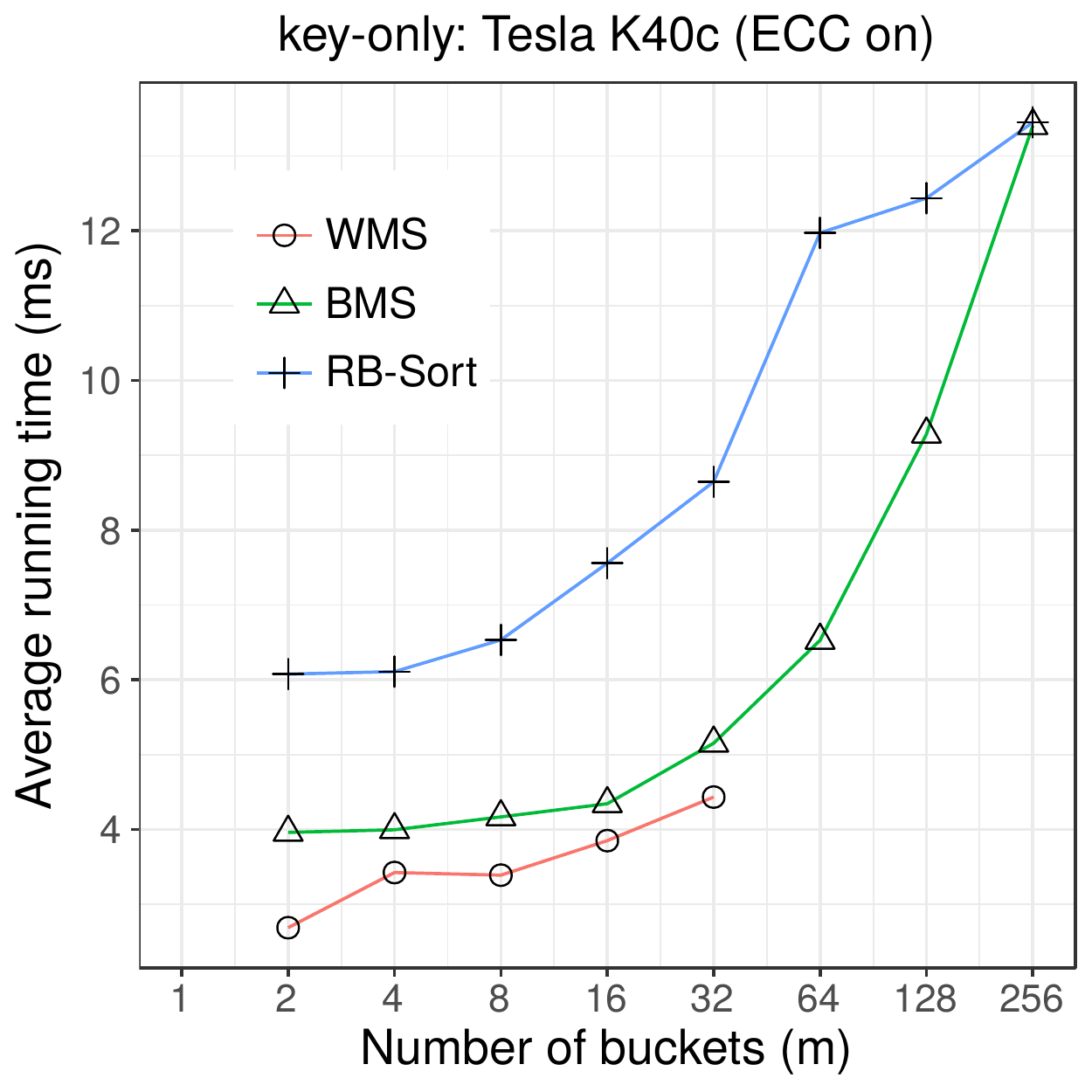}
}
\subfloat[Key-only: K40c (ECC off)]{
        \includegraphics[width=0.32\linewidth]{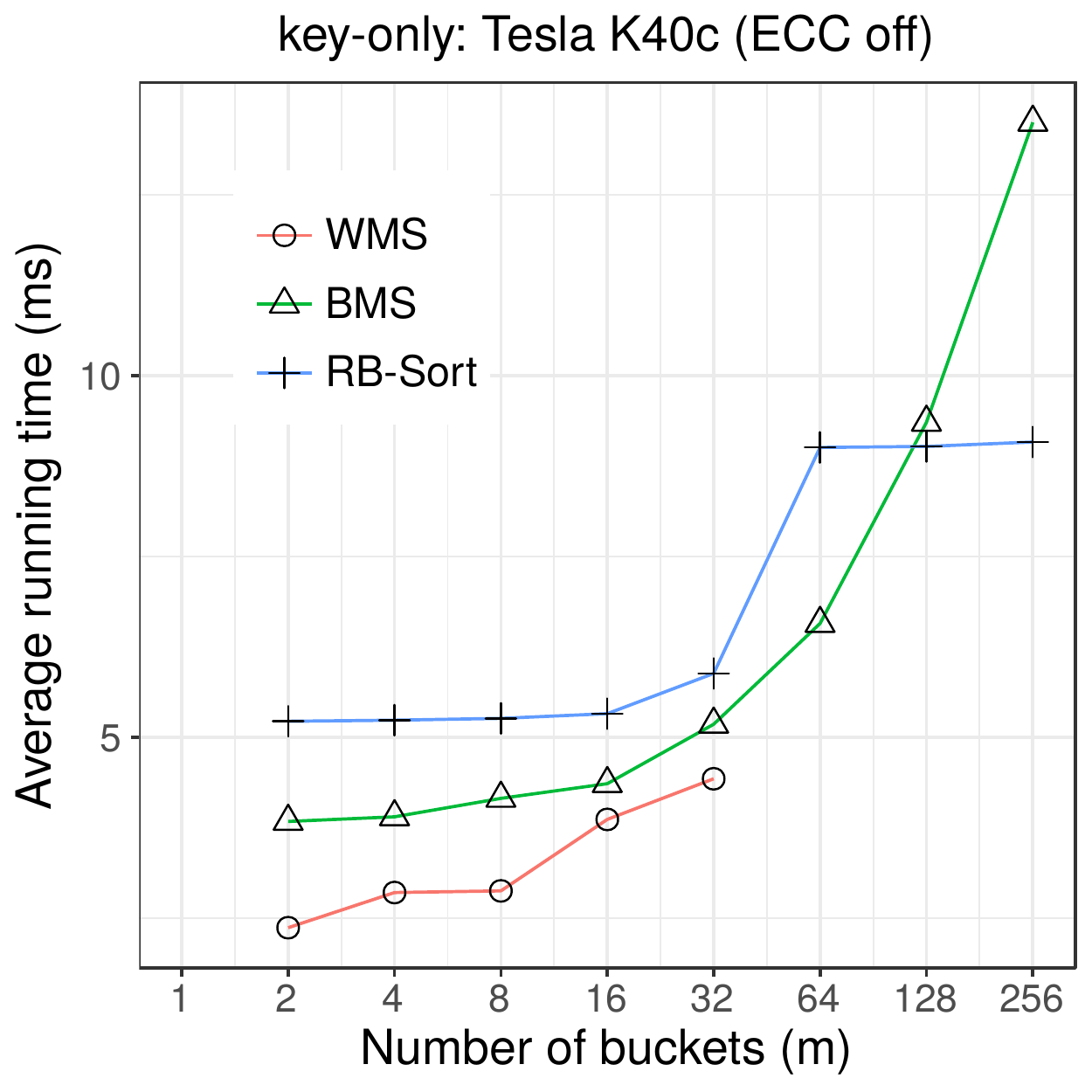}
}
\subfloat[Key-only: GTX 1080]{
        \includegraphics[width=0.32\linewidth]{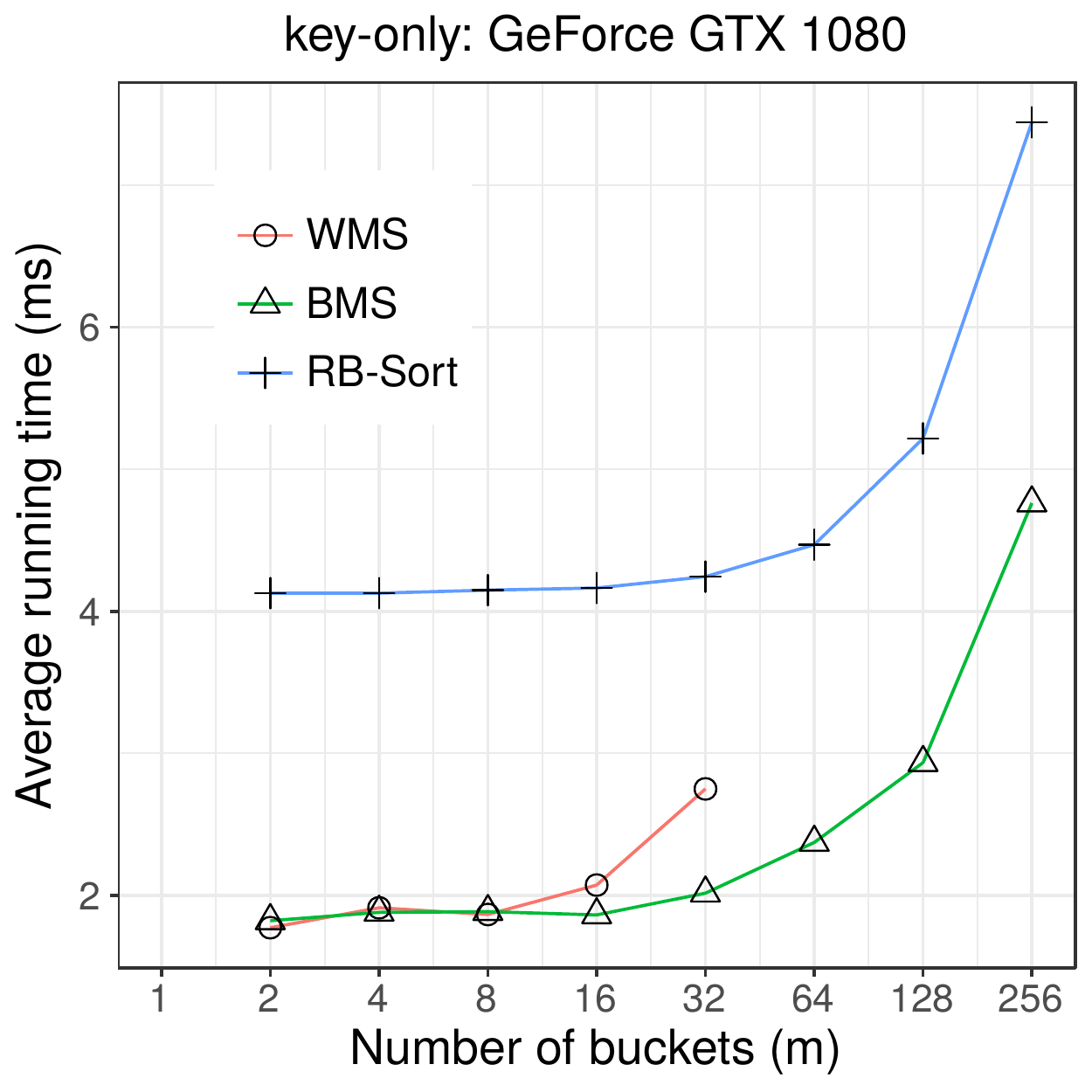}
}\\
\subfloat[Key-value: K40c (ECC on)]{
        \includegraphics[width=0.32\linewidth]{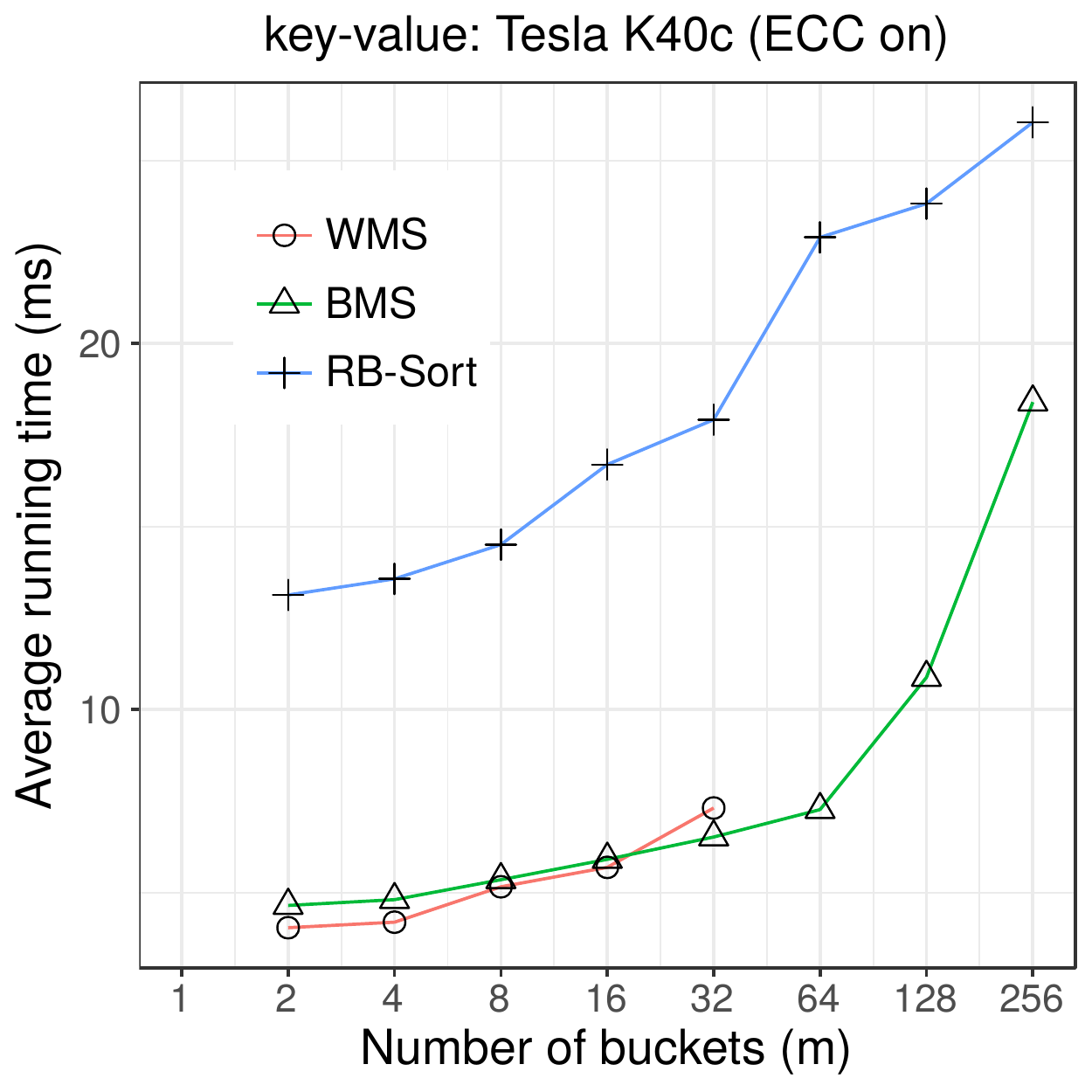}
}
\subfloat[Key-value: K40c (ECC off)]{
        \includegraphics[width=0.32\linewidth]{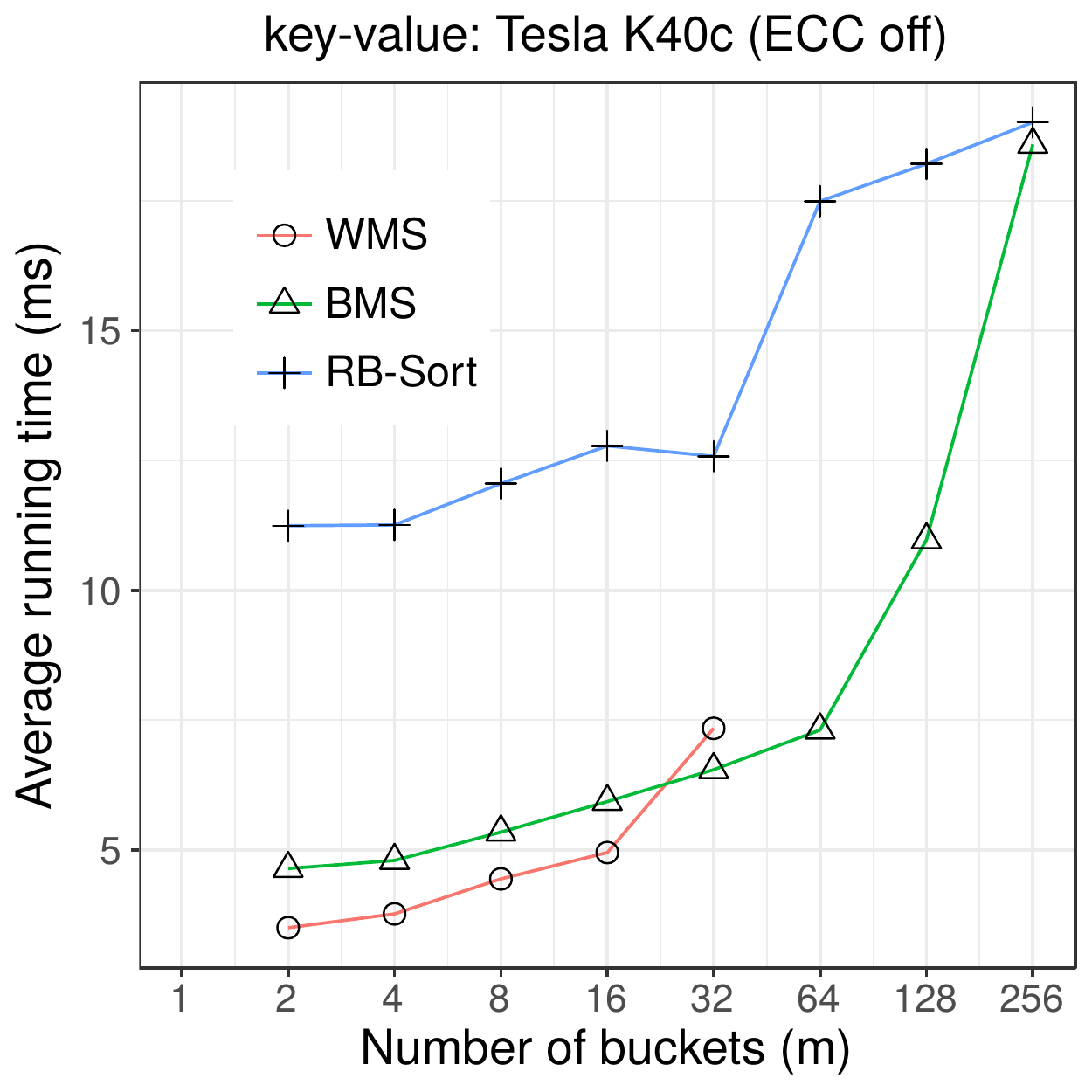}
}
\subfloat[Key-value: GTX 1080]{
        \includegraphics[width=0.32\linewidth]{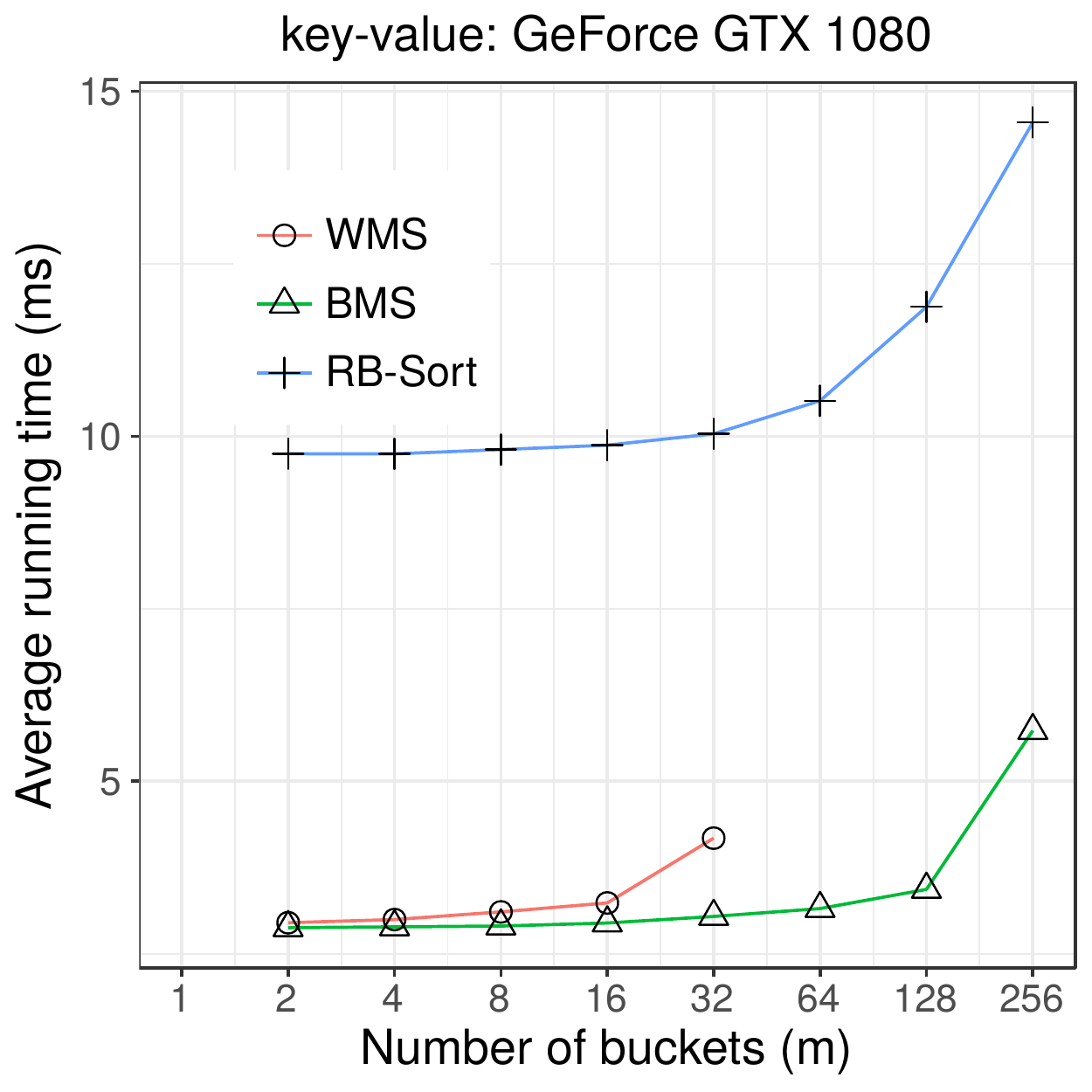}
}
\caption{Average running time (ms) versus number of buckets for all multisplit methods: (a,b,c) key-only, 32~M elements (d,e,f) key-value, 32~M elements.} \label{fig:avg_time}
\end{figure}

Figure~\ref{fig:avg_time} shows the average running time of our multisplit algorithms versus the number of buckets ($m$).
For small $m$, BMS has the best locality (at the cost of substantial local work), but WMS achieves fairly good locality coupled with simple local computation; it is the fastest choice for small $m$ ($\leq 32$ [key-only, Tesla K40c], $\leq 16$ [key-value, Tesla K40c], and $\leq2$ [key-only, GeForce GTX 1080]).
For larger $m$, the superior memory locality of BMS coupled with a minimized global scan cost makes it the best method overall.

Our multisplit methods are also almost always superior to the RB-sort method (except for the $m \geq 128$ key-only case on Tesla K40c with ECC off).
This is partly because of the extra overheads that we introduced for bucket identification and creating the label vector, and packing/unpacking stages for key-value multisplit.
Even if we ignore these overheads, since RB-sort performs its operations and permutations over the label vector as well as original key (key-value) elements, its data movements are more expensive compared to all our multisplit methods that instead only process and permute original key (key-value) elements.\footnote{In our comparisons against our own multisplit methods, RB-sort will be the best sort-based multisplit method as long as our bucket identifier cannot be interpreted as a selection of some consecutive bits in its key's binary representation (i.e., $\defn{f}(u) = (u \gg k) \& (2^{r}-1)$ for some $k$ and $r$). Otherwise, these cases can be handled directly by a radix sort over a selection of bits (from the $k$-th bit until the $(k+r)$-th bit) and do not require the extra overhead that we incur in RB-sort (i.e., sorting certain bits from input keys is equivalent to a stable multisplit solution). We will discuss this more thoroughly in Section~\ref{subsec:multisplit_sort}.}

\begin{figure}
\centering
\subfloat[Key-only: K40c (ECC on)]{
        \includegraphics[width=0.32\linewidth]{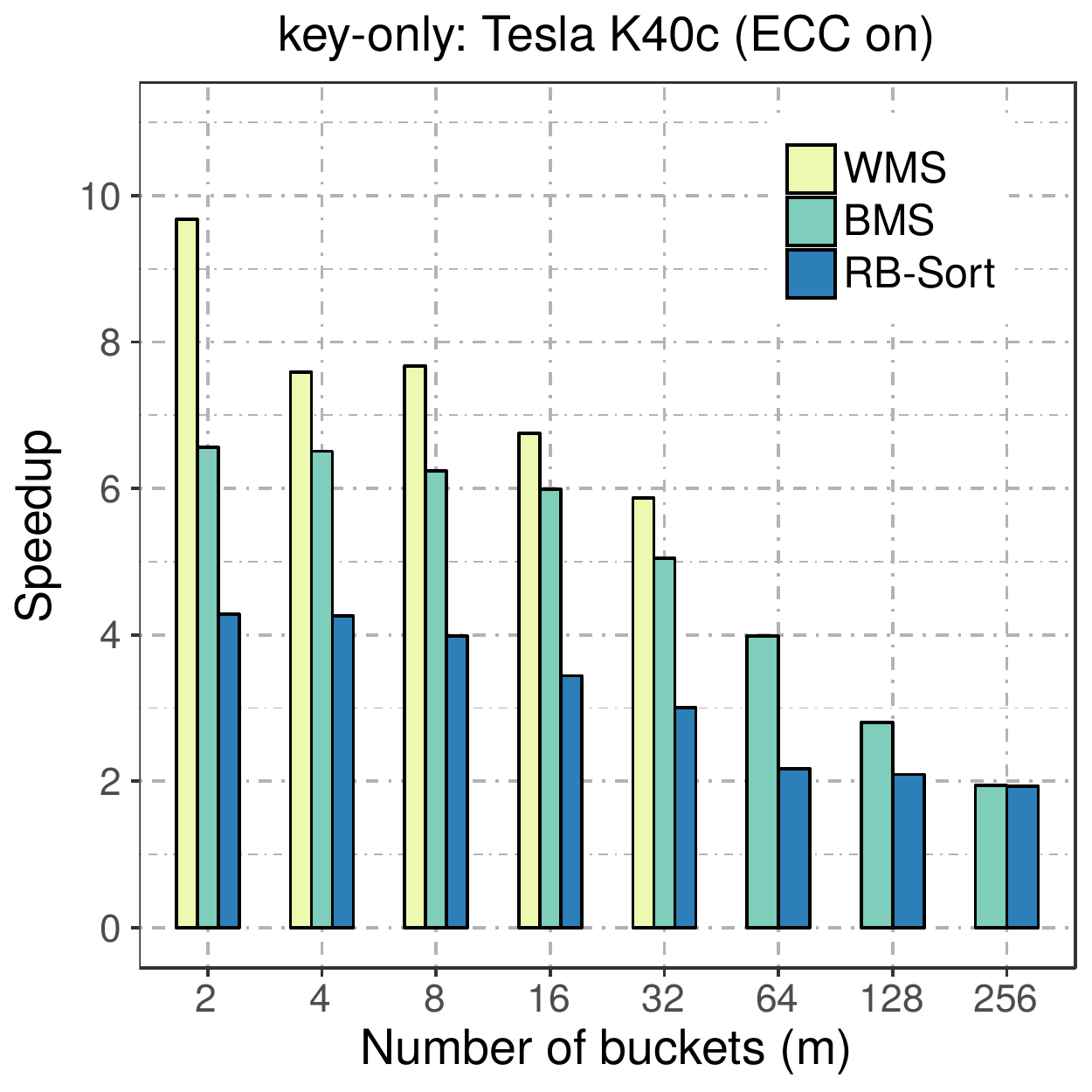}
}
\subfloat[Key-only: K40c (ECC off)]{
        \includegraphics[width=0.32\linewidth]{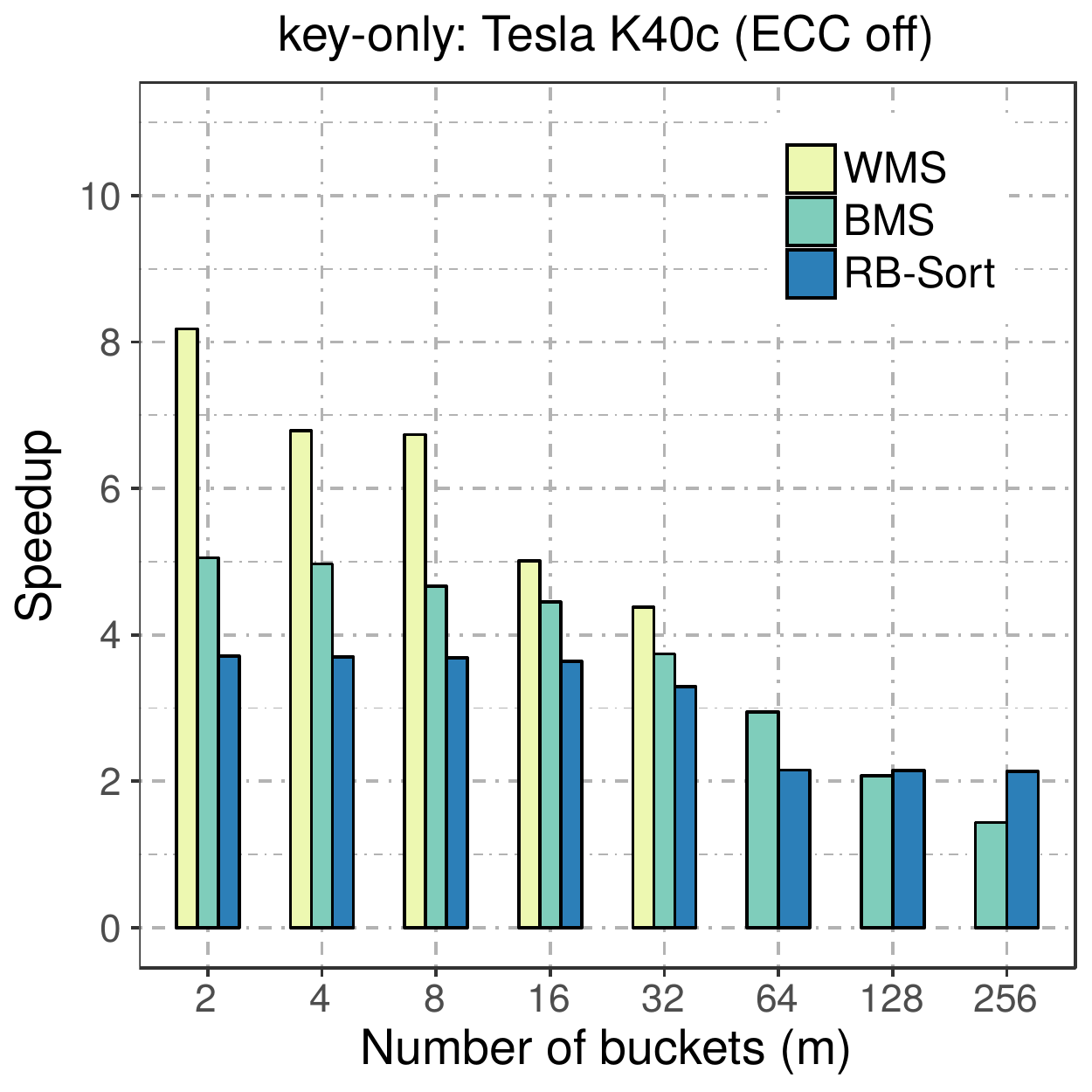}
}
\subfloat[Key-only: GTX 1080]{
        \includegraphics[width=0.32\linewidth]{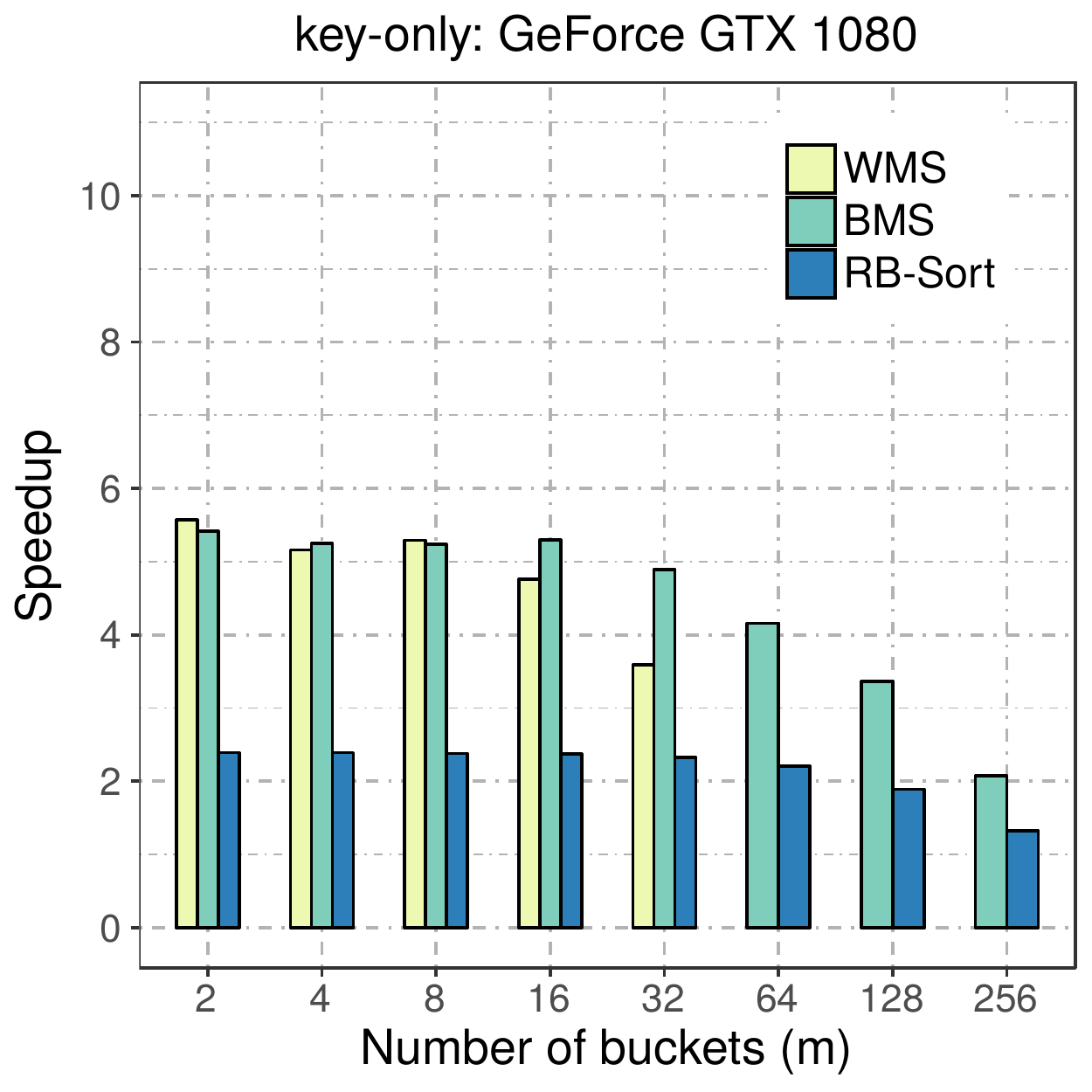}
}\\
\subfloat[Key-value: K40c (ECC on)]{
        \includegraphics[width=0.32\linewidth]{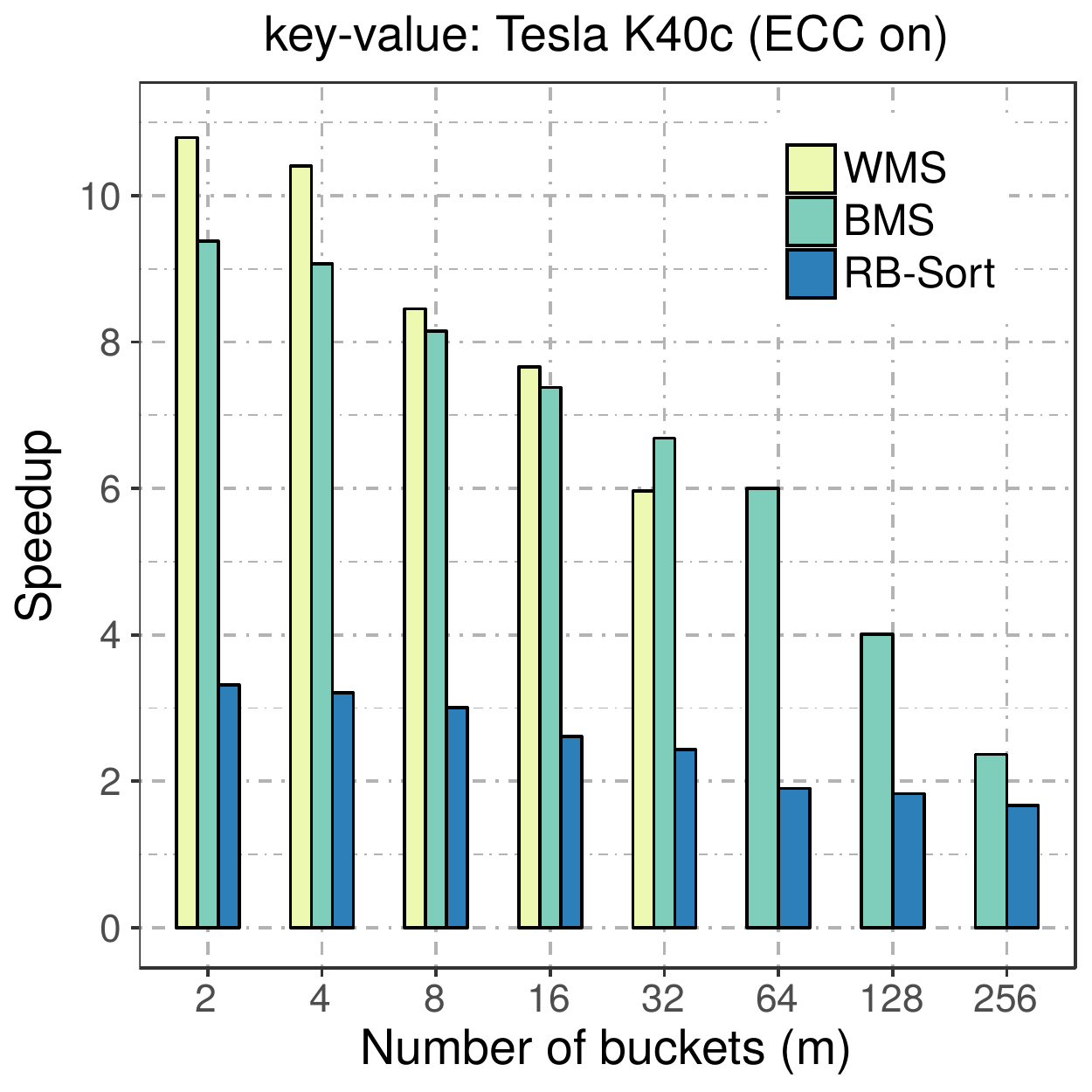}
}
\subfloat[Key-value: K40c (ECC off)]{
        \includegraphics[width=0.32\linewidth]{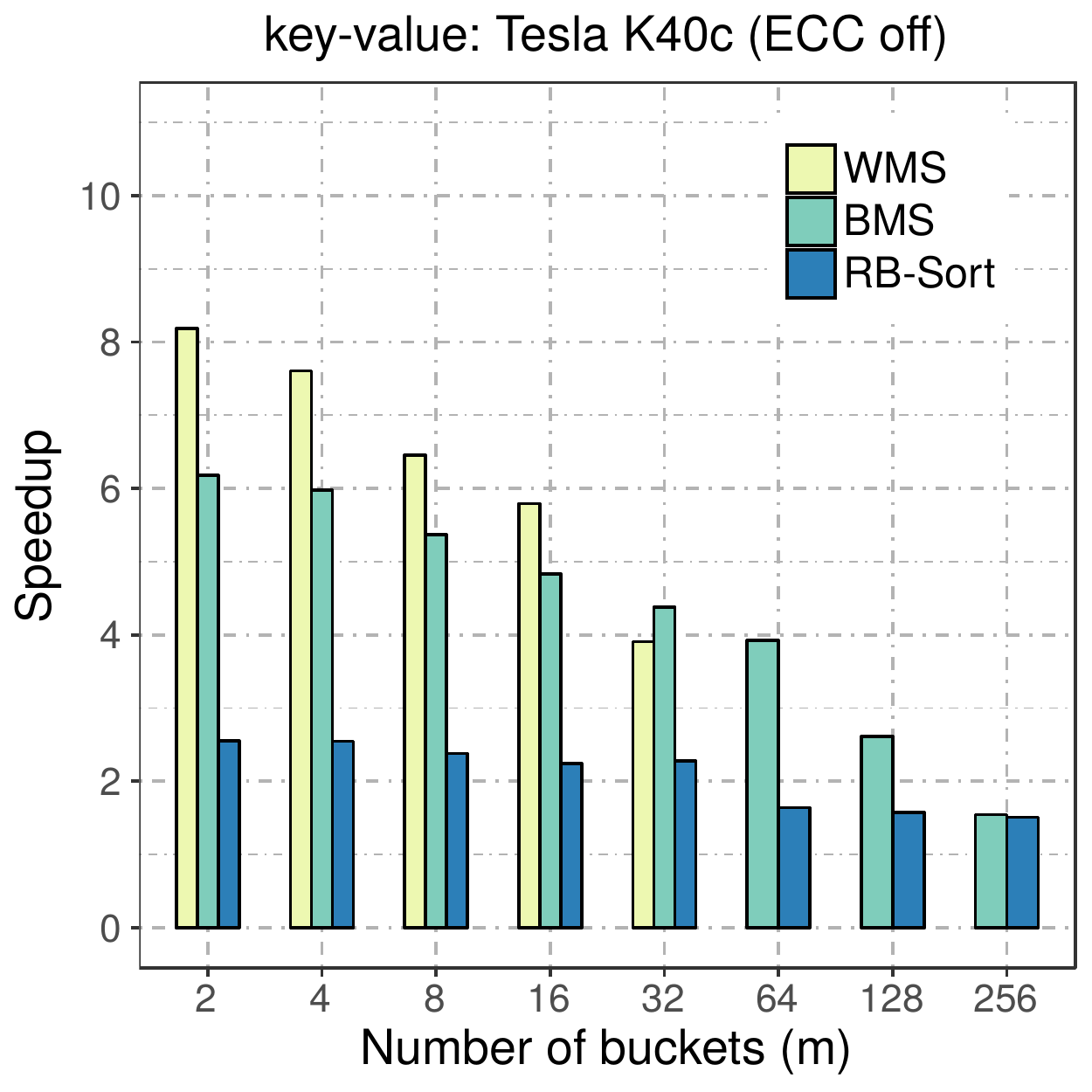}
}
\subfloat[Key-value: GTX 1080]{
        \includegraphics[width=0.32\linewidth]{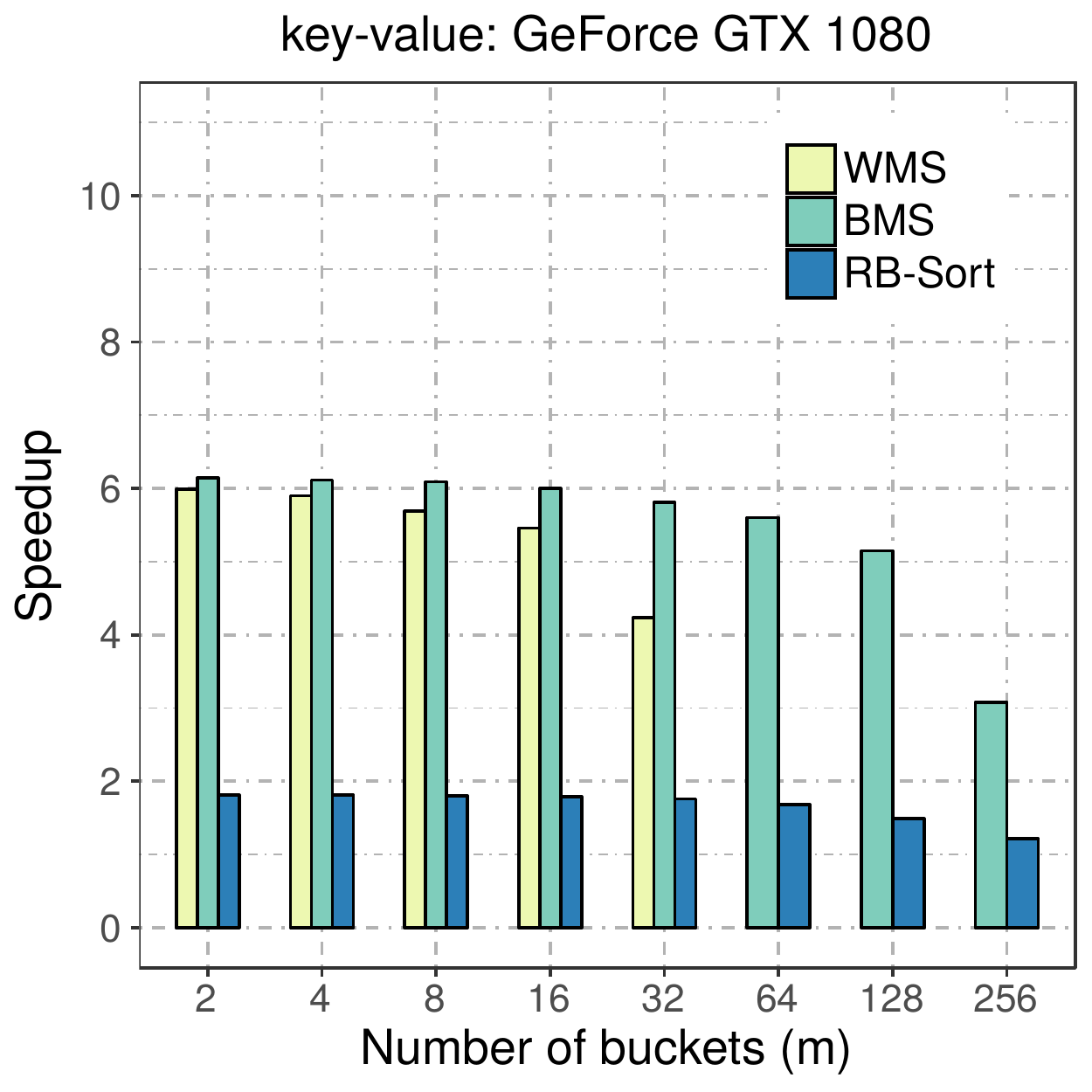}
}
\caption{Achieved speedup against regular radix sort versus number of buckets for all multisplit methods: (a,b,c) key-only, (d,e,f) key-value. Both scenarios are over 32~M elements random elements uniformly distributed among buckets, and delta-bucket identifiers.} \label{fig:speedup}
\end{figure}

For our user-defined delta-buckets and with a uniform distribution of keys among all 32-bit integers, by comparing Table~\ref{table:reference}~and~Table~\ref{table:timing} it becomes clear that our multisplit method outperforms radix sort by a significant margin. Figure~\ref{fig:speedup} shows our achieved speedup against the regular 32-bit radix sort performance (Table~\ref{table:reference}).
We can achieve up to 9.7x (and 10.8x) for key-only (and key-value) multisplits against radix sort.

\subsubsection{Processing rate, and multisplit speed of light}\label{sec:perf_rate}
It is instructive to compare any implementation to its ``speed of light'': a processing rate that could not be exceeded.
For multisplit's speed of light, we consider that computations take no time and all memory accesses are fully coalesced. Our parallel model requires one single global read of all elements before our global scan operation to compute histograms. We assume the global scan operation is free. Then after the scan operation, we must read all keys (or key-value pairs) and then store them into their final positions. For multisplit on keys, we thus require 3 global memory accesses per key; 5 for key-value pairs.
Our Tesla K40c has a peak memory bandwidth of 288~GB/s, so the speed of light for keys, given the many favorable assumptions we have made for it, is 24~Gkeys/s, and for key-value pairs is 14.4~G~pairs/s.
Similarly, our GTX 1080 has 320~GB/s memory bandwidth and similar computations give us a speed of light of 26.6 G~keys/s for key-only case and 16~Gpairs/s for key-value pairs.

Table~\ref{table:ms_rate} shows our processing rates for 32M keys and key-value pairs using delta-buckets and with keys uniformly distributed among all buckets.
WMS has the highest peak throughput (on 2 buckets): 12.48~Gkeys/s on Tesla K40c (ECC on), 14.15~Gkeys/s on Tesla K40c (ECC off), and 18.93~Gkeys/s on GeForce GTX 1080.
We achieve more than half the speed of light performance (60\% on Tesla K40c and 71\% on GeForce GTX 1080) with 2 buckets.
As the number of buckets increases, it is increasingly more costly to sweep all input keys to compute final permutations for each element.
We neglected this important part in our speed of light estimation.
With 32 buckets, we reach 7.57 G~keys/s on Tesla K40c and 16.64 G~keys/s on GeForce GTX 1080. While this is less than the 2-bucket case, it is still a significant fraction of our speed of light estimation (32\% and 63\% respectively).

The main obstacles in achieving the speed of light performance are 1)~non-coalesced memory writes and 2)~the non-negligible cost that we have to pay to sweep through all elements and compute permutations.
The more registers and shared memory that we have (fast local storage as opposed to the global memory), the easier it is to break the whole problem into larger subproblems and localize required computations as much as possible. This is particularly clear from our results on the GeForce GTX 1080 compared to the Tesla K40c, where our performance improvement is proportionally more than just the GTX 1080's global memory bandwidth improvement (presumably because of more available shared memory per SM).
\begin{table}
\centering
\tiny
\resizebox{\columnwidth}{!}{
\begin{tabular}{lll cccccccc}
\toprule
& & & \multicolumn{8}{c}{Throughput (speedup against radix-sort)} \\
\cmidrule{4-11}
& & & \multicolumn{8}{c}{Number of buckets (m)} \\
\cmidrule{4-11}
& & Method  & 2 & 4 & 8 & 16 & 32 & 64 & 128 & 256 \\
\midrule
    \multirow{8}{*}{\begin{turn}{90}K40c (ECC on)\end{turn}}
    & \multirow{4}{*}{\begin{turn}{90}\tiny key-only\end{turn}}
    & DMS
        & 8.79 (6.8 x) &  8.36 (6.5 x) &  6.91 (5.4 x) &  6.90 (5.4 x) &  3.26 (2.5 x) & -- & -- & -- \\ 
    & & WMS 
        & \textbf{12.48 (9.7 x)} &  \textbf{9.79 (7.6 x)} &  \textbf{9.90 (7.7 x)} &  \textbf{8.71 (6.8 x)} &  \textbf{7.57 (5.9 x)} & -- & -- & -- \\ 
    & & BMS
        & 8.47 (6.6 x) &  8.39 (6.5 x) &  8.05 (6.2 x) &  7.72 (6.0 x) &  6.51 (5.0 x) &  \textbf{5.14 (4.0 x)} &  \textbf{3.61 (2.8 x)} &  \textbf{2.50 (1.9 x)} \\
    & & RB-sort  
        & 5.52 (4.3 x) &  5.49 (4.3 x) &  5.14 (4.0 x) &  4.44 (3.4 x) &  3.88 (3.0 x) &  2.80 (2.2 x) &  2.70 (2.1 x) &  2.50 (1.9 x) \\
\cmidrule{2-11}
    & \multirow{4}{*}{\begin{turn}{90}\tiny key-value\end{turn}} 
    & DMS
        & 6.90 (9.0 x) &  6.31 (8.2 x) &  5.62 (7.3 x) &  5.62 (7.3 x) &  1.94 (2.5 x) & -- & -- & -- \\
    & & WMS 
        & \textbf{8.31 (10.8 x)} &  \textbf{8.01 (10.4 x)} &  \textbf{6.51 (8.5 x)} &  \textbf{5.90 (7.7 x)} &  4.59 (6.0 x) & -- & -- & -- \\ 
    & & BMS
        & 7.22 (9.4 x) &  6.98 (9.1 x) &  6.27 (8.1 x) &  5.68 (7.4 x) &  \textbf{5.15 (6.7 x)} &  \textbf{4.62 (6.0 x)} &  \textbf{3.09 (4.0 x)} &  \textbf{1.82 (2.4 x)} \\ 
    & & RB-sort  
        & 2.56 (3.3 x) &  2.47 (3.2 x) &  2.31 (3.0 x) &  2.01 (2.6 x) &  1.87 (2.4 x) &  1.47 (1.9 x) &  1.41 (1.8 x) &  1.29 (1.7 x) \\  
\midrule
\midrule
    \multirow{8}{*}{\begin{turn}{90}K40c (ECC off)\end{turn}}
    & \multirow{4}{*}{\begin{turn}{90}\tiny key-only\end{turn}} 
    & DMS
        & 8.99 (5.2 x) &  8.52 (4.9 x) &  6.98 (4.0 x) &  4.94 (2.9 x) &  3.26 (1.9 x) & -- & -- & -- \\
    & & WMS 
        & \textbf{14.15 (8.2 x)} &  \textbf{11.74 (6.8 x)} &  \textbf{11.65 (6.7 x)} &  \textbf{8.68 (5.0 x)} &  \textbf{7.57 (4.4 x)} & -- & -- & -- \\ 
    & & BMS
        & 8.74 (5.1 x) &  8.59 (5.0 x) &  8.07 (4.7 x) &  7.69 (4.4 x) &  6.47 (3.7 x) &  \textbf{5.10 (2.9 x)} &  3.59 (2.1 x) &  2.48 (1.4 x)  \\ 
    & & RB-sort
        & 6.42 (3.7 x) &  6.40 (3.7 x) &  6.37 (3.7 x) &  6.30 (3.6 x) &  5.70 (3.3 x) &  3.72 (2.2 x) &  \textbf{3.72 (2.1 x)} &  \textbf{3.69 (2.1 x)} \\ 
\cmidrule{2-11}
    & \multirow{4}{*}{\begin{turn}{90}\tiny key-value\end{turn}} 
    & DMS 
        & 8.99 (7.7 x) &  7.05 (6.0 x) &  5.71 (4.9 x) &  3.98 (3.4 x) &  1.96 (1.7 x) & -- & -- & -- \\
    & & WMS 
        & \textbf{9.58 (8.2 x)} &  \textbf{8.90 (7.6 x)} &  \textbf{7.55 (6.5 x)} &  \textbf{6.78 (5.8 x)} &  4.57 (3.9 x) & -- & -- & -- \\ 
    & & BMS
        & 7.23 (6.2 x) &  6.99 (6.0 x) &  6.28 (5.4 x) &  5.66 (4.8 x) &  \textbf{5.13 (4.4 x)} &  \textbf{4.59 (3.9 x)} &  \textbf{3.06 (2.6 x)} &  \textbf{1.81 (1.5 x)} \\ 
    & & RB-sort
        & 2.98 (2.6 x) &  2.98 (2.5 x) &  2.78 (2.4 x) &  2.63 (2.2 x) &  2.67 (2.3 x) &  1.92 (1.6 x) &  1.84 (1.6 x) &  1.76 (1.5 x) \\  
\midrule
\midrule
    \multirow{8}{*}{\begin{turn}{90}GTX 1080  \end{turn}}
    & \multirow{4}{*}{\begin{turn}{90}\tiny key-only\end{turn}} 
    & DMS
    & 17.67 (5.2 x) &  14.38 (4.2 x) &  11.00 (3.2 x) &  7.73 (2.3 x) &  5.54 (1.6 x) & -- & -- & -- \\
    & & WMS 
        & \textbf{18.93 (5.6 x)} &  17.54 (5.2 x) &  \textbf{17.98 (5.3 x)} &  16.18 (4.8 x) &  12.20 (3.6 x) & -- & -- & -- \\ 
    & & BMS
        & 18.42 (5.4 x) &  \textbf{17.84 (5.2 x)} &  17.79 (5.2 x) &  \textbf{18.01 (5.3 x)} &  \textbf{16.64 (4.9 x)} &  \textbf{14.14 (4.2 x)} &  \textbf{11.43 (3.4 x)} &  \textbf{7.05 (2.1 x)} \\ 
    & & RB-sort
        & 8.13 (2.4 x) &  8.13 (2.4 x) &  8.09 (2.4 x) &  8.06 (2.4 x) &  7.91 (2.3 x) &  7.51 (2.2 x) &  6.43 (1.9 x) &  4.51 (1.3 x) \\ 
\cmidrule{2-11}
    & \multirow{4}{*}{\begin{turn}{90}\tiny key-value\end{turn}} 
    & DMS
        & 11.17 (5.9 x) &  9.75 (5.1 x) &  7.07 (3.7 x) &  4.95 (2.6 x) &  3.51 (1.8 x) & -- & -- & -- \\
    & & WMS 
        & 11.38 (6.0 x) &  11.21 (5.9 x) &  10.81 (5.7 x) &  10.37 (5.5 x) &  8.04 (4.2 x) & -- & -- & -- \\ 
    & & BMS
        & \textbf{11.67 (6.1 x)} &  \textbf{11.62 (6.1 x)} &  \textbf{11.57 (6.1 x)} &  \textbf{11.40 (6.0 x)} &  \textbf{11.04 (5.8 x)} &  \textbf{10.64 (5.6 x)} &  \textbf{9.78 (5.1 x)} &  \textbf{5.85 (3.1 x)}  \\ 
    & & RB-sort
        & 3.44 (1.8 x) &  3.44 (1.8 x) &  3.42 (1.8 x) &  3.40 (1.8 x) &  3.34 (1.8 x) &  3.19 (1.7 x) &  2.83 (1.5 x) &  2.31 (1.2 x) \\  

\bottomrule
\end{tabular}
}
  \caption{Multisplit with delta-buckets and $2^{25}$ random keys uniformly distributed among $m$ buckets. Achieved processing rates (throughput) are shown in Gkeys/s (or Gpairs/s for key-value pairs). In parenthesis speedup against regular CUB's radix-sort over input elements are shown. }\label{table:ms_rate}
\end{table}

\subsubsection{Performance on different GPU microarchitectures}\label{subsec:perf_architecture}
In our design we have not used any (micro)architecture-dependent optimizations and hence we do not expect radically different behavior on different GPUs, other than possible speedup differences based on the device's capability.
Here, we briefly discuss some of the issues related to hardware differences that we observed in our experiments.

\paragraph{Tesla K40c} It is not yet fully disclosed whether disabling ECC (which is a hardware feature and requires reboot after modifications) has any direct impact besides available memory bandwidth (such as available registers, etc.).
For a very small number of buckets, our local computations are relatively cheap and hence having more available bandwidth (ECC off compared to ECC on) results in better overall performance (Table~\ref{table:ms_rate}).
The performance gap, however, decreases as the number of buckets increases.
This is mainly because of computational bounds due to the increase in ballot, shuffle, and numerical integer operations as $m$ grows.

CUB's radix sort greatly improves on Tesla K40c when ECC is disabled (Table~\ref{table:reference}), and because of it, RB-sort improves accordingly.
CUB has particular architecture-based fine-grained optimizations, and we suspect it is originally optimized for when ECC is disabled to use all hardware resources to exploit all available bandwidth as much as possible. We will discuss CUB further in Section~\ref{subsec:multisplit_sort}.
RB-sort's speedups in Fig.~\ref{fig:speedup} are relatively less for when ECC is disabled compared to when it is enabled. The reason is not because RB-sort performs worse (Table~\ref{table:ms_rate} shows otherwise), but rather because  CUB's regular radix sort (that we both use in RB-sort and compare against for speedup computations) improves when ECC is disabled (Table~\ref{table:reference}).

\paragraph{GeForce GTX 1080}
This GPU is based on NVIDIA's latest ``Pascal'' architecture. It both increases global memory bandwidth (320~GB/s) and appears to be better at hiding memory latency caused by non-coalesced memory accesses. The GTX 1080 also has more available shared memory per SM, which results in more resident thread-blocks within each SM.
As a result, it is much easier to fully occupy the device, and our results (Table~\ref{table:ms_rate}) show this.
\subsection{Performance for more than 256 buckets}\label{subsec:perf_more}
So far, we have only characterized problems with $m \leq 256$ buckets.
As we noted in Section~\ref{sec:radix}, we expect that as the number of buckets increases, multisplit converges to a sorting problem and we should see the performance of our multisplits and sorting-based multisplits converge as well.

The main obstacle for efficient implementation of our multisplits for large  bucket counts is the limited amount of shared memory available on GPUs for each thread-block.
Our methods rely on having privatized portions of shared memory with $m$ integers per warp (total of $32m$ bits/warp).
As a result, as $m$ increases, we require more shared storage, which limits the number of resident thread-blocks per SM, which limits our ability to hide memory latency and hurts our performance.
Even if occupancy was not the main issue, with the current GPU shared memory sizes (48~KB per SM for Tesla K40c, and 96~KB per SM to be shared by two blocks on GeForce GTX 1080), it would only be physically possible for us to scale our multisplit up to about $m = 12\text{k}/{N_\text{warp}}$ (at most 12k buckets if we use only one warp per thread-block).

In contrast, RB-sort does not face this problem. Its labeling stage (and the packing/unpacking stage required for key-value pairs) are independent of the number of buckets. However, the radix sort used for RB-sort's sorting stage is itself internally scaled by a factor of $\log m$, which results in an overall logarithmic dependency for RB-sort vs.\ the number of buckets.

\paragraph{Solutions for larger multisplits ($m>256$)}
Our solution for handling more buckets is similar to how radix sort handles the same scalability problem: iterative usage of multisplit over $m' \leq 256$ buckets. However, this is not a general solution and it may not be possible for any general bucket identifier.
For example, for delta-buckets with 257 buckets, we can treat the first 2 buckets together as one single super-bucket which makes the whole problem into 256 buckets. Then, by two rounds of multisplit 1)~on our new 256 buckets, 2)~on the initial first 2 buckets (the super-bucket), we can have a stable multisplit result.
This approach can potentially be extended to any number of buckets, but only if our bucket identifier is suitable for such modifications.
There are some hypothetical cases for which this approach is not possible (for instance, if our bucket identifier is a random hash function, nearby keys do not necessarily end up in nearby buckets).

Nevertheless, if iterative usage is not possible, it is best to use RB-sort instead, as it appears to be quite competitive for a very large number of buckets. As a comparison with regular radix sort performance, on Tesla K40c (ECC on), RB-sort outperforms radix sort up to almost 32k keys and 16k key-value pairs. However, we reiterate that unlike RB-sort, which is always a correct solution for multisplit problems, direct usage of radix sort is not always a possible solution.

\subsection{Initial key distribution over buckets}\label{subsec:perf_distribution}
So far we have only considered scenarios in which initial key elements were uniformly distributed over buckets (i.e., a uniform histogram).
In our implementations we have considered small subproblems (warp-sized for WMS and block-sized for BMS) compared to the total size of our initial key vector.
Since these subproblems are relatively small, having a non-uniform distribution of keys means that we are more likely to see empty buckets in some of our subproblems; in practice, our methods would behave as if there were fewer buckets for those subproblems.
All of our \emph{computations} (e.g., warp-level histograms) are data-independent and, given a fixed bucket count, would have the same performance for any distribution. However, our \emph{data movement}, especially after reordering, would benefit from having more elements within fewer buckets and none for some others (resulting in better locality for coalesced global writes). Consequently, the uniform distribution is the worst-case scenario for our methods.

As an example of a non-uniform distribution, consider the binomial distribution.
In general $B(m-1,p)$ denotes a binomial distribution over $m$ buckets with a probability of success $p$. For example, the probability that a key element belongs to bucket $0\leq k < m$ is $\binom{m-1}{k}p^k {(1-p)}^{m-k-1}$.
This distribution forces an unbalanced histogram as opposed to the uniform distribution.
Note that by choosing $p=0.5$, the expected number of keys within the $k$th bucket will be $n\binom{m-1}{k} 2^{1-m}$.
For example, with $n=2^{25}$ elements and $m=256$ total buckets, there will be on average almost 184 empty buckets (72\%).
Such an extreme distribution helps us evaluate the sensitivity of our multisplit algorithms to changes in input distribution.

\begin{figure}
  \centering
  \subfloat[Key-only]
  {
    \includegraphics[width = 0.4\linewidth]{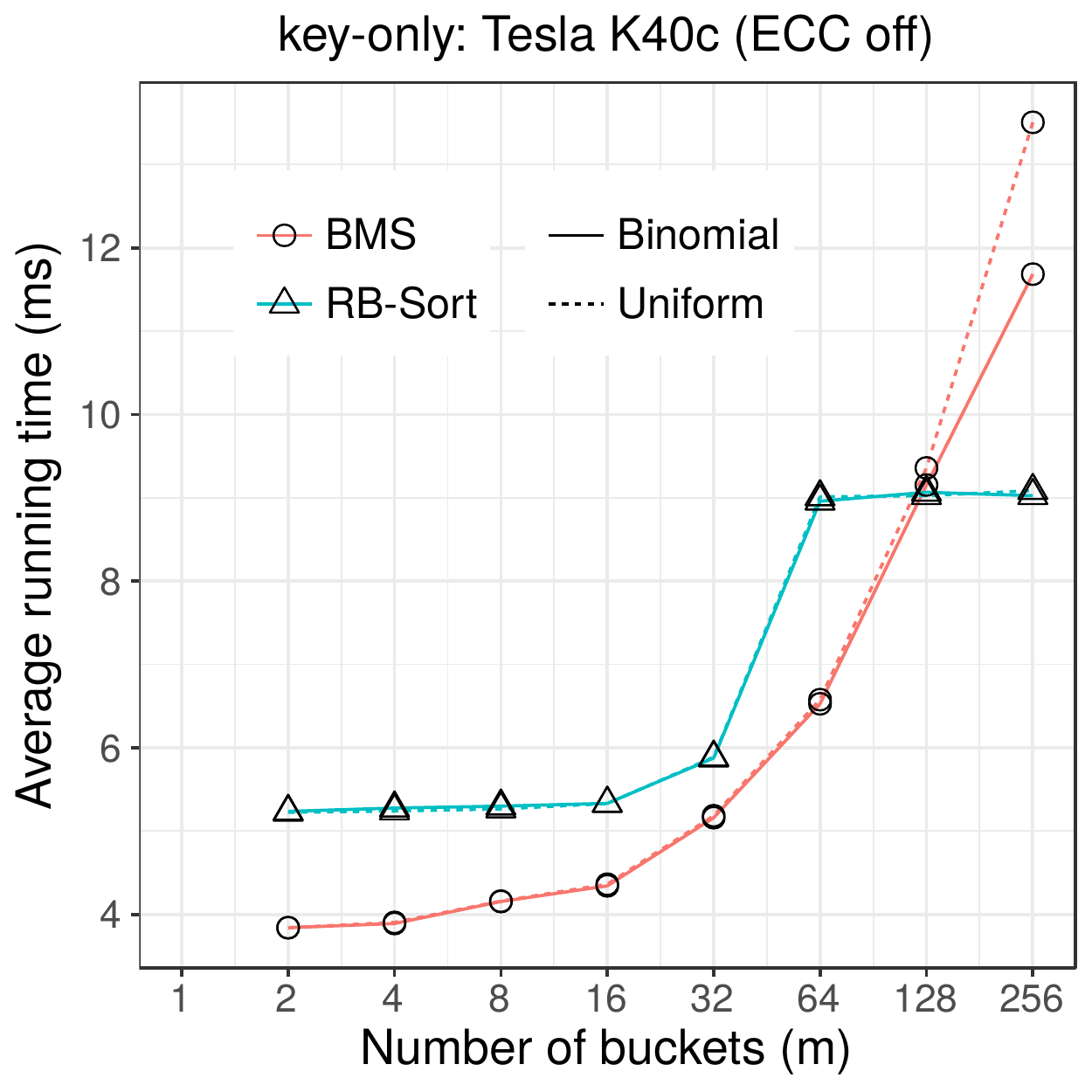}
    \label{fig:dist_k}
  }
  \subfloat[Key-value]
  {
    \includegraphics[width = 0.4\linewidth]{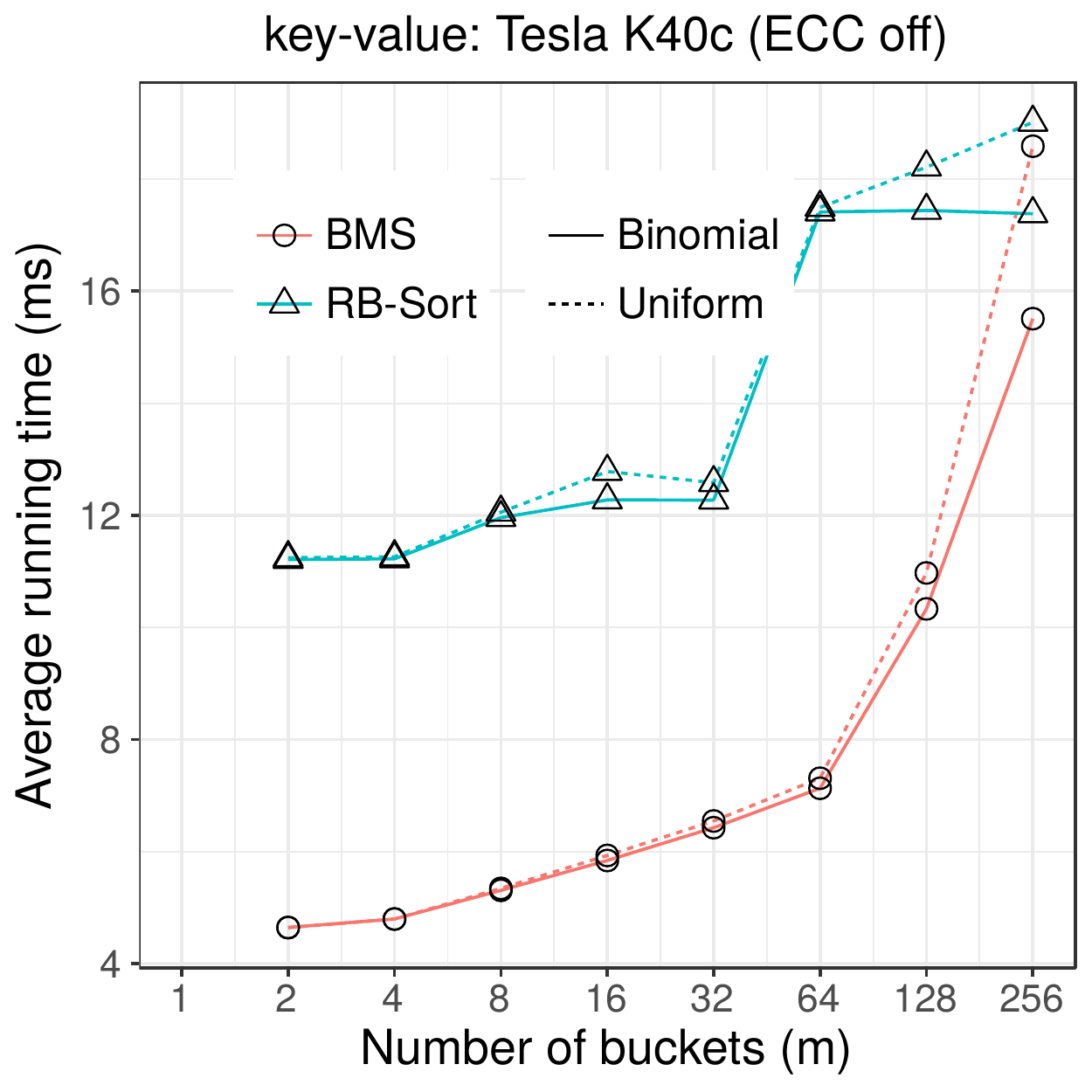}
    \label{fig:dist_kv}
  }
  \caption{Average running time (ms) vs.\ number of buckets ($m$) for two different initial key distributions: (a) a uniform distribution and (b) the binomial distribution $B(m-1,0.5)$.}\label{fig:binomial}
\end{figure}

Figure~\ref{fig:binomial} shows the average running time versus the number of buckets for BMS and RB-sort with binomial and uniform distributions, on our Tesla K40c (ECC off).
There are two immediate observations here.
First, as the number of buckets increases, both algorithms become more sensitive to the input distribution of keys.
This is mainly because, on average, there will be more empty buckets and our data movement will resemble situations where there are essentially much fewer number of buckets than the actual $m$.
Second, the sensitivity of our algorithms increases in key-value scenarios when compared to key-only scenarios, mainly because data movement is more expensive in the latter. As a result, any improvement in our data movement patterns (here caused by the input distribution of keys) in a key-only multisplit will be almost doubled in a key-value multisplit.

To get some statistical sense over how much improvement we are getting, we ran multiple experiments with different input distributions, with delta-buckets as our bucket identifiers, and on different GPUs. Table~\ref{table:distribution} summarizes our results with $m=256$ buckets.
In this table, we also consider a milder distribution where $\alpha n$ of total keys are uniformly distributed among buckets and the rest are within one random bucket ($\alpha$-uniform).
BMS achieves up to 1.24x faster performance on GeForce GTX 1080 when input keys are distributed over buckets in the binomial distribution. Similarly, RB-sort achieve up to 1.15x faster on Tesla K40c (ECC off) with the binomial distribution.
In general, compared to our methods, RB-sort seems to be less sensitive to changes to the input distribution.


\begin{table}
\centering
\scriptsize
\begin{tabular}{lll cc | cc | cc}
\toprule
Type & Method & Distribution & \multicolumn{2}{c}{Tesla K40c (ECC on)} & \multicolumn{2}{c}{Tesla K40c (ECC off)} & \multicolumn{2}{c}{GeForce GTX 1080} \\
\midrule
    \multirow{6}{*}{\begin{turn}{90}Key-only\end{turn}}
    & \multirow{3}{*}{BMS}
    & Uniform
        &  2.50 &  & 2.48 &  & 7.05 &   \\
    & & 0.25-uniform
        &   2.64 & 1.06x & 2.61 & 1.05x & 7.36  & 1.05x    \\
    & & Binomial
        &   2.89 & 1.15x & 2.87  & 1.16x & 7.89 & 1.11x    \\
\cmidrule{2-9}
    & \multirow{3}{*}{RB-sort}
    & Uniform
        &  2.50 &  & 3.69 &  & 4.51 &   \\
    & & 0.25-uniform
        &  2.69  & 1.08x  & 3.71  & 1.00x  & 4.56  & 1.01x   \\
    & & Binomial
        &   2.80  & 1.12x  & 3.72  & 1.01x  & 4.95  & 1.10x    \\
\cmidrule{1-9}
\cmidrule{1-9}
\multirow{6}{*}{\begin{turn}{90}Key-value\end{turn}}
    & \multirow{3}{*}{BMS}
    & Uniform
        &  1.82 &  & 1.81 &  & 5.85 &   \\
    & & 0.25-uniform
        &   1.99  & 1.10x & 2.00  & 1.11x  & 6.71  & 1.15x    \\
    & & Binomial
        &   2.18  & 1.20x  & 2.16  & 1.20x  & 7.28 & 1.24x    \\
\cmidrule{2-9}
    & \multirow{3}{*}{RB-sort}
    & Uniform
        & 1.29 & & 1.77 &  & 2.31 &   \\
    & & 0.25-uniform
        &  1.45  & 1.13x  & 1.81  & 1.02x  & 2.33  & 1.01x    \\
    & & Binomial
        &   1.48  & 1.15x   & 1.93  & 1.9x  & 2.52  & 1.09x    \\

\bottomrule
\end{tabular}
\caption{Processing rate (billion elements per second) as well as speedup against the uniform distribution for delta-bucket multisplit with different input distributions. All cases are with $m=256$ buckets.}\label{table:distribution}
\end{table}

\section{Multisplit, a useful building block}\label{sec:multisplit_app} 
\subsection{Building a radix sort}\label{subsec:multisplit_sort}
In Section~\ref{sec:perf_eval}, we concentrated on the performance evaluation of our multisplit methods with user-defined bucket identifiers, and in particular the delta-bucket example.
In this section, we focus on identity buckets and how we can modify them to implement our own version of radix sort.

\paragraph{Multisplit with identity buckets} Suppose we have identity buckets, meaning that each key is identical to its bucket ID (i.e., $\defn{f}(u_i) = u_i = j$, for $0\leq j < m$).
In this case, sorting keys (at least the first $\lceil \log m \rceil$ bits of keys) turns out to be equivalent to the stable multisplit problem.
In this case, there is no need for the extra overheads inherent in RB-sort; instead, a direct radix sort can be a competitive alternate solution.

\paragraph{Radix sort} In Section~\ref{sec:radix} we briefly discussed the way radix sort operates.
Each round of radix sort sorts a group of bits in the input keys until all bits have been consumed.
For CUB, for example, in the Kepler architecture (e.g., Tesla K40c) each group consists of 5 consecutive bits, while for the more recent Pascal architecture (e.g., GeForce GTX 1080), each group is 7 bits.\footnote{These stats are for CUB 1.6.4.}

\paragraph{Multisplit-sort} Each round of radix sort is essentially bucketing its output based on the set of bits it considers in that round. If we define our buckets appropriately, multisplit can do the same.
Suppose we have $\defn{f}_k(u) = (u \gg kr) \& (2^{r}-1)$, where $\gg$ denotes bitwise shift to the right and $\&$ is a bitwise AND operator. $r$ denotes our radix size (i.e., the size of the group of bits to be considered in each iteration).
Then, with $0 \leq k < \lceil 32/r\rceil$ iterations of multisplit with $\defn{f}_k$ as each iteration's bucket identifier, we have built our own radix sort.

\paragraph{High level similarities and differences with CUB radix sort}
At a high level, multisplit-sort is similar to CUB's radix sort.
In CUB, contiguous chunks of bits from input keys are sorted iteratively. Each iteration includes an up-sweep where bin counts are computed, a scan operation to compute offsets, and a down-sweep to actually perform the stable sorting on a selected chunk of bits (similar roles to our pre-scan, scan and post-scan stages in our multisplit).
The most important differences are 1)~shared memory privatization (CUB's thread-level versus our warp-level), which decreases our shared memory usage compared to CUB, and 2)~our extensive usage of warp-wide intrinsics versus CUB's more traditional register-level computations.

\subsubsection{Performance Evaluation}
Our multisplit-based radix sort has competitive performance to CUB's radix sort. Our experiments show that, for example on the GeForce GTX 1080, our key-only sort can be as good as 0.9x of CUB's performance, while our key-value sort can be as good as 1.1x faster than CUB's.

\paragraph{Multisplit with identity buckets}
We first compare our multisplit methods (WMS and BMS) with identity buckets to  CUB's radix sort over $\lceil \log m \rceil$ bits.
Multisplit and CUB's performance are roughly similar.
Our multisplit methods are usually better for a small number of buckets, while CUB's performance is optimized around the number of bits that it uses for its internal iterations (5 bits for Kepler, 7 bits for Pascal).
Table~\ref{table:sort_bits_rate} shows our achieved throughput (billion elements sorted per second) as a function of the number of bits per key.
\begin{table}
\centering
\tiny
\resizebox{\columnwidth}{!}{
\begin{tabular}{lll cccccccc}
\toprule
& & & \multicolumn{8}{c}{Throughput (speedup against CUB's)} \\
\cmidrule{4-11}
& & & \multicolumn{8}{c}{Number of bits in each key} \\
\cmidrule{4-11}
& & Method  & 1 & 2 & 3 & 4 & 5 & 6 & 7 & 8 \\
\midrule 
    \multirow{6}{*}{\begin{turn}{90}K40c (ECC on)\end{turn}}
    & \multirow{3}{*}{\begin{turn}{0}\tiny key-only\end{turn}} 
    & {WMS} 
        & \textbf{14.08 (1.12 x)} &  \textbf{13.88 (1.12 x)} &  12.17 (0.99 x) &  10.12 (0.94 x) &  7.67 (1.05 x) & -- & -- & -- \\ 
    & & {BMS}
        & 13.84 (1.10 x) &  12.94 (1.04 x) &  12.10 (0.99 x) &  \textbf{10.88 (1.01 x)} &  \textbf{9.45 (1.29 x)} &  \textbf{6.89 (1.18 x)} &  4.55 (0.79 x) &  2.69 (0.50 x) \\ 
    & & { CUB } 
        & 12.56 (1x) &  12.45 (1x) &  \textbf{12.28 (1x)} &  10.75 (1x) &  7.33 (1x) &  5.83 (1x) &  \textbf{5.74 (1x)} &  \textbf{5.42 (1x)} \\ 
\cmidrule{2-11}
    & \multirow{3}{*}{\begin{turn}{0}\tiny key-value\end{turn}} 
    & {WMS} 
        & \textbf{8.62 (1.13 x)} &  \textbf{7.94 (1.04 x)} &  6.58 (0.96 x) &  5.37 (0.94 x) &  4.56 (0.95 x) & -- & -- & -- \\
    & & {BMS}
        & 7.79 (1.02 x) &  7.85 (1.03 x) &  \textbf{7.55 (1.10 x)} &  \textbf{7.16 (1.25 x)} &  \textbf{6.76 (1.41 x)} &  \textbf{5.23 (1.60 x)} &  \textbf{3.41 (1.09 x)} &  1.92 (0.67 x) \\ 
    & & {CUB}  
        & 7.61 (1x) &  7.60 (1x) &  6.87 (1x) &  5.72 (1x) &  4.79 (1x) &  3.27 (1x) &  3.12 (1x) &  \textbf{2.86 (1x)} \\
\midrule
\midrule
    \multirow{6}{*}{\begin{turn}{90}K40c (ECC off)\end{turn}}
    & \multirow{3}{*}{\begin{turn}{0}\tiny key-only\end{turn}} 
    & WMS 
        & \textbf{17.57 (1.35 x)} &  \textbf{16.47 (1.26 x)} &  \textbf{13.36 (1.04 x)} &  10.18 (0.80 x) &  7.80 (0.68 x) & -- & -- & -- \\ 
    & & BMS
        &  15.26 (1.17 x) &  13.89 (1.06 x) &  12.76 (0.99 x) &  \textbf{10.91 (0.85 x)} &  9.49 (0.82 x) &  6.85 (1.06 x) &  4.53 (0.70 x) &  2.68 (0.42 x) \\ 
    & & CUB
        & 13.05 (1x) &  13.06 (1x) &  12.86 (1x) &  12.76 (1x) &  \textbf{11.54 (1x)} &  \textbf{6.50 (1x)} &  \textbf{6.46 (1x)} &  \textbf{6.44 (1x)} \\ 
\cmidrule{2-11}
    & \multirow{3}{*}{\begin{turn}{0}\tiny key-value\end{turn}} 
    & WMS 
        & \textbf{10.31 (1.13 x)} &  \textbf{9.67 (1.07 x)} &  7.73 (0.86 x) &  6.21 (0.70 x) &  4.53 (0.59 x) & -- & -- & -- \\ 
    & & BMS
        &  9.01 (0.99 x) &  9.00 (1.00 x) &  8.50 (0.94 x) &  7.69 (0.86 x) &  6.78 (0.88 x) &  \textbf{5.20 (1.16 x)} &  3.31 (0.74 x) &  1.92 (0.43 x) \\ 
    & & CUB
        & 9.14 (1x) &  9.00 (1x) &  \textbf{9.00 (1x)} &  \textbf{8.92 (1x)} &  \textbf{7.69 (1x)} &  4.50 (1x) &  \textbf{4.46 (1x)} &  \textbf{4.47 (1x)} \\
\midrule
\midrule
    \multirow{6}{*}{\begin{turn}{90}GTX 1080  \end{turn}}
    & \multirow{3}{*}{\begin{turn}{0}\tiny key-only\end{turn}} 
    & WMS 
        & 18.87 (1.15 x) &  17.70 (1.07 x) &  16.76 (1.03 x) &  16.33 (0.96 x) &  12.56 (0.71 x)  & -- & -- & -- \\ 
    & & BMS
        & \textbf{19.37 (1.18 x)} &  \textbf{19.15 (1.15 x)} &  \textbf{18.99 (1.17 x)} &  \textbf{18.61 (1.09 x)} &  \textbf{17.89 (1.02 x)} &  16.83 (0.93 x) &  13.41 (0.92 x) &  8.17 (0.94 x)  \\ 
    & & CUB
        &  16.35 (1x) &  16.59 (1x) &  16.24 (1x) &  17.05 (1x) &  17.62 (1x) &  \textbf{18.05 (1x)} &  \textbf{14.50 (1x)} &  \textbf{8.65 (1x)} \\ 
\cmidrule{2-11}
    & \multirow{3}{*}{\begin{turn}{0}\tiny key-value\end{turn}} 
    & WMS 
        & 11.39 (1.01 x) &  11.04 (1.00 x) &  10.73 (0.96 x) &  10.05 (0.90 x) &  8.29 (0.76 x) & -- & -- & -- \\ 
    & & BMS
        &  \textbf{11.68 (1.03 x)} &  \textbf{11.61 (1.05 x)} &  \textbf{11.58 (1.04 x)} &  \textbf{11.41 (1.02 x)} &  \textbf{11.18 (1.03 x)} &  \textbf{10.89 (1.07 x)} &  \textbf{10.20 (1.23 x)} &  \textbf{6.42 (1.20 x)} \\ 
    & & CUB
        & 11.30 (1x) &  11.06 (1x) &  11.18 (1x) &  11.14 (1x) &  10.88 (1x) &  10.20 (1x) &  8.30 (1x) &  5.34 (1x) \\ 
\bottomrule
\end{tabular}
}
  \caption{Multisplit with identity buckets. $2^{25}$ random keys are uniformly distributed among buckets. Achieved throughput (Gkeys/s or Gpairs/s) are shown as well as the achieved speedup against CUB's radix sort (over limited number of bits).}\label{table:sort_bits_rate}
\end{table}

There are several important remarks to make:
\begin{itemize}
\item Our multisplit methods outperform CUB for up to 4 bits on the Tesla K40c and up to 6 bits on the GeForce GTX 1080. We note CUB is highly optimized for specific bit counts: 5-bit radixes on Kepler (Tesla K40c) and 7-bit radixes on Pascal (GeForce GTX 1080).
\item By comparing our achieved throughputs with those of delta-buckets in Table~\ref{table:ms_rate}, it becomes clear that the choice of bucket identifier can have an important role in the efficiency of our multisplit methods.
  In our delta-bucket computation, we used integer divisions, which are expensive computations.
  For example, in our BMS, the integer division costs 0.72x, 0.70x, and 0.90x geometric mean decrease in our key-only throughput for Tesla K40c (ECC on), Tesla K40c (ECC off) and GeForce GTX 1080 respectively. The throughput decrease for key-value scenarios is 0.87x, 0.82x and 0.98x respectively. The GeForce GTX 1080 is less sensitive to such computational load variations. Key-value scenarios also require more expensive data movement so that the cost of the bucket identifier is relatively less important.
\item On the GeForce GTX 1080, our BMS method is always superior to WMS. This GPU appears to be better at hiding the latency of BMS's extra synchronizations, allowing the marginally better locality from larger subproblems to become the primary factor in differentiating performance.
\end{itemize}

\paragraph{Multisplit-sort}
Now we turn to characterizing the performance of sort using multisplit with identity buckets.
It is not immediately clear what the best radix size ($r$) is for achieving the best sorting performance. As a result, we ran all choices of $4\leq r \leq 8$.
Because our bucket identifiers are also relatively simple (requiring one shift and one AND), our multisplit performance should be close to that of identity buckets.

Since BMS is almost always superior to WMS for $r\geq 4$ (Table~\ref{table:sort_bits_rate}), we have only used BMS in our implementations.
For sorting 32-bit elements, we have used $\lfloor 32/r \rfloor$ iterations of r-bit BMS followed by one last BMS for the remaining bits. For example, for $r = 7$, we run 4 iterations of 7-bit BMS then one iteration of 4-bit BMS\@.
Table~\ref{table:sort} summarizes our sort results.

\begin{table}
\centering
\scriptsize
\begin{tabular}{l c ccc | ccc | ccc}
\toprule
& & \multicolumn{3}{c}{K40c (ECC on)} & \multicolumn{3}{c}{K40c (ECC off)} & \multicolumn{3}{c}{GeFroce GTX 1080} \\  
\cmidrule(r){3-5} \cmidrule{6-8} \cmidrule(l){9-11}
Method & $r$ & time & throughput & speedup & time & throughput & speedup & time & throughput & speedup \\
\midrule
\multirow{5}{*}{Our sort (key-only)}
	& 4
		& 25.84 & 1.299 & 1.01x 		& 26.00 & 1.290 & 0.75x 				& 14.11 & 2.368 & 0.70x \\ 
	& 5
		& \textbf{24.82} & \textbf{1.35} & \textbf{1.05x} 	& \textbf{24.81} & \textbf{1.352} & \textbf{0.78x} 	& 12.65 & 2.654 & 0.78x \\
	& 6
		& 26.18 & 1.282 & 0.99x 	& 26.41 & 1.271 & 0.74x 	& 11.44 & 2.933 & 0.86x \\
	& 7
		& 34.59 & 0.970 & 0.75x 		& 34.93 & 0.961 & 0.56x 		& \textbf{11.21} & \textbf{2.994} & \textbf{0.88x} \\ 
	& 8 
		& 50.17 & 0.669 & 0.52x 		& 50.58 & 0.663 & 0.38x 		& 14.59 & 2.300 & 0.68x \\

		\midrule
		\multicolumn{2}{c}{CUB (key-only)} & 25.99 & 1.291 & -- & 19.42 & 1.728 & -- & 9.88 & 3.397 & -- \\
		\midrule
		\midrule
\multirow{5}{*}{Our sort (key-value)}
	& 4
		& 36.86 & 0.910 & 1.18x 		& 35.17 & 0.954 & 0.81x 		& 14.11 & 1.441 & 0.75x \\ 
	& 5
		& 34.90 & 0.962 & 1.25x 		& \textbf{34.32} & \textbf{0.978} & \textbf{0.83x} 		& 12.65 & 1.619 & 0.85x  \\
	& 6
		& \textbf{34.79} & \textbf{0.97} & \textbf{1.26x} 		& 34.70 & 0.967 & 0.82x 		& 18.01 & 1.852 & 0.97x  \\
	& 7
		& 43.90 & 0.764 & 1.00x 		& 44.59 & 0.753 & 0.64x 		& \textbf{15.97} & \textbf{2.101} & \textbf{1.10x}  \\ 
	& 8 
		& 67.47 & 0.497 & 0.65x 		& 67.94 & 0.494 & 0.42x 		& 18.01 & 1.863 & 0.97x \\

		\midrule
		\multicolumn{2}{c}{CUB (key-value)} & 43.73 & 0.767 & -- & 28.58 & 1.174 & -- & 17.56 & 1.911 & -- \\

\bottomrule
\end{tabular}
\caption{Our Multisplit-based radix sort is compared to CUB. We have used various number of bits ($r$) per iteration of mutlisplit for our various sort implementations. Average running time (ms) for sorting $2^{25}$ random 32-bit elements, achieved throughput (Gkeys/s for key-only, and Gpairs/s for key-value), and speedup against CUB's radix sort.}\label{table:sort}
\end{table}

By looking at our achieved throughputs (sorting rates), we see that our performance increases up to a certain radix size, then decreases for any larger $r$. This optimal radix size is different for each different GPU and depends on numerous factors, for example available bandwidth, the efficiency of warp-wide ballots and shuffles, the occupancy of the device, etc.
For the Tesla K40c, this crossover point is earlier than the GeForce GTX 1080 (5 bits compared to 7 bits).
Ideally, we would like to process more bits (larger radix sizes) to have fewer total iterations. But, larger radix sizes mean a larger number of buckets ($m$) in each iteration, requiring more resources (shared memory storage, more register usage, and more shuffle usage), yielding an overall worse performance per iteration.

\paragraph{Comparison with CUB}
CUB is a carefully engineered and highly optimized library. For its radix sort, it uses a persistent thread style of programming~\cite{Gupta:2012:ASO}, where a fixed number of thread-blocks (around 750) are launched, each with only 64 threads (thus allowing many registers per thread).
Fine-tuned optimizations over different GPU architectures enables CUB's implementation to efficiently occupy all available resources in the hardware, tuned for various GPU architectures.

The major difference between our approach with CUB has been our choice of privatization. CUB uses thread-level privatization, where each thread keeps track of its local processed information (e.g., computed histograms) in an exclusively assigned portion of shared memory. Each CUB thread processes its portion of data free of contention, and later, combines its results with those from other threads.
However, as CUB considers larger radix sizes, it sees increasing pressure on each block's shared memory usage. The pressure on shared memory becomes worse when dealing with key-value sorts as it now has to store more elements into shared memory than before.

In contrast to CUB's thread privatization, our multisplit-based implementations instead target warp-wide privatization. An immediate advantage of this approach is that we require smaller privatized exclusive portions of shared memory because we only require a privatized portion per warp rather than per thread.
The price we pay is the additional cost of warp-wide communications (shuffles and ballots) between threads, compared to CUB's register-level communication within a thread.

The reduced shared memory usage of our warp privatization becomes particularly valuable when sorting key-value pairs.
Our key-value sort on GeForce GTX 1080 shows this advantage: when both approaches use 7-bit radixes, our multisplit-sort achieves up to a 1.10x higher throughput than CUB\@.
On the other hand, CUB demonstrates its largest performance advantage over our implementation (ours has 0.78x the throughput of CUB's) for key-only sorts on Tesla K40c (ECC off). In this comparison, our achieved shared memory benefits do not balance out our more costly shuffles.


Our multisplit-based radix sort proves to be competitive to CUB's radix sort, especially in key-value sorting. 
For key-only sort, our best achieved throughputs are 1.05x, 0.78x, and 0.88x times the throughput that CUB provides for Tesla K40c (ECC on), Tesla K40c (ECC off), and GeForce GTX 1080, respectively.
For key-value sorting and with the same order of GPU devices, our multisplit-based sort provides 1.26x, 0.83x, and 1.10x times more throughput than CUB, respectively. 
Our highest achieved throughput is 3.0 Gkeys/s (and 2.1 Gpairs/s) on a GeFroce GTX 1080, compared to CUB's 3.4 Gkeys/s (and 1.9 Gpairs/s) on the same device. 

\paragraph{Future of warp privatized methods}
We believe the exploration of the difference between thread-level and warp-level approaches has implications beyond just multisplit and its extension to sorting.
In general, any future hardware improvement in warp-wide intrinsics will reduce the cost we pay for warp privatization, making the reduction in shared memory size the dominant factor. We advocate further hardware support for warp-wide voting with a generalized ballot that returns multiple 32-bit registers, one for each bit of the predicate.
Another useful addition that would have helped our implementation is the possibility of shuffling a dynamically addressed register from the source thread.
This would enable the user to share lookup tables among all threads within a warp, only requesting the exact data needed at runtime rather than delivering every possible entry so that the receiver can choose.

\subsection{The Single Source Shortest Path problem}\label{sec:app_sssp}
In Section~\ref{sec:intro} we argued that an efficient multisplit primitive would have helped Davidson et al.~\cite{Davidson:2014:WPG:nourl} in their delta-stepping formulation of the Single Source Shortest Path (SSSP) problem on the GPU\@.
In this section, we show that by using our multisplit implementation, we can achieve significant speedups in SSSP computation, especially on highly connected graphs with low diameters.

\subsubsection{The Single Source Shortest Path (SSSP) problem}
Given an arbitrary graph $G = (V, E)$, with non-negative weights assigned to each edge and a source vertex $s\in V$, the SSSP problem finds the minimum cost path from the source to every other vertex in the graph.
As described in Section~\ref{sec:intro}, Dijkstra's~\cite{Dijkstra:1959:ANO} and Bellman-Ford-Moore's~\cite{Bang-Jensen:2009:DTA} algorithms are two classical approaches to solve the SSSP problem.
In the former, vertices are organized in a single priority queue and are processed sequentially from the lowest to the highest weight.
In the latter, for each vertex we process all its neighbors (i.e., processing all edges). This can be done in parallel and is repeated over multiple iterations until convergence.
Dijkstra is highly work-efficient but essentially sequential and thus unsuitable for parallelization.
Bellman-Ford-Moore is trivially parallel but does much more work than necessary (especially for highly connected graphs).

As an alternative algorithm between these two extremes (sequential processing of all vertices vs.\ processing all edges in parallel), delta-stepping allows the selective processing of a subset of vertices in parallel~\cite{Meyer:2003:DAP}.
In this formulation, nodes are put into different buckets (based on their assigned weights) and buckets with smaller weights are processed first.
Davidson et al.~\cite{Davidson:2014:WPG:nourl} proposed multiple GPU implementations based on the delta-stepping formulation. Their two most prominent implementations were based on a \emph{Near-Far} strategy and a \emph{Bucketing} strategy.
Both divide vertices into multiple buckets, which can be processed in parallel.
Both use efficient load-balancing strategies to traverse all vertices within a bucket.
Both iterate over multiple rounds of processing until convergence is reached.
The main difference between the two is in the way they organize the vertices to be processed next (work frontiers):
\begin{description}
        \item[Near-Far strategy] In this strategy the work queue is prioritized based on a variable splitting distance. In every iteration, only those vertices less than this threshold (the \emph{near set}) are processed. Those falling beyond the threshold are appended to a \emph{far pile}. Elements in the far pile are ignored until work in the near set is completely exhausted.
        When all work in the near set is exhausted (this could be after multiple relaxation phases), this strategy increases the splitting distance (by adding an incremental weight $\Delta$ to it) and removes invalid elements from the far pile (those which have been updated with similar distances), finally splitting this resulting set into a new near set and far pile. This process continues until both the near set and far pile are empty (the convergence criterion).
        \item[Bucketing strategy] In this strategy, vertices are partitioned into various buckets based on their weights (Davidson et al.\ reported the best performance resulted from 10 buckets). This strategy does a more fine-grained classification of vertices compared to Near-Far, resulting in a greater potential reduction in work queue size and hence less work necessary to converge. The downside, however, is the more complicated bucketing process, which due to lack of an efficient multisplit primitive was replaced by a regular radix sort in the original work. As a result of this expensive radix sort overhead, Near-Far was more efficient in practice~\cite{Davidson:2014:WPG:nourl}.
\end{description}

\subsubsection{Multisplit-SSSP} Now that we have implemented an efficient multisplit GPU primitive in this paper, we can use it in the Bucketing strategy explained above to replace the costly radix sort operation. We call this new Bucketing implementation \emph{Multisplit-SSSP}\@.
Our Multisplit-SSSP should particularly perform well on highly connected graphs with relatively large out degrees and smaller diameters (such as in social graphs), causing fewer iterations and featuring large enough work fronts to make multisplit particularly useful. However, graphs with low average degrees and large diameters (such as in road networks) require more iterations over smaller work frontiers, resulting in high kernel launch overheads (because of repetitive multisplit usage) without large enough work frontiers to benefit from the efficiency of our multisplit.
We note that this behavior for different graph types is not limited to our SSSP implementation; GPU graph analytics in general demonstrate their best performance on highly connected graphs with low diameters~\cite{Wang:2016:GAH:nourl}.

\subsubsection{Performance Evaluation}\label{subsubsec:perf_eval_sssp}
In this part, we quantitatively evaluate the performance of our new Multisplit-SSSP compared to Davidson et al.'s Bucketing and Near-Far approaches.
Here, we choose a set of graph datasets listed in Table~\ref{table:graphs}.\footnote{All matrices except for rmat are downloaded from University of Florida Sparse Matrix Collection~\cite{Davis:2011:UOF}. Rmat was generated with parameters $(a,b,c,d) = (0.5, 0.1, 0.1, 30)$.}
For those graphs that are not weighted, we randomly assign a non-negative integer weight between 0 and 1000 to each edge.

Table~\ref{table:sssp_results} shows the convergence time for Near-Far, Bucketing, and Multisplit-SSSP (in million traversed edges per second, MTEPS), with Multisplit-SSSP's speedup against Near-Far.
Multisplit-SSSP is always better than Bucketing, on both devices and on every graph we tested (up to 9.8x faster on Tesla K40c and 9.1x faster on the GeForce GTX 1080).
This behavior was expected because of the performance superiority of our multisplit compared to a regular radix-sort (Fig.~\ref{fig:speedup}).

Against Near-Far, our performance gain depends on the type of graph.
As we expected, on highly connected graphs with low diameters (such as rmat), we achieve up to 1.58x and 2.17x speedup against Near-Far, on the Tesla K40c and GeForce GTX 1080 respectively.
However, for high diameter graphs such as road networks (e.g., belgium\_osm), we are closer to Near-Far's performance: Multisplit-SSSP is slower than Near-Far on Tesla K40c (0.93x) and marginally faster on GeForce GTX 1080 (1.04x).
Road graphs have significantly higher diameter and hence more iterations. As a result, the extra overhead in each phase of Multisplit-SSSP on large diameters can become more important than the saved operations due to fewer edge re-relaxations.

\begin{table}
\centering
\scriptsize
\begin{tabular}{lccc}
\toprule
Graph Name & Vertices & Edges & Avg. Degree  \\
\midrule
cit-Patents~\cite{Hall:2001:NPC} & 3.77~M & 16.52~M & 8.8  \\
flickr~\cite{Davis:2011:UOF}  & 0.82~M & 9.84~M & 24.0 \\
belgium\_osm~\cite{Kobitzsh:2010:DIMACS}  & 1.44~M & 1.55~M & 2.2  \\
rmat~\cite{Chakrabarti:2004:RAR} & 0.8~M & 4.8~M & 12.0 \\
\bottomrule
\end{tabular}
\caption{Datasets used for evaluating our SSSP algorithms.}\label{table:graphs}
\end{table}

\begin{table}
\centering \scriptsize
\resizebox{\columnwidth}{!}{
\begin{tabular}{l cc ccc | cc ccc}
\toprule
& \multicolumn{5}{c}{Tesla K40c (ECC on)} & \multicolumn{5}{c}{GeForce GTX 1080} \\
\cmidrule(r){2-6}\cmidrule(l){7-11}
Graph Name & Near-Far & Bucketing & \multicolumn{3}{c}{Multisplit-SSSP} & Near-Far & Bucketing & \multicolumn{3}{c}{Multisplit-SSSP} \\
\cmidrule(r){1-1} \cmidrule(r){2-2} \cmidrule(r){3-3} \cmidrule(r){4-6} \cmidrule(l){7-7} \cmidrule(l){8-8} \cmidrule(l){9-11}
-- & time (ms) & time (ms) & time (ms) & MTEPS & speedup & time (ms) & time (ms) & time (ms) & MTEPS & speedup \\
\midrule
cit-Patents & 458.4 & 3375.9 & 343.1 & 96.3 & 1.34x & 444.2 & 3143.0 & 346.8 & 95.2 & 1.28x \\
flickr                          & 96.0 & 163.0 & 64.5 & 305.2 & 1.49x & 66.7 & 111.1 & 36.5 & 539.0 & 1.83x \\
belgium\_osm    & 561.4 & 3588.0 & 604.5 & 5.12 & 0.93x & 443.8 & 3014.2 & 427.0 & 7.3 & 1.04x \\
rmat & 20.9 & 28.7 & 13.2 & 727.3 & 1.58x & 12.17 & 14.9 & 5.8 & 1655.2 & 2.17x \\
\bottomrule
\end{tabular}
}
\caption{Near-Far, Bucketing, and our new Multisplit-SSSP methods over various datasets. Speedups are against the Near-Far strategy (which appears to be always better than the Bucketing strategy).}\label{table:sssp_results}
\end{table}


\subsection{GPU Histogram}\label{subsec:multisplit_histogram}
All three of our multisplit methods from  Section~\ref{sec:impl_details}   (DMS, WMS and BMS) have a pre-scan stage, where we compute bucket histograms for each subproblem using our warp-wide ballot-based voting scheme (Algorithm~\ref{alg:warp_histogram}).
In this section, we explore the possibility of using our very same warp-wide histogram to compute a \emph{device-wide} (global) histogram.
We define our histogram problem as follows: The inputs are $n$ unordered input elements (of any data type) and a bucket identifier $\defn{f}$ that assigns each input element to one of $m$ distinct buckets (bins). The output is an array of length $m$ representing the total number of elements within each bucket.

\paragraph{Our Histogram} In the pre-scan stage of our multisplit algorithms, we store histogram results for each subproblem so that we can perform a global scan operation on them, then we use this result in our post-scan stage to finalize the multisplit.
In the GPU Histogram problem, however, we no longer need to report per-subproblem histogram details. Instead, we only must sum all subproblems' histograms together to form the output global histogram.
Clearly, we would prefer the largest subproblems possible to minimize the cost of the final global summation. So, we base our implementation on our BMS method (because it always addresses larger subproblems than the other two).
We have two main options for implementation. 1)~Store our subproblem results into global memory and then perform a segmented reduction, where each bucket represents a segment. 2)~Modify our pre-scan stage to atomically add histogram results of each subproblem into the final array.
Based on our experiments, the second option appears to be more efficient on both of our devices (Tesla K40c and GeForce GTX 1080).

\paragraph{Experimental setup}
In order to evaluate a histogram method, it is common to perform an extensive set of experiments with various distributions of inputs to demonstrate the performance and consistency of that method.
A complete and thorough benchmarking of all possible distributions of inputs is beyond the scope of this short section.
Nevertheless, just to illustrate the potentials in our histogram implementation, we continue this section with a few simple experimental scenarios.
Our goal is to explore whether our warp-wide histogram method can potentially be competitive to others (such as CUB's histogram), under what conditions, and most importantly why.
To achieve this goal, we consider the following two scenarios:
\begin{enumerate}
        \item Even Histogram: Consider a set of evenly spaced splitters $\{s_0, s_1, \dots, s_{m}\}$ such that each two consecutive splitters bound a bucket ($m$ buckets, each with a width of $|s_i - s_{i-1}| = \Delta$). For each real number input $s_0 < x < s_m$, we can easily identify its bucket as $\lfloor(x-s_0)/\Delta\rfloor$.
        \item Range Histogram: Consider a set of arbitrarily ranged splitters $\{s_0, s_1, \dots, s_{m}\}$ such that each two consecutive splitters bound a bucket. For each real number input $s_0 < x < s_m$, we must perform a binary search (i.e., an upper bound operation) on splitters to find the appropriate bucket (requires at most $\lceil \log m \rceil$ searches).
\end{enumerate}

We generate $n=2^{25}$ random floating point numbers uniformly distributed between 0 and 1024. For the Even histogram experiment, splitters are fixed based on the number of buckets $m$. For Range histogram experiment, we randomly generate a set of $m-1$ random splitters ($s_0$ and $s_m$ are already fixed).
We use our histogram method to compute the global histogram for $m\leq 256$ buckets. We compare against CUB Histogram, which supports equivalents  (\texttt{HistogramEven} and \texttt{HistogramRange}) to our Even and Range test scenarios.
We have repeated our experiments over 100 independent random trials.

\subsubsection{Performance Evaluation}
Table~\ref{table:histogram_rate} shows our achieved average processing rate (in billion elements per second) as well as our speedup against CUB, and for different hardware choices.
For the Even Histogram, we observe that we are better than CUB for $m\leq 128$ buckets on Tesla K40c, but only marginally better for $m\leq 8$ on GeForce GTX 1080. For Range Histogram, we are always better than CUB for $m\leq 256$ on both devices.

\begin{table}
\centering
\scriptsize
\resizebox{\columnwidth}{!}{
\begin{tabular}{lll cccccccc}
\toprule
& &  & \multicolumn{8}{c}{Number of buckets (bins)} \\
\cmidrule{4-11}
Example & GPU & Method & 2 & 4 & 8 & 16 & 32 & 64 & 128 & 256 \\
\midrule
  \multirow{9}{*}{\begin{turn}{90}Even Histogram\end{turn}}
  & \multirow{3}{*}{Tesla K40c (ECC on)}
  & Our Histogram
      & \textbf{45.3} &  \textbf{42.5} &  \textbf{45.4} &  \textbf{37.8} &  \textbf{32.1} &  \textbf{26.5} &  \textbf{24.2} &  18.8 \\
  & & CUB
        &  13.7&  14.9&  16.9&  19.1&  21.4&  21.8&  20.8&  \textbf{19.5} \\
  & & Speedup vs.\ CUB
                  &  3.30x &  2.86x &  2.69x &  1.98x &  1.50x &  1.21x &  1.16x &  0.96x \\
\cmidrule{2-11}
  & \multirow{3}{*}{Tesla K40c (ECC off)}
  & Our Histogram
      & \textbf{53.0}&  \textbf{47.2}&  \textbf{48.1}&  \textbf{38.3}&  \textbf{32.3}&  \textbf{26.5}&  \textbf{24.0}&  18.7 \\
  & & CUB
        &  13.6&  14.7&  16.7&  18.9&  21.3&  21.8&  20.7&  \textbf{19.5} \\
  & & Speedup vs.\ CUB
                  &  3.90x &  3.20x &  2.88x &  2.03x &  1.52x &  1.21x &  1.16x &  0.96x \\
                  \cmidrule{2-11}
  & \multirow{3}{*}{GeForce GTX 1080}
  & Our Histogram
      &  \textbf{61.0}&  \textbf{61.1}&  \textbf{60.9}&  60.7&  60.2&  45.2&  59.6&  52.7 \\
  & & CUB
        &   60.5&  60.7&  60.5&  \textbf{60.7}&  \textbf{61.1}&  \textbf{60.6}&  \textbf{60.3}&  \textbf{60.9} \\
  & & Speedup vs.\ CUB
                  & 1.01x &  1.01x &  1.01x &  1.00x &  0.98x &  0.75x &  0.99x &  0.87x\\
\midrule
\midrule
  \multirow{9}{*}{\begin{turn}{90}Range Histogram\end{turn}}
  & \multirow{3}{*}{Tesla K40c (ECC on)}
  & Our Histogram
    &  \textbf{28.0}&  \textbf{22.1}&  \textbf{18.4}&  \textbf{14.6}&  \textbf{11.9}& \textbf{ 9.0}& \textbf{ 7.7}& \textbf{ 7.3} \\
  & & CUB
        &   8.7&  6.8&  6.2&  5.8&  5.7&  5.5&  5.2&  4.8 \\
  & & Speedup vs.\ CUB
                  & 3.21x &  3.26x &  2.96x &  2.51x &  2.09x &  1.63x &  1.50x &  1.51x \\
\cmidrule{2-11}
  & \multirow{3}{*}{Tesla K40c (ECC off)}
  & Our Histogram
    & \textbf{27.6}&  \textbf{22.2}&  \textbf{17.8}&  \textbf{14.5}&  \textbf{11.7}& \textbf{ 8.7}& \textbf{ 7.6}& \textbf{ 7.1}  \\
  & & CUB
        &  8.4&  6.8&  6.2&  5.8&  5.6&  5.4&  5.1&  4.8  \\
  & & Speedup vs.\ CUB
                & 3.29x &  3.28x &  2.89x &  2.50x &  2.10x &  1.61x &  1.50x &  1.50x\\
                  \cmidrule{2-11}
  & \multirow{3}{*}{GeForce GTX 1080}
  & Our Histogram
      &  \textbf{56.7}&  \textbf{51.4}&  \textbf{45.4}&  \textbf{39.8}&  \textbf{33.9}&  \textbf{28.4}&  \textbf{24.8}&  \textbf{20.0}\\
  & & CUB
        &  42.4 &  35.2 &  30.9 &  27.1 &  24.4 &  22.3 &  19.3 &  14.8 \\
  & & Speedup vs.\ CUB
                  & 1.34x &  1.46x &  1.47x &  1.47x &  1.39x &  1.28x &  1.29x &  1.35x \\
\bottomrule
\end{tabular}
}
\caption{Histogram computation over two examples of even bins (Even Histogram) and customized bins (Range Histogram). Procesing rates (in billion elements per second) are shown for our Histogram and CUB as well as our achieved speedup. Experiments are repeated on three different hardware settings.}\label{table:histogram_rate}
\end{table}

\paragraph{Even Histogram}
CUB's even histogram  is designed to operate with any number of buckets (even $m\gg256$).
This generality has consequences in its design choices.
For example, if an implementation generalizes to any number of buckets (especially large $m>256$), it is not possible to privatize histogram storage for each thread (which requires size proportional to $m$) and then combine  results to compute a block-level histogram solution.
(This is a different scenario than radix sort, because radix sort can choose the number of bits to process on each iterations. CUB's radix sort only supports histograms of size up to 128 buckets [at most 7 bits] on Pascal GPUs).
As a result, CUB uses shared memory atomics to directly compute histograms in shared memory.
Then these intermediate results are atomically added again into global memory to form the output global histogram. (Since CUB is mostly bounded by atomic operations, ECC does not have much effect on its performance on Tesla K40c.)

On the other hand, our focus here is on a limited number of buckets.
As a result, by using warp-level privatization of histograms, we avoid shared memory atomics within blocks, resulting in better performance for the histogram bucket counts that we target.
As the number of buckets increases, we gradually increase pressure on our shared memory storage until CUB's histogram becomes the better choice.

Atomic operation performance has improved significantly on the Pascal architecture (GeForce GTX 1080), which helps CUB's histogram performance. On this architecture, Table~\ref{table:histogram_rate} shows that that we are barely better than CUB for a very small number of buckets ($m\leq 8$), and then we witness a decrease in our performance because of the shared-memory pressure that we explained above.

\paragraph{Range Histogram}
We see better performance for range histograms for two main reasons.
First, bucket identification in this case requires a binary search, which is much more expensive than the simpler floating point multiplications required for the even histogram.
Our histogram implementation is relatively insensitive to expensive bucket-computation operations because they help us hide any extra overheads caused by our extensive shuffle and ballot usage. A thread-level approach like CUB's would also have to do the same set of expensive operations (in this case, expensive memory lookups), but then they would lose the comparative benefit they would otherwise gain from faster local register-level computations.

Another reason for our better performance is again because of CUB's generality. Since CUB must operate with an arbitrary number of splitters, it does not store those splitters into shared memory, which means every binary search is directly performed in global memory. On the other hand, for $m \leq 256$, we can easily store our splitters into shared memory in order to avoid repetitive global memory accesses.

\paragraph{Summary} Specializing a histogram computation to support only a limited number of buckets allows potential performance increases over more general histogram implementations. For some applications, this may be desirable. For other applications---and this is likely CUB's design goal---consistent performance across an arbitrary number of buckets may be more important. Nonetheless, multisplit was the key building block that made these performance increases possible.

\section{Conclusion}\label{sec:conclusion}
The careful design and analysis of our GPU multisplit implementations allow us to provide significant performance speedups for multisplit operations over traditional sort-based methods. 
The generality of our multisplit algorithm let us extend it further into other interesting applications, such as building a competitive GPU radix sort, getting significant improvements in Single Source Shortest Path problem, and providing a promising GPU histogram mostly suitable for small number of buckets. 
Beyond simply demonstrating the design and implementation of a family of fast and efficient multisplit primitives, we offer three main lessons that are broadly useful for parallel algorithm design and implementation:
Considering a warp-synchronous programming by leveraging warp-wide hardware intrinsics and promoting warp-level privatization of memory, where applicable, can potentially brings interesting and efficient implementations;    
Minimize global (device-wide) operations, even at the cost of increased local computation; the benefit of more coalesced memory accesses outweighs the cost of local reordering.

\begin{acks}
The authors would like to thank Duane Merrill for his valuable comments on CUB and our paper drafts.
Also, thanks to NVIDIA for providing the GPUs that made this research possible.
We appreciate the funding support from UC Lab Fees Research Program Award 12-LR-238449, DFG grant ME 2088/3-1, MADALGO (Center for Massive Data Algorithmics), NSF awards CCF-1017399 and OCI-1032859, Sandia LDRD award \#13-0144, and a 2016--17 NVIDIA Graduate Fellowship.

\end{acks}

\bibliographystyle{ACM-Reference-Format}
\bibliography{all,temp}




\end{document}